\documentclass[paper]{JHEP3}
\usepackage{cite}
\usepackage{amssymb,amsmath}
\usepackage{graphicx}
\graphicspath{{figures/}}

\def\h{\mathrm{H}}
\def\Q2{\left(Q^{2}\right)}
\def\e{\epsilon}
\def\eps{\epsilon}
\def\d{{\rm d}}
\def\JET{J}
\def\Li{\hbox{Li}}
\def\dd{\mathcal{D}}
\def\h{\mathrm{H}}
\def\l({\left(}
\def\r){\right)}
\def\cf{C_{F}}
\def\ca{C_{A}}
\def\nf{N_{F}}
\def\nl{\nonumber\\}

\title{Antenna subtraction at NNLO with hadronic initial states:
initial-final configurations}
\author{Alejandro Daleo, Aude Gehrmann-De Ridder\\Institute for Theoretical Physics, ETH, CH-8093 Z\"urich,
Switzerland}
\author{Thomas Gehrmann, Gionata Luisoni\\Institut f\"ur Theoretische Physik, Universit\"at Z\"urich,
Wintherturerstrasse 190,\\CH-8057 Z\"urich, Switzerland}

\abstract{We extend the antenna subtraction method to include
initial states containing one hadron at NNLO. We present results
for all the necessary subtraction terms,
antenna functions, for the master integrals
required to integrate them over the relevant phase space and
finally for the integrated antennae themselves.
Where applicable, our results are cross-checked against the known
NNLO coefficient functions for deep inelastic scattering processes.}
\preprint{{ZU-TH 17/09}}

\begin{document}
\allowdisplaybreaks

\section{Introduction}

Final states containing hadronic jets are produced at large rates at high
energy particle colliders. Since the distribution of these jets relates
directly to the dynamics of the hard parton-level scattering process,
jet observables can be used to probe elementary particle reactions.
Owing to their large production cross sections, various jet observables
can be measured to a high statistical accuracy, thus making them ideal for
precision studies~\cite{dissertori}: for example three-jet production in
$e^+e^-$ annihilation to determine the strong coupling constant $\alpha_s$
and for
quantum chromodynamics (QCD)
 studies, two-plus-one-jet production in deep inelastic scattering
as a probe of the gluon distribution in the proton and to measure $\alpha_s$,
single jet inclusive production and vector-boson-plus-jet production
at hadron colliders as benchmark reactions, and to measure parton
distributions. Experimental data on these observables often attain an
accuracy of a few per cent or better, such that meaningful precision studies
must rely on theoretical predictions accurate to the same level. In
perturbative QCD, this precision usually requires corrections at
next-to-next-to-leading order (NNLO).

NNLO calculations of observables with $n$ jets in the final state require
several ingredients: the two-loop corrected $n$-parton matrix elements, the
one-loop corrected $(n+1)$-parton matrix elements, and the tree-level
$(n+2)$-parton matrix elements. For most massless jet observables of
phenomenological interest, the two-loop matrix elements have been
computed some time ago~\cite{twol,3jme}, while the other two types
of matrix
elements are usually known from calculations of next-to-leading order (NLO)
corrections to $(n+1)$ jet production~\cite{mcfm,fourj}.

The one-loop $(n+1)$-parton matrix elements
contribute to $n$ jet observables at NNLO if one of the
partons involved becomes unresolved (soft or collinear)~\cite{onelstr}.
In these cases, the infrared
singular parts of the matrix elements need to be extracted and integrated
over the phase space appropriate to the unresolved configuration
to make the infrared pole structure explicit. Methods for the
extraction of soft and collinear limits of one-loop matrix elements are
worked out in detail in the
literature~\cite{onelstr,oneloopsoft,onelstr1,onelstr2,twolstr}.
Likewise, the $(n+2)$-parton matrix elements
contain double real radiation singularities corresponding to two
partons becoming simultaneously soft and/or
collinear~\cite{audenigel,campbell,cg,campbellandother}.
To determine the
contribution to NNLO jet observables from these configurations, one has to
find two-parton subtraction terms which coincide with the full matrix element
and are still sufficiently simple to be integrated analytically in order
to cancel  their  infrared pole structure with the two-loop virtual and
the one-loop single-unresolved contributions. Often starting from systematic
methods for subtraction at NLO~\cite{kunszt,cs,ant,nlosub},
several NNLO subtraction methods have been
proposed in the
literature~\cite{Kosower:2002su,nnlosub2,nnlosub3,nnlosub4,nnlosub5}, and are
worked out
to a varying level of sophistication.

For observables  with partons only in the final state,
an NNLO subtraction formalism, antenna subtraction,
 has been derived in~\cite{ourant}. The antenna subtraction formalism
constructs the subtraction terms from antenna functions. Each
antenna function encapsulates all singular limits due to the
emission of one or two unresolved partons between two colour-connected
hard radiator partons. This construction exploits the  universal factorization
of matrix elements and phase space in all unresolved limits.
The antenna functions are derived
systematically from physical matrix elements~\cite{our2j}. This formalism
has been applied in the derivation of NNLO corrections to three-jet
production in electron-positron annihilation~\cite{our3j,weinzierljet}
and related event shapes~\cite{ourevent,weinzierlevent}, which
were used subsequently in precision determinations of the strong coupling
constant~\cite{ouras,bechernew,davisonwebber,bethkeas,power}.
The formalism
can be extended to include parton showers at higher orders~\cite{antshower},
thereby offering a process-independent matching of fixed-order calculations
and logarithmic resummations~\cite{bechernew,resumall},
 which is done on a case-by-case
basis for individual observables~\cite{gionata} up to now. The formalism
can be extended to include massive fermions~\cite{ritzmann}.
For processes
with initial-state partons, antenna subtraction has been fully worked out
only to NLO so far~\cite{hadant}. In this case, one encounters two new types
of antenna functions, initial-final antenna function with one radiator parton
in the initial state, and initial-initial antenna functions with both
radiator partons in the initial state.
In this work, we derive all NNLO initial-final antenna
functions and perform their integration over the appropriately
factorized phase space. These functions form part of the full set of
antenna functions needed for NNLO calculations of hadron collider processes,
and are,
together with the already known final-final antenna subtraction terms,
 sufficient for NNLO calculations of  jet observables in deeply
inelastic lepton-hadron scattering.

Other approaches to perform NNLO calculations  of exclusive observables with initial
state partons are  the use of sector decomposition and a subtraction method based on the
transverse momentum structure of the final state.  The sector decomposition
algorithm~\cite{secdec}  analytically decomposes both phase space and loop integrals into
their  Laurent expansion in dimensional regularization, and  performs a subsequent
numerical  computation of the coefficients of this expansion. Using this formalism,  NNLO
results were  obtained for Higgs  production~\cite{babishiggs}     and vector boson
production~\cite{babisdy} at hadron colliders. Both reactions  were equally computed
independently using an NNLO subtraction  formalism exploiting the specific transverse
momentum  structure of these observables~\cite{grazzinihiggs}.

This paper is structured as follows: in Section~\ref{sec:ant}, we
construct the subtraction terms required at NNLO
for initial-final configurations with two unresolved partons at
tree-level or one unresolved parton at one loop. They require
two new types of antenna functions: tree-level $2\to 3$  and
one-loop $2\to 2$, each with one parton and one
off-shell neutral current in the initial state. The phase space
mappings relevant to these antenna subtraction terms are
discussed in Section~\ref{sec:map}. The analytic integration of the
new initial-final antenna functions is described in Section~\ref{sec:int},
it proceeds through a reduction to a set of master integrals, which we
computed and collect in Appendix~\ref{app:mi}. The integrated
antenna functions are listed in Sections~\ref{sec:qi} and~\ref{sec:gi} for the
quark-initiated and gluon-initiated antennae. A strong check
on these results is provided by the rederivation of NNLO coefficient functions
for deep inelastic scattering, which we carry out in
Section~\ref{sec:discoeff}. Finally, we conclude with an outlook in
Section~\ref{sec:conc}.

\section{Initial-final antenna subtraction at NNLO}
\label{sec:ant}

Antenna subtraction of initial-final configurations at NLO is derived in detail
in~\cite{hadant}.
Subtraction terms in the case of one hard parton in the initial state
are built in the same fashion as for the final-final case (formula (2.5)
in \cite{ourant}).
We have the following subtraction term
associated to a hard radiator parton $i$ with momentum $p$ in the initial state:
\begin{eqnarray}\label{eq:subif}
&&\d\hat{\sigma}^{S,(if)}(p,r)={\cal N}\sum_{m+1}\d\Phi_{m+1}(k_1,\dots,k_{m+1};p,r)
  \,\frac{1}{S_{m+1}}\nonumber\\
&&\times\sum_{j} X^{0}_{i,jk}
  \left|{\cal M}_m(k_1,\dots,K_{K},\dots,k_{m+1};x p,r)\right|^2\,
  J^{(m)}_{m}(k_1,\dots,K_{K},\dots,k_{m+1})\,.
\end{eqnarray}
The additional momentum $r$ stands for the momentum of the second
incoming particle, for example,
a virtual boson in DIS, or a second incoming parton in a hadronic
collision process.
This contribution has to be appropriately convoluted with
the parton distribution function $f_i$.
The tree antenna $X^{0}_{i,jk}$, depending only on
the original momenta $p$, $k_j$ and $k_k$, contains all the
configurations in which parton $j$ becomes unresolved.
The $m$-parton amplitude depends only on redefined on-shell momenta
$k_1,\dots,K_{K},\dots,$ and on the momentum fraction $x$.
 In the case where the second incoming particle is a parton, there is an additional convolution
with the
parton distribution of parton $r$ and corresponding subtraction terms
associated with it.

The jet function, $J^{(m)}_{m}$, in (\ref{eq:subif}) depends on the
momenta $k_j$ and $k_k$ only through $K_K$. Thus, provided a
suitable factorization of the phase space, one can perform the
integration of the antennae analytically. Due to the hard particle in the
initial state, the factorization of phase space is not as straightforward
as for final-final antennae.

The phase space can be factorized in an $m$-parton phase space
convoluted with a two particle phase space:
\begin{eqnarray}
\d\Phi_{m+1}(k_1,\dots,k_{m+1};p,r)&=&\d\Phi_{m}(k_1,\dots,K_{K},\dots,k_{m+1};x p,r)\nonumber\\
&\times&\frac{Q^2}{2\pi}\d\Phi_{2}(k_j,k_k;p,q)\frac{\d x}{x}\,,
\end{eqnarray}
where $Q^2 = -q^2=-(k_j+k_k-p)^2$.
Replacing the phase space in (\ref{eq:subif}), we can explicitly carry out the integration of the
antenna factors over the two particle phase space. When combining the integrated subtraction terms
with virtual contributions and mass factorization terms, it turns
out to be convenient to normalize the
integrated antennae as follows
\begin{equation}\label{eq:aint}
{\cal X}^0_{i,jk}=\frac{1}{C(\epsilon)}\int \d\Phi_2 \frac{Q^2}{2\pi} X^0_{i,jk}\,,
\end{equation}
where
\begin{equation}
C(\epsilon)=\left({4\pi}\right)^{\epsilon}\,\frac{e^{-\epsilon\gamma_E}}{8\pi^2}\,.
\end{equation}
The integrated form of the subtraction term is then
\begin{eqnarray}
\d\hat{\sigma}^{S,(if)}(p,r)&=&\sum_{m+1}\sum_{j}
\frac{{\cal N}}{S_{m+1}}\int\frac{\d x}{x}\,
{\cal X}^0_{i,jk}(x,Q^2)\,\d\Phi_{m}(k_1,\dots,K_{K},\dots,k_{m+1};x\,p,r)\nonumber\\
&&\times\left|{\cal M}_m(k_1,\dots,K_{K},\dots,k_{m+1};x\,p,r)\right|^2
J^{(m)}_{m}(k_1,\dots,K_{K},\dots,k_{m+1})\,.
\end{eqnarray}
Finally, the subtraction term has to be convoluted with the parton
distribution functions to give the corresponding contribution to the
hadronic cross section. The explicit poles in the integrated form
cancel the corresponding ones in the virtual and
mass factorization
contributions. To carry out the explicit cancellation of poles, it is
convenient to recast, by a simple change of variables, the integrated
subtraction term, once convoluted with the
parton distribution functions (PDFs), in the following form
\begin{eqnarray}
d{\sigma}^{S,(if)}(p,r)&=&\sum_{m+1}\sum_{j}
\frac{S_{m}}{S_{m+1}}\int\frac{\d\xi_1}{\xi_1}\int\frac{\d\xi_2}{\xi_2}\int_{\xi_1}^{1}\frac{\d x}{x}
f_{i/1}\left(\frac{\xi_1}{x}\right)\,f_{b/2}\left(\xi_2\right)\nonumber\\
&&\times C(\epsilon)\,{\cal X}^0_{i,jk}(x)\,\d\hat{\sigma}^B(\xi_1 H_1,\xi_2 H_2)\,.
\end{eqnarray}
This convolution has already the appropriate
structure and mass factorization can be carried out explicitly leaving a finite
contribution. The remaining phase space integration, implicit in
the Born cross section, $\d\hat{\sigma}^B$, and the convolutions can
be safely done numerically. When considering reactions with only one
incoming hadron, the second PDF has to be replaced by a Dirac delta. Reactions
with two hadrons will require additional subtractions containing initial-final
antennae involving the second parton in the initial state and initial-initial
antennae as well. This case is discussed to NLO in~\cite{hadant}.

At NNLO, two types of contributions to $m$-jet observables require
subtraction: the tree-level  $m+2$ parton matrix elements (where one or
two partons can become unresolved), and
the one-loop $m+1$ parton matrix elements (where one parton can become
unresolved). The corresponding
subtraction terms are denoted by  ${\rm d}\sigma^{S}_{NNLO}$
and  $\d \sigma^{VS,1}_{NNLO}$.
Final-final antenna subtraction terms for both cases are
constructed in~\cite{ourant}.

In ${\rm d}\sigma^{S}_{NNLO}$, we have to distinguish four
different types of unresolved configurations:
\begin{itemize}
\item[(a)] One unresolved parton but the experimental observable selects only
$m$ jets;
\item[(b)] Two colour-connected unresolved partons (colour-connected);
\item[(c)] Two unresolved partons that are not colour connected but share a common
radiator (almost colour-unconnected);
\item[(d)] Two unresolved partons that are well separated from each other
in the colour chain (colour-unconnected).
\end{itemize}
Among those, configuration (a) is properly
accounted for by a single tree-level three-parton antenna function
like used already at NLO. Configuration (b) requires a
tree-level four-parton antenna function (two unresolved partons emitted
between a pair of hard partons), while (c) and (d) are accounted for by
products of two tree-level three-parton antenna functions.
With radiator parton $i$ in the initial state,
the subtraction terms for these configurations read:
\begin{eqnarray}
{\rm d}\sigma_{NNLO}^{S,a}
&=&  {\cal N}\,\sum_{m+2}{\rm d}\Phi_{m+2}(k_{1},\ldots,k_{m+2};p,r)
\frac{1}{S_{{m+2}}} \nonumber \\
&\times& \,\Bigg [ \sum_{j}\;X^0_{i,jk}\,
|{\cal M}_{m+1}(k_{1},\ldots,{K}_{K},\ldots,k_{m+2};xp,r)|^2\,
\nonumber \\ &&
\hspace{3cm} \times
\JET_{m}^{(m+1)}(k_{1},\ldots,{K}_{K},\ldots,k_{m+2})\;
\Bigg
]\;,\label{eq:sub2a}\\
{\rm d}\sigma_{NNLO}^{S,b}
&=&  {\cal N}\,\sum_{m+2}{\rm d}\Phi_{m+2}(k_{1},\ldots,k_{m+2};p,r)
\frac{1}{S_{{m+2}}} \nonumber \\
&\times& \,\Bigg [ \sum_{jk}\;\left( X^0_{i,jkl}
- X^0_{i,jk} X^0_{I,Kl} - X^0_{jkl} X^0_{i,JL} \right)\nonumber \\
&\times&
|{\cal M}_{m}(k_{1},\ldots,{K}_{L},\ldots,k_{m+2};xp,r)|^2\,
\JET_{m}^{(m)}(k_{1},\ldots,{K}_{L},\ldots,k_{m+2})\;
\Bigg]\;,
\label{eq:sub2b}
\end{eqnarray}
\begin{eqnarray}
{\rm d}\sigma_{NNLO}^{S,c1}
&= & - {\cal N}\,\sum_{m+2}{\rm d}\Phi_{m+2}(k_{1},\ldots,k_{m+2};p,r)
\frac{1}{S_{{m+2}}} \nonumber \\
&\times& \,\Bigg[  \sum_{j,l}\;X^0_{i,jk}\;x^0_{mlK}\,
|{\cal M}_{m}(k_{1},\ldots,{K}_{K},{K}_{M},\ldots,k_{m+2};xp,r)|^2\,\nonumber \\
&&\hspace{3cm}\times
\JET_{m}^{(m)}(k_{1},\ldots,K_K,{K}_{M},\ldots,k_{m+2})\;
\phantom{\Bigg]}
\nonumber \\
&& \,+ \sum_{j,l}\;X^0_{klm}\;x^0_{i,jK}\,
|{\cal M}_{m}(k_{1},\ldots,K_{K},K_{M},\ldots,k_{m+2});xp,r|^2\,\nonumber \\
&&\hspace{3cm}\times
\JET_{m}^{(m)}(k_{1},\ldots,K_K,K_{M},\ldots,k_{m+2})\;
\Bigg
]\;,
\label{eq:sub2c1}\\
{\rm d}\sigma_{NNLO}^{S,c2}
&= & - {\cal N}\,\sum_{m+2}{\rm d}\Phi_{m+2}(k_{1},\ldots,k_{m+2};p,r)
\frac{1}{S_{{m+2}}} \nonumber \\
&\times& \,\Bigg[  \sum_{j,l}\;X^0_{i,jk}\;x^0_{I,lm}\,
|{\cal M}_{m}(k_{1},\ldots,{K}_{K},{K}_{M},\ldots,k_{m+2};xp,r)|^2\,\nonumber \\
&&\hspace{3cm}\times
\JET_{m}^{(m)}(k_{1},\ldots,K_K,{K}_{M},\ldots,k_{m+2})\;
\phantom{\Bigg]}
\nonumber \\
&& \,+ \sum_{j,l}\;X^0_{i,lm}\;x^0_{I,jk}\,
|{\cal M}_{m}(k_{1},\ldots,K_{K},K_{M},\ldots,k_{m+2});xp,r|^2\,\nonumber \\
&&\hspace{3cm}\times
\JET_{m}^{(m)}(k_{1},\ldots,K_K,K_{M},\ldots,k_{m+2})\;
\Bigg
]\;,
\label{eq:sub2c2}\\
{\rm d}\sigma_{NNLO}^{S,d}
&= & - {\cal N}\,\sum_{m+2}{\rm d}\Phi_{m+2}(k_{1},\ldots,k_{m+2};p,r)
\frac{1}{S_{{m+2}}} \nonumber \\
&\times& \,\Bigg [ \sum_{j,o}\;X^0_{i,jk}\;X^0_{nop}\,
|{\cal M}_{m}(k_{1},\ldots,{K}_{K},\ldots,{K}_{N},{K}_P,\ldots,k_{m+2};xp,r)|^2\,\nonumber \\
&\times&
\JET_{m}^{(m)}(k_{1},\ldots,{K}_{K},\ldots,{K}_{N},{K}_P,\ldots,k_{m+2})\;\Bigg
]\;.
\label{eq:sub2d}
\end{eqnarray}
As before, the original momenta of the $(m+2)$-parton phase space are
denoted by
$j,k,\ldots$, while the combined momenta obtained from a phase space
mapping are labelled by $J,K,\ldots$. Only the combined momenta appear in the
jet function. $x^0_{abc}$ is a three-parton
sub-antenna function containing only  limits where parton $b$ is unresolved
with respect to parton $a$, but not limits where parton $b$ is unresolved
with respect to parton $c$.
${\rm d}\sigma_{NNLO}^{S,c1}$ applies if the common radiator is in the final
state, while ${\rm d}\sigma_{NNLO}^{S,c2}$ applies if the common radiator is
in the initial state. The only genuinely new ingredient here is the
four-parton initial-final antenna function $X^0_{i,jkl}$, which can be
obtained by crossing the corresponding final-final antenna functions,
and has to be integrated analytically over the appropriate
antenna phase space. The resulting integrated antenna function
exploits the phase space factorization
\begin{eqnarray}
\d\Phi_{m+2}(k_1,\dots,k_{m+2};p,r)&=&\d\Phi_{m}(k_1,\dots,K_{L},\dots,k_{m+2};x p,r)\nonumber\\
&\times&\frac{Q^2}{2\pi}\d\Phi_{3}(k_j,k_k,k_l;p,q)\frac{\d x}{x}\,,
\end{eqnarray}
yielding the integrated antenna
\begin{equation}
{\cal X}^0_{i,jkl}=\frac{1}{[C(\epsilon)]^2}\int \d\Phi_3 \frac{Q^2}{2\pi} X^0_{i,jkl}\,.
\end{equation}

In all products of two three-parton antenna functions,
the analytic integration has to be performed only over the outmost antenna
function, yielding the integrated NLO antenna functions in the appropriate
(final-final or initial-final) kinematics.

The
one-loop single unresolved subtraction term $\d \sigma^{VS,1}_{NNLO}$
must account for three types of singular contributions:
\begin{itemize}
\item[(a)] Explicit infrared poles of the virtual one-loop
$(m+1)$ parton matrix element.
\item[(b)] Single unresolved limits of the  virtual one-loop
$(m+1)$ parton matrix element.
\item[(c)] Terms common to both above contributions,
which are oversubtracted.
\end{itemize}

With radiator parton $i$ in the initial state,
the subtraction terms for these configurations read:
\begin{eqnarray}
{\rm d}\sigma_{NNLO}^{VS,1,a}
&=&   {\cal N}\,\sum_{m+1}{\rm d}\Phi_{m+1}(k_{1},\ldots,k_{m+1};p,r)
\frac{1}{S_{{m+1}}} \nonumber \\
&\times& \,\Bigg [
\sum_{ik}\;  - \int \frac{\d x}{x}\,  {\cal X}^0_{i,jk}(x,-t_{ik}) \,
|{\cal M}_{m+1}(k_{1},\ldots,{k}_{k},\ldots,k_{m+1};xp,r)|^2\,
\nonumber \\
&& \hspace{3cm} \times
\JET_{m}^{(m+1)}(k_{1},\ldots,{k}_{k},\ldots,k_{m+1})\;
\Bigg
]\;, \label{eq:subv2a}\\
{\rm d}\sigma_{NNLO}^{VS,1,b}
&= & {\cal N}\,\sum_{m+1}{\rm d}\Phi_{m+1}(k_{1},\ldots,k_{m+1};p,r)
\frac{1}{S_{{m+1}}} \nonumber \\
&\times& \,\sum_{j} \Bigg [X^0_{i,jk}\,
|{\cal M}^1_{m}(k_{1},\ldots,\tilde{K}_{K},\ldots,k_{m+1};xp,r)|^2\,
\JET_{m}^{(m)}(k_{1},\ldots,\tilde{K}_{K},\ldots,k_{m+1})\;
\nonumber \\
&&\phantom{\sum_{j} }+\;X^1_{i,jk}\,
|{\cal M}_{m}(k_{1},\ldots,\tilde{K}_{K},\ldots,k_{m+1};xp,r)|^2\,
\JET_{m}^{(m)}(k_{1},\ldots,\tilde{K}_{K},\ldots,k_{m+1})\;\Bigg
]\;,\nonumber \\
\label{eq:subv2b}\\
{\rm d}\sigma_{NNLO}^{VS,1,c1}
&=&   {\cal N}\,\sum_{m+1}{\rm d}\Phi_{m+1}(k_{1},\ldots,k_{m+1};p,r)
\frac{1}{S_{{m+1}}} \nonumber \\
&\times& \,\Bigg [ \sum_{ik}\; \int \frac{\d x}{x}\,
{\cal X}^0_{i,jk}(x,-t_{ik}) \,
\sum_o X^0_{nop} \,
|{\cal M}_{m}(k_{1},\ldots,k_k,\ldots,K_{N},K_{P},\ldots,k_{m+1};xp,r)|^2\,
\nonumber \\
&& \hspace{3cm} \times
\JET_{m}^{(m)}(k_{1},\ldots,k_k,\ldots,K_{N},K_{P},\ldots,k_{m+1})\;
\Bigg
],\\
{\rm d}\sigma_{NNLO}^{VS,1,c2}
&=&   {\cal N}\,\sum_{m+1}{\rm d}\Phi_{m+1}(k_{1},\ldots,k_{m+1};p,r)
\frac{1}{S_{{m+1}}} \nonumber \\
&\times& \,\Bigg [ \sum_{np}\;
{\cal X}^0_{nop}(s_{np}) \,
\sum_j X^0_{i,jk} \,
|{\cal M}_{m}(k_{1},\ldots,K_K,\ldots,k_{n},k_{p},\ldots,k_{m+1};xp,r)|^2\,
\nonumber \\
&& \hspace{3cm} \times
\JET_{m}^{(m)}(k_{1},\ldots,K_K,\ldots,k_{n},k_{k},\ldots,k_{m+1})\;
\Bigg
],
\label{eq:subv2c1}
\end{eqnarray}
In here, $X^1_{i,jk}$ denotes a one-loop three-parton initial-final
antenna function, which is the only new ingredient. These antenna functions
can be obtained by crossing from their final-final counterparts, listed
in~\cite{ourant}, and have to be integrated over the appropriate
phase space:
\begin{equation}
{\cal X}^1_{i,jk}=\frac{1}{C(\epsilon)}\int \d\Phi_2 \frac{Q^2}{2\pi}
X^1_{i,jk}\,.
\end{equation}
\TABLE[t]{
\renewcommand{\arraystretch}{1.2}
\begin{tabular}{p{3cm}p{4cm}p{4cm}}
\hline
Quark initiated&tree level&one loop\\
\hline
\underline{quark-quark}&&\\
$q\rightarrow gq$&$A^{0}_{q,gq}$&$A^{1}_{q,gq}$, $\tilde{A}^{1}_{q,gq}$,
$\hat{A}^{1}_{q,gq}$\\
$q\rightarrow ggq$&$A^{0}_{q,ggq}$,$\tilde{A}^{0}_{q,ggq}$&\\
$q\rightarrow q^{\prime}\bar{q}^{\prime}q$&
              $B^{0}_{q,q^{\prime}\bar{q}^{\prime}q}$&\\
$q^\prime\rightarrow q\bar{q}q^\prime$&
              $B^{0}_{q^{\prime},q\bar{q}q^{\prime}}$&\\
$q\rightarrow q\bar{q}q$&${C}^{0}_{q,q\bar{q}q}$, ${C}^{0}_{\bar{q},\bar{q}q\bar{q}}$, ${C}^{0}_{\bar{q},q\bar{q}\bar{q}}$&\\[3mm]
\hline
\underline{quark-gluon}&&\\
$q\rightarrow gg$&$D^{0}_{q,gg}$&$D^{1}_{q,gg}$, $\hat{D}^{1}_{q,gg}$\\
$q\rightarrow ggg$&$D^{0}_{q,ggg}$&\\
$q\rightarrow q^{\prime}\bar{q}^{\prime}$&
              $E^{0}_{q,q^{\prime}\bar{q}^{\prime}}$&
              $E^{1}_{q,q^{\prime}\bar{q}^{\prime}}$,
              $\tilde{E}^{1}_{q,q^{\prime}\bar{q}^{\prime}}$,
              $\hat{E}^{1}_{q,q^{\prime}\bar{q}^{\prime}}$\\
$q\rightarrow q^{\prime}\bar{q}^{\prime}g$&
              $E^{0}_{q,q^{\prime}\bar{q}^{\prime}g}$,
              $\tilde{E}^{0}_{q,q^{\prime}\bar{q}^{\prime}g}$&\\
$q^\prime\rightarrow q^\prime {q}$&
              $E^{0}_{q^\prime,{q}^{\prime}q}$&
              $E^{1}_{q^\prime,{q}^{\prime}q}$,
              $\tilde{E}^{1}_{q^\prime,{q}^{\prime}q}$,
              $\hat{E}^{1}_{q^\prime,{q}^{\prime}q}$\\
$q^\prime\rightarrow q^\prime qg$&
              $E^{0}_{q^\prime,q^{\prime}qg}$,
              $\tilde{E}^{0}_{q^\prime,q^{\prime}qg}$&\\[3mm]
\hline
\underline{gluon-gluon}&&\\
$q\rightarrow qg$&
              $G^{0}_{q,qg}$&
              $G^{1}_{q,qg}$, $\tilde{G}^{1}_{q,qg}$, $\hat{G}^{1}_{q,qg}$
              \\
$q\rightarrow qgg$&
              $G^{0}_{q,qgg}$, $\tilde{G}^{0}_{q,qgg}$
              &\\
$q\rightarrow qq^{\prime}\bar{q}^{\prime}$&
              $H^{0}_{q,qq^{\prime}\bar{q}^{\prime}}$
              &\\[3mm]
\hline
\end{tabular}
\caption{List of tree level and one loop antenna functions for
the initial-final configurations with a quark in the initial state.\label{tab:one}}
}

The  subtraction terms $\d \sigma^{S}_{NLO}$,
$\d \sigma^{S}_{NNLO}$
and $\d \sigma^{VS,1}_{NNLO}$ require three different types of
antenna functions corresponding to the different pairs of hard partons
forming the antenna: quark-antiquark, quark-gluon and gluon-gluon antenna
functions.
 We derived these antenna functions~\cite{our2j}
for final-final kinematics in
a systematic manner from physical matrix elements known to possess the
correct limits. For the initial-final kinematics, one parton is crossed into the initial state. Special care has to be taken in these crossings, if we
start from
those final-final antenna functions which contain
more than one quark/antiquark or more than one gluon
in the final state. In the case of
more than one quark/antiquark pair of different flavour (final-final
antenna functions $B_4^0$, $E_3^0$, $E_3^1$-type, $E_4^0$-type), we have to
distinguish the crossing of the primary quark $q$ (which is coupled
to the external current) and the secondary quark $q^{\prime}$ (which is not
coupled to the external current). The identical flavour antenna function
$C_4^0 (1_q,3_q,4_{\bar q},2_{\bar q})$ is constructed
from the interference of
the four-quark amplitudes with the antiquark-momenta interchanged, and
it contains only the $(3_q,4_{\bar q},2_{\bar q})$ triple collinear limit.
Consequently, it is symmetric in the two antiquark momenta, but not in
the two quark momenta, and has thus three different crossings:
$C^0_{q,q\bar q q}$ (either antiquark crossed), $C^0_{\bar q,\bar q q \bar q}$
(quark $(1_q)$ crossed) and  $C^0_{\bar q,q \bar q  \bar q}$ (quark $(3_q)$
crossed).
Crossing one of several gluons into the initial state is unambiguous for
most antenna functions owing to their symmetry properties. The only exception
is the quark-gluon antenna function $D_4^0 (1_q,3_g,4_g,5_g)$, where
gluons $(3_g)$ and $(5_g)$ are colour-connected to the quark, while gluon
$(4_g)$ is not. We thus distinguish two crossings, $D^0_{g,ggq}$ (gluon
 $(3_g)$ or $(5_g)$ crossed) and $D^0_{g',ggq}$ (gluon $(4_g)$ crossed).
We list all NLO and NNLO
initial-final antenna functions with an initial state
quark in Table~\ref{tab:one} and with an initial state gluon in
Table~\ref{tab:two}.

\TABLE[t]{
\renewcommand{\arraystretch}{1.2}
\begin{tabular}{p{3cm}p{4cm}p{4cm}}
\hline
Gluon initiated&tree level&one loop\\
\hline
\underline{quark-quark}&&\\
$g\rightarrow q\bar{q}$&$A^{0}_{g,q\bar{q}}$&$A^{1}_{g,q\bar{q}}$, $\tilde{A}^{1}_{g,q\bar{q}}$,
$\hat{A}^{1}_{g,q\bar{q}}$\\
$g\rightarrow gq\bar{q}$&$A^{0}_{g,gq\bar{q}}$, $\tilde{A}^{0}_{g,gq\bar{q}}$&\\[3mm]
\hline
\underline{quark-gluon}&&\\
$g\rightarrow gg$&$D^{0}_{g,gq}$&$D^{1}_{g,gq}$, $\hat{D}^{1}_{g,gq}$\\
$g\rightarrow ggq$&$D^{0}_{g,ggq}$, ${D}^{0}_{g^\prime,ggq}$&\\
$g\rightarrow qq^{\prime}\bar{q}^{\prime}$&$E^{0}_{g,qq^{\prime}\bar{q}^{\prime}}$,
              $\tilde{E}^{0}_{g,qq^{\prime}\bar{q}^{\prime}}$&\\[3mm]
\hline
\underline{gluon-gluon}&&\\
$g\rightarrow gg$&
              $F^{0}_{g,gg}$&
              $F^{1}_{g,gg}$,$\hat{F}^{1}_{g,gg}$
              \\
$g\rightarrow ggg$&
              $F^{0}_{g,ggg}$&\\
$g\rightarrow q\bar{q}$&
              $G^{0}_{g,q\bar{q}}$&
              $G^{1}_{g,q\bar{q}}$, $\tilde{G}^{1}_{g,q\bar{q}}$, $\hat{G}^{1}_{g,q\bar{q}}$
              \\
$g\rightarrow q\bar{q}g$&
              $G^{0}_{g,q\bar{q}g}$, $\tilde{G}^{0}_{g,q\bar{q}g}$
              &\\[3mm]
\hline
\end{tabular}
\caption{List of tree level and one loop antenna functions for
the initial-final configurations with a gluon in the initial state.\label{tab:two}}
}

It was shown in~\cite{weinzierljet} that these antenna subtraction terms
result in an oversubtraction of large-angle soft gluon radiation. To correct
for this oversubtraction, one introduces the soft antenna function
\begin{equation}
S_{ajc} = 2\frac{s_{ac}}{s_{aj}s_{cj}}\,,
\end{equation}
where $a$ and $c$ label arbitrary hard partons.
Those soft factors are associated with an
antenna phase space mapping $(i,j,k)\to(I,K)$ (final-final) or
$(p,j,k)\to(xp,K)$ (initial-final). In contrast to all previous
subtraction terms, the hard momenta $a$, $c$
do not need to be equal to the hard momenta of partons $i$, $k$ in the
antenna phase space - they can be arbitrary on-shell momenta in the initial
or final state.

If parton $(a)$ is in the initial state, and $(c,i,j,k)$ are in the final state,
the integral of each of the soft antenna function over
the antenna phase space can be written as
\begin{eqnarray}
{\cal S}_{a,c;ik} &=& \int \d \Phi_{X_{ijk}} S_{ajc} \nonumber \\
&=& \left( s_{IK} \right)^{-\e}\;  \frac{\Gamma^2(1-\e)e^{\e \gamma}}
{\Gamma(1-3 \e)}\;
\left(-\frac{2}{\e}\right)
 \left[-\frac{1}{\e} +\ln\left(x_{ac,IK}\right)
+ \e\,\Li_{2}\left(-\frac{1-x_{ac,IK}}{x_{ac,IK}}
\right) \right]\;,\nonumber \\
\end{eqnarray}
where we have defined
\begin{equation}
x_{ac,IK} = \frac{t_{ac}s_{IK}}{(t_{aI}+t_{aK})(s_{cI}+s_{cK})}\;.
\end{equation}
If parton $(i)$ is in the initial state, while $(a,c,j,k)$ are in the final
state, we obtain the following integral:
\begin{eqnarray}
{\cal S}_{ac;i,k} &=& \frac{1}{C(\e)}\,\int \d \Phi_{2_{jk}} \frac{Q^2}{2\pi}
\,S_{ajc} \nonumber \\
&=&  \left( Q^2 \right)^{-\e}\; \frac{\Gamma^2(1-\e)\Gamma(1+e) e^{\e \gamma}}
{\Gamma(1-2 \e)}\, \left(-\frac{2}{\e}\right)\, x^{1+2\e}\, (1-x)^{-1-2\e}
y_{ac,iK}^{-\e}\;,
\end{eqnarray}
where we have defined
\begin{equation}
y_{ac,iK} = \frac{s_{ac}Q^2}{\left(s_{aK}+(1-x)s_{ai}\right)\,
\left(s_{cK}+(1-x)s_{ci}\right)}\;.
\end{equation}

\section{Phase space mappings}
\label{sec:map}
As discussed above, the construction of subtraction terms requires
mapping the original set of momenta onto a reduced set. The
mappings interpolate between the different soft and collinear limits
which the subtraction term regulates. Appropriate mappings for the
initial-final configurations, both for single and double unresolved
configurations have been discussed in~\cite{hadant}, and are only briefly
summarized here.

The proper subtraction of infrared singularities requires that the momentum mapping
satisfy
\begin{eqnarray}
&&x p\rightarrow p\,,\qquad K_{K}\rightarrow k_k\qquad\mbox{when $j$ becomes soft}\,,\nonumber\\
&&x p\rightarrow p\,,\qquad K_{K}\rightarrow k_j+k_k
\qquad\mbox{when $j$ becomes collinear with $k$}\,,\nonumber\\
&&x p\rightarrow p-k_j\,,\qquad K_{K}\rightarrow k_k
\qquad\mbox{when $j$ becomes collinear with $i$}\,.
\end{eqnarray}
In this way, infrared singularities are subtracted locally, except for
angular correlations, {\em before convoluting with the
parton distributions}. That is,
matrix elements and subtraction terms are convoluted together with PDFs.
In addition, the redefined momentum, $K_{K}$, must be on shell and
momentum must be conserved, $p-k_j-k_k=x p-K_{K}$, for the phase space to
factorize as above. This is accomplished by:
\begin{eqnarray}\label{eq:ifmappingNLO}
x&=&\frac{s_{1j}+s_{1k}-s_{jk}}{s_{1j}+s_{1k}}\,,\nonumber \\
K_{K}&=&k_j+k_k-(1-x)p\,,
\end{eqnarray}
where $s_{1j}=(p-k_j)^2$, etc. If parton $j$ becomes soft or collinear to
parton $k$, $x\rightarrow 1$. If parton $j$ becomes collinear with the
initial state parton $i$, $x=1-z$ with $z$ the fraction of
the momentum $p$ carried
by parton $j$.

This mapping is  easily generalized
to deal with more than one parton becoming unresolved.
As explained above, the building blocks for the double real radiation
in the initial-final
situation are colour-ordered
four-parton antenna functions $X_{i,jkl}$, with
one radiator parton $i$ (with momentum $p$)
in the initial state, two unresolved partons $j,k$
and one radiator parton $l$ in the final state:
\begin{eqnarray}\label{eq:ifmappingNNLO}
x&=&\frac{s_{1j}+s_{1k}+s_{1l}-s_{jk}-s_{jl}-s_{kl}}{s_{1j}+s_{1k}+s_{1l}}\,,
\nonumber \\
K_{L}&=&k_j+k_k+k_l-(1-x)p\,,
\end{eqnarray}
where $k_j$, $k_k$ and $k_l$ are the three final-state momenta involved in
the subtraction term.
It satisfies the appropriate limits in all double singular configurations:
\begin{enumerate}
\item $j$ and $k$ soft: $x\rightarrow 1$, $K_L\rightarrow k_l$,
\item $j$ soft and $k_k\parallel k_l$: $x\rightarrow 1$, $K_L\rightarrow k_k+k_l$,
\item $k_j=zp\parallel p$ and $k_k$ soft: $x\rightarrow 1-z$, $K_L\rightarrow
  k_l$,
\item $k_j=zp\parallel p$ and $k_k\parallel k_l$: $x\rightarrow 1-z$, $K_L\rightarrow
  k_k+k_l$,
\item $k_j\parallel k_k\parallel k_l$: $x\rightarrow 1$, $K_L\rightarrow
  k_j+k_k+k_l$,
\item $k_j+k_k=zp\parallel p$: $x\rightarrow 1-z$, $K_L\rightarrow
  k_l$,
\end{enumerate}
where partons $j$ and $k$ can be interchanged in all cases.

The construction of NNLO antenna subtraction terms requires moreover that
all single unresolved limits of the four-parton antenna function
$X_{i,jkl}$
have to be subtracted, (\ref{eq:sub2b}),
such that the resulting subtraction term is active only
in its double unresolved limits. A systematic subtraction of these single
unresolved limits by products of two three-parton antenna functions can be
performed only if the NNLO phase space mapping turns into an NLO phase space
mapping in its single unresolved limits.

In the limits where parton $j$ becomes unresolved, we denote the parameters
of the reduced NLO phase space mapping
(\ref{eq:ifmappingNLO}) by ${x}^\prime$ and $K_L^\prime$.
We find for
(\ref{eq:ifmappingNNLO}):
\begin{enumerate}
\item $j$ becomes soft: $$
x \to
\frac{s_{1k}+s_{1l}-s_{kl}}{s_{1k}+s_{1l}} = {x}^\prime \,,\qquad
K_{L} \to k_k+k_l-(1-x)p = K_L^\prime \,. $$
\item $k_j\parallel k_k$, $k_j+k_k = K_K$: $$
x \to
\frac{s_{1K}+s_{1l}-s_{Kl}}{s_{1K}+s_{1l}} = {x}^\prime \,,\qquad
K_{L} \to k_K+k_l-(1-x)p = K_L^\prime\,. $$
\item $k_j = zp \parallel p$: $$
x \to \frac{(1-z)(s_{1k}+s_{1l})-s_{kl}}{s_{1k}+s_{1l}} = (1-z) x^\prime
\,,\qquad K_{L} \to k_k+k_l-(1-x^\prime)(1-z) p = K_L^\prime.
$$
\end{enumerate}
It can be seen that in the first two limits, the NLO mapping
involves the original incoming momentum $p$, while in the last limit
(initial state collinear emission), it involves the rescaled incoming
momentum $(1-z)p$.
To subtract all three single
unresolved limits of parton $j$ between emitter partons $i$ and $k$
from $X_{i,jkl}$, one needs to subtract from it the
product of two three-parton
antenna functions $X_{i,jk}\cdot X_{I,Kl}$. The phase space mapping
relevant to these terms is the iteration of two NLO phase space
mappings. Analytical integration of terms with this mapping
is required only over the phase space appropriate to the first
antenna function.

Equally, parton $k$ can become unresolved. Expressing the reduced NLO
phase space mapping by  ${x}^{\prime\prime}$ and $K_L^{\prime\prime}$.
We find for
(\ref{eq:ifmappingNNLO}):
\begin{enumerate}
\item $k$ becomes soft: $$
x \to
\frac{s_{1j}+s_{1l}-s_{jl}}{s_{1j}+s_{1l}} = {x}^{\prime\prime} \,,\qquad
K_{L} \to k_j+k_l-(1-x)p = K_L^{\prime\prime} \,. $$
\item $k_k\parallel k_j$, $k_j+k_k = K_K$: $$
x \to
\frac{s_{1K}+s_{1l}-s_{Kl}}{s_{1K}+s_{1l}} = {x}^{\prime\prime} \,,\qquad
K_{L} \to k_K+k_l-(1-x)p = K_L^{\prime\prime}\,. $$
\item $k_k\parallel k_l$, $k_l+k_k = K_K$: $$
x \to
\frac{s_{1K}+s_{1j}-s_{Kj}}{s_{1K}+s_{1j}} = {x}^{\prime\prime} \,,\qquad
K_{L} \to k_K+k_j-(1-x)p = K_L^{\prime\prime}\,. $$
\end{enumerate}
In all limits, the reduced NLO mapping involves the original incoming momentum
$p$. Consequently, the three single
unresolved limits of parton $k$ between emitter partons $j$ and $l$
can be subtracted from
$X_{i,jkl}$
by a product of a final-final and an initial-final three-parton
antenna function $X_{jkl}\cdot X_{i,JL}$. The phase space mapping
relevant to these terms is the product of an NLO final-final phase space
mapping with an initial-final mapping. Integration of the
 final-final antenna phase space yields a constant, not involving an extra
convolution.

\section{Integration of initial-final antenna functions at NNLO}
\label{sec:int}
The initial-final antenna functions all have the scattering kinematics
\begin{displaymath}
q + p_i \to p_1+p_2 (+p_3)\;,
\end{displaymath}
where
\begin{displaymath}
q^2 = -Q^2 <0\,, \quad p_i^2=0\,, z=\frac{Q^2}{2\,q\cdot p_i} \,,
\quad p_1^2=p_2^2=p_3^2 =0\,,
\end{displaymath}
and $p_3$ is present only for the NNLO real radiation antenna functions.
Consequently, integration over the final-state two-parton or three-parton
phase space yields a result which depends only on $Q^2$ and $z$. From
dimensional counting, one can immediately conclude that the
dependence on $Q^2$ is only multiplicative, according to the mass
dimension of the integral.

The NNLO double
real radiation antenna functions $X^0_{i,jkl}$
have to be integrated over the
inclusive three-parton final state phase space.  The NNLO one-loop
single real radiation antenna functions $X^1_{i,jk}$ are integrated over the
inclusive two-parton final state phase space, and over the loop momentum.
For both types of integration, we employ the by-now standard technique
of reduction to master integrals. The master integrals are then
computed from their differential equations.

To perform the reduction, we first express all phase space integrals as
loop integrals with cut propagators~\cite{babis}. Consequently, all NNLO
integrals are expressed as cuts of two-loop four-point functions with two
off-shell legs in forward scattering kinematics. Using integration-by-parts
(IBP,~\cite{chet}) and Lorentz invariance (LI,~\cite{gr}) identities among
the integrals of any given topology, the large  number of different
integrals can be expressed in terms of a small  number of master-integrals.
This reduction is performed iteratively, based on the lexicographic ordering
of the integrals, expressed by the Laporta algorithm~\cite{laporta}.

After carrying out the reduction, one finds nine master integrals
for the NNLO double real radiation antenna functions, described in
Section~\ref{sec:realmasters} below, and listed in
Appendix~\ref{app:realmasters}. For the NNLO one-loop single real radiation
antennae, one finds six master integrals, which we describe in
Section~\ref{sec:virtmasters} and list in Appendix~\ref{app:virtmasters}.

We computed
these integrals both directly and  by
using the differential equation
technique~\cite{gr,kotikov,remiddi}.
 To derive the differential equations for each master integral,
we employ
\begin{eqnarray}
Q^2\frac{\partial }{\partial Q^2} &=& \frac{1}{2}\,
q^\mu\frac{\partial }{\partial q^\mu} + \frac{1}{2}\,
p_i^\mu\frac{\partial }{\partial p_i^\mu}\,,\\
z\frac{\partial }{\partial z} &=& -p_i^\mu\frac{\partial }{\partial p_i^\mu}\
\end{eqnarray}
to carry out the differentiations at the integrand level.
The boundary conditions required for the solution of the differential
equations are either obtained from self-consistency conditions on the
integrals, or by explicit evaluation. The explicit evaluation
is very similar to the evaluation
of inclusive four-point phase space integrals, described in~\cite{ourphase}.

Some of the more involved  master integrals can be related to phase space
integrals computed by Zijlstra and van Neerven in the context of
the NNLO corrections to deep inelastic structure
functions~\cite{zv}.
Where appropriate, we compared our results to the expressions in the
appendix of~\cite{zv}, finding full agreement.
Explicit expressions for all master integrals are listed in
Appendix~\ref{app:realmasters}--\ref{app:virtmasters}.

All master integrals contain multiplicative factors of the form
$(1-z)^{-\e}$ or $(1-z)^{-2\e}$, which regulate soft endpoint singularities
in initial state convolution integrals. These factors must be left as such
in the master integrals, and can be expanded in the form of distributions
\begin{equation}
(1-z)^{-1-\e}\,=\, -\frac{1}{\e} \, \delta(1-z) + \sum_{n}
\frac{(-\e)^{n}}{n!}\mathcal{D}_{n}(z)\,,
\end{equation}
with
$$\mathcal{D}_{n}(z)=\l(\frac{\ln^{n}\l(1-z\r)}{1-z}\r)_{+}\,,$$
only after being inserted into the integrated antenna functions.
All other terms in the master integrals can be expanded, yielding
Harmonic Polylogarithms (HPLs,\cite{hpl}) of argument $z$.

\subsection{Tree-level $2\to 3$ antenna functions}
\label{sec:realmasters}

For the tree-level $2\to 3$ antenna functions, we have the 'DIS-like'
process
$$ q+p_i \to p_1+p_2+p_3\;.$$
There are 12 propagators, including the three that are cut
in the phase space integration ($D_{10}$, $D_{11}$, $D_{12}$):
\begin{eqnarray}
D_1 &=& (p_i-p_1)^2\;, \nonumber \\
D_2 &=& (q-p_1)^2\;, \nonumber \\
D_3 &=& (p_2+p_3)^2\;, \nonumber \\
D_4 &=& (p_i-p_2)^2\;, \nonumber \\
D_5 &=& (q-p_2)^2 \;,\nonumber \\
D_6 &=& (p_1+p_3)^2\;, \nonumber \\
D_7 &=& (p_i-p_3)^2\;, \nonumber \\
D_8 &=& (q-p_3)^2\;, \nonumber \\
D_9 &=& (p_1+p_2)^2\;, \nonumber \\
D_{10} &=& p_1^2\;, \nonumber \\
D_{11} &=& p_2^2\;, \nonumber \\
D_{12} &=& p_3^2 \;.
\end{eqnarray}
To perform the reduction to master integrals, we impose momentum conservation
$p_3 = q+p_i -p_1-p_2$, set $p_i^2 = 0$, $q^2 = -Q^2$ and drop any integral
where $D_{10}$, $D_{11}$ and $D_{12}$ are not in the denominator. After
labelling the inclusive phase space integral as $I[0]$, the convention
for naming the master integrals follows the labelling of the numerators, i.e.\
\begin{equation}
I[1,2,5] = \int\,
\frac{[\d p_1]\,[\d p_2]\,[\d p_3]}{D_1\, D_2\, D_5} \,
\delta^d(q+p_i - p_1-p_2 -p_3)\;,
\end{equation}
where
\begin{equation}
[ \d p ]  =  \frac{\d^d p}{(2\pi)^d}  \delta^+ (p^2)\;.
\end{equation}

When squaring the $2\rightarrow 3$ antennae, we find at
most 4 propagators, plus the
3 cut ones. All the integrals can be reduced to the set of
9 master integrals shown in Figure~\ref{fig:rmasters}.
\FIGURE[t]{
\includegraphics[width=0.60\textwidth]{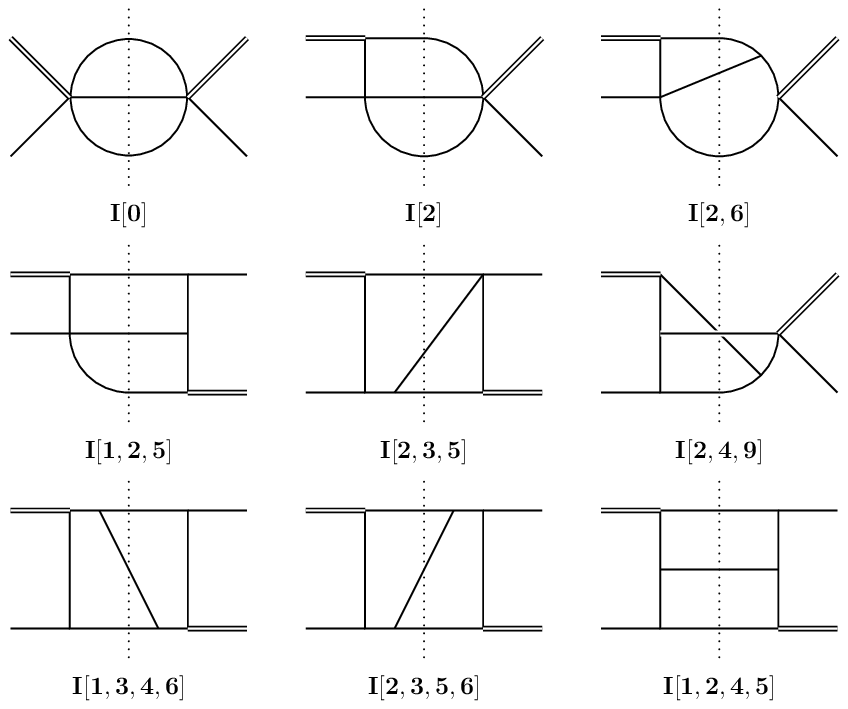}
\caption{Master integrals for the phase space integration of the
tree level initial-final antennae at NNLO. The double line in the
external states represents the off-shell momentum, $q$ with $q^2=-Q^2$,
the single one is the incoming parton. All internal lines are massless. The
cut propagators are the ones intersected by the dotted line.\label{fig:rmasters}}}
\begin{table}[t]
\[
\begin{array}{c|c|c|c}
\text{master}&\text{deepest pole}&\text{behaviour at}\,x=1&\text{known to order}\\
\hline
I[0]&\eps^0&(1-x)^{1-2\eps}&\text{all}\\
I[2]&\eps^0&(1-x)^{-2\eps}&\text{all},\,\eps^5\\
I[2,6]&\eps^{-1}&(1-x)^{-2\eps}&\text{all},\,\eps^5\\
I[1,2,5]&\eps^{-2}&(1-x)^{-2\eps}&\eps^3\\
I[2,3,5]&\eps^{-2}&(1-x)^{-1-2\eps}&\eps^3\\
I[2,4,9]&\eps^{-3}&(1-x)^{-2\eps}&\eps^3\\
I[1,3,4,6]&\eps^{-3}&(1-x)^{-1-2\eps}&\text{all},\,\eps^5\\
I[2,3,5,6]&\eps^{-3}&(1-x)^{-1-2\eps}&\eps^1\\
I[1,2,4,5]&\eps^{-2}&(1-x)^{-2\eps}&\eps^1
\end{array}
\]
\caption{Summary of the main properties of the three particles phase space master integrals}
\label{tab:mastersR}
\end{table}
All the masters,
except $I[1,2,4,5]$, have been computed by direct integration and by the differential
equations method, supplemented, where necessary,
 by a direct calculation at $x=1$ after
factorizing the leading singularity.
The $x=1$ boundary conditions for $I[1,3,4,6]$ and
$I[2,3,5,6]$ were checked numerically using sector
decomposition~\cite{secdec}.
$I[1,2,4,5]$ has been computed only using the differential equations method.
The master
integrals $I[1,2,5]$, $I[2,3,5]$, $I[2,3,5,6]$ and $I[1,2,4,5]$ agree up
to order $\epsilon^0$ with the results in~\cite{zv}.  We summarize
in Table~\ref{tab:mastersR} some of the properties of the master integrals.

\subsection{One-loop $2\to 2$ antenna functions}
\label{sec:virtmasters}
When interfering the one loop $2\rightarrow 2$ antennae with the corresponding
tree level ones, we can combine the loop and phase space integrations. The
partonic process in this case is
$$
q+p_i \to p_1+p_2\;.
$$
Denoting the loop momentum by $k$, we can identify four topologies, two planar
(Topology 1 and 2) and two non-planar (Topology 3 and 4). The topologies
are defined in Table~\ref{tab:virttopo}. Topologies 1 and 2 only differ
in the propagator $D_5$, as is also the case for Topologies 3 and 4.
Subtopologies of  the non-planar integrals can be expressed by the
planar topologies.
 \begin{table}[t]
\[
\begin{array}{c|c|c|c|c}
& \text{Topology 1}&\text{Topology 2}&\text{Topology 3} &\text{Topology 4}\\
\hline
D_1 & k^2 & k^2 & k^2 & k^2 \\
D_2 & (k+p_i)^2 & (k+p_i)^2 & (k+p_i)^2 & (k+p_i)^2 \\
D_3 & (k+p_i-p_1)^2 & (k+p_i-p_1)^2 & (k+p_i-p_1)^2 & (k+p_i-p_1)^2 \\
D_4 & (k-q)^2 & (k-q)^2 & (k+p_2)^2 & (k+p_2)^2 \\
D_5 & (p_i-p_i)^2 & (q-p_1)^2 & (p_i-p_1)^2 & (q-p_1)^2 \\
D_6 & p_1^2 & p_1^2 & p_1^2 & p_1^2 \\
D_7 & p_2^2 & p_2^2 & p_2^2 & p_2^2
\end{array}
\]
\caption{Definition of the topologies for the combined phase space
and loop integration.}
\label{tab:virttopo}
\end{table}

\FIGURE[t]{
\includegraphics[width=0.60\textwidth]{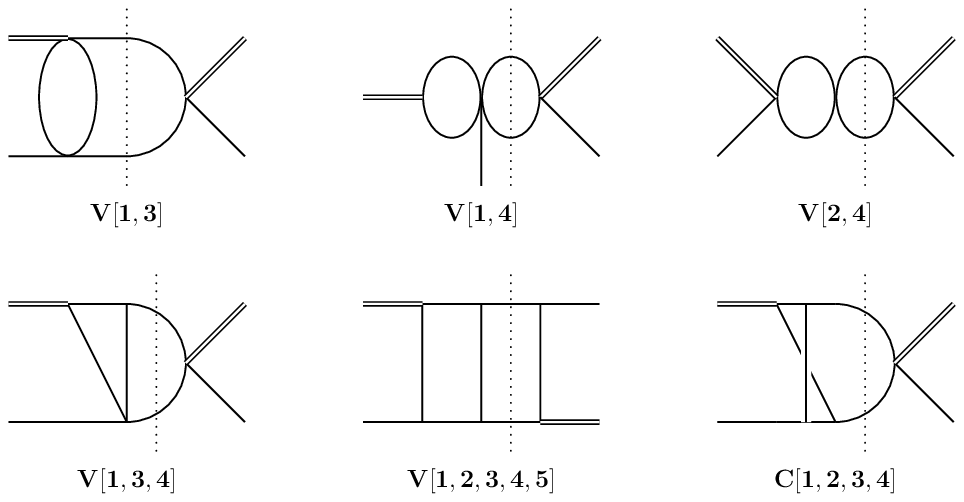}
\caption{Master integrals for the loop plus phase space integration of the
one loop initial-final antennae at NNLO. The double line in the
external states represents the off-shell momentum, $q$ with $q^2=-Q^2$,
the single one is the incoming parton. All internal lines are massless. The
cut propagators are the ones intersected by the dotted line.}
\label{fig:Vmasters}}
All the
resulting integrals can again be reduced to a small set of masters. In this
case we find only six of them. They are shown in Figure~\ref{fig:Vmasters}.
The notation for the master integrals follows the propagator definitions
of Topology 2 in the planar case (denoted by $V$)
and of Topology 3 in the non-planar case (denoted by $C$).

The one loop plus phase-space master integrals have been computed using
differential equations in external invariants, together with a direct
calculation at $x=1$ after factorizing the leading singularity at $x=1$.
Integrals $V[1,2,3,4]$, $V[1,2,3,4,5]$ and $C[1,2,3,4]$
have been checked, up to order $\epsilon^0$, against a direct
analytic calculation.
We summarize
in table~\ref{tab:mastersV} some of the properties of the master integrals.
Explicit expressions for the master integrals are listed in
Appendix~\ref{app:virtmasters}.
\begin{table}[t]
\[
\begin{array}{c|c|c|c}
\text{master}&\text{deepest pole}&\text{behaviour at}\,x=1&\text{known to order}\\
\hline
V[1,3]&\eps^{-1}&(1-x)^{-\eps}&\text{all},\,\eps^5\\
V[1,4]&\eps^{-1}&(1-x)^{-\eps}&\text{all},\,\eps^5\\
V[2,4]&\eps^{-1}&(1-x)^{-2\eps}&\text{all},\,\eps^5\\
V[1,3,4]&\eps^{-1}&(1-x)^{-\eps}&\eps^3\\
V[1,2,3,4,5]&\eps^{-3}&(1-x)^{-1-2\eps}&\eps^3\\
C[1,2,3,4]&\eps^{-3}&(1-x)^{-\eps}&\eps^3
\end{array}
\]
\caption{Summary of the main properties of the one loop plus two particles phase
space master integrals}
\label{tab:mastersV}
\end{table}

We always integrate the unrenormalized $2\to 2$ one-loop squared matrix
elements, divided by a normalization factor $C(\e)$, relevant to
a particular antenna function, which we denote as $X^{1,U}_{i,jk} $. The antenna
function is obtained after renormalization and subtraction of the corresponding
tree-level antenna function multiplied with the one-loop correction to the hard
radiator pair.
Renormalization of the one-loop antenna functions is always carried out in the $\overline{{\rm MS}}$-scheme
at fixed renormalization scale $\mu^2=Q^2$. It amounts to a renormalization of the strong
coupling constant and (in the case of the quark-gluon and gluon-gluon antenna functions)
to a renormalization of the effective operators used to couple an external current to
the partonic radiators. The relation between renormalized
and unrenormalized one-loop squared matrix elements is as follows:
\begin{eqnarray}
X^{1,R}_{i,jk} &=& X^{1,U}_{i,jk} -  \frac{b_0}{\e}\, X^0_{i,jk} - \frac{\eta_0}{\e}\, X^0_{i,jk}
\label{eq:renorm1}\,,\\
\tilde{X}^{1,R}_{i,jk} &=& \tilde{X}^{1,U}_{i,jk} \label{eq:renorm2}\,, \\
\hat{X}^{1,R}_{i,jk} &=& X^{1,U}_{i,jk} -  \frac{b_{0,F}}{\e}\, X^0_{i,jk} - \frac{\eta_{0,F}}{\e}\, X^0_{i,jk}\,,
\end{eqnarray}
where
\begin{equation}
b_0 = \frac{11}{6}\;, \quad b_{0,F}= -\frac{1}{3}
\end{equation}
are the colour-ordered coefficients of the one-loop QCD $\beta$-function:
\begin{equation}
\beta_0 = b_0 N + b_{0,F} \nf\,.
\end{equation}
The renormalization constants for the effective operators are
\begin{eqnarray*}
 \eta_0 = 0\,, \quad \eta_{0,F}=0 \quad && \mbox{for $X=A$}\,,\\
 \eta_0 = b_0+\frac{3}{2} \,, \quad \eta_{0,F}=b_{0,F} \quad && \mbox{for $X=D,E$}\,,\\
 \eta_0 = 2\,b_0\,, \quad \eta_{0,F}=2\, b_{0,F} \quad && \mbox{for $X=F,G$}\,.
 \end{eqnarray*}

The one-loop antenna functions are obtained from the renormalized
one-loop  squared matrix elements by subtracting from them the product of
the tree-level antenna function with the
the virtual one-loop hard radiator vertex
correction~\cite{onelstr,ourant}:
\begin{eqnarray}
X^1_{i,jk} &=& X^{1,R}_{i,jk}
- \mbox{Re}\left((-1)^{-\e} {\cal X}_2^1\right)\, X^0_{i,jk} \,,\\
\tilde{X}^1_{i,jk} &=& \tilde{X}^{1,R}_{i,jk}  - \mbox{Re}\left((-1)^{-\e} \tilde{{\cal X}}_2^1\right)\,
X^0_{i,jk}\,, \\
\hat{X}^1_{i,jk} &=& X^{1,U}_{i,jk}
- \mbox{Re}\left((-1)^{-\e} \hat{{\cal X}}_2^1\right)\, X^0_{i,jk} \,.
\end{eqnarray}
 The one-loop corrections to the hard radiator vertex
are listed in~\cite{ourant}
 for ${\cal A}_2^1$, ${\cal D}_2^1$, $\hat{\cal D}_2^1$, ${\cal F}_2^1$
 and  $\hat{\cal F}_2^1$.
 From these, the remaining functions follow:
 \begin{eqnarray}
 \tilde{\cal A}_2^1&=& {\cal A}_2^1\,, \qquad \hat{\cal A}_2^1=0\,,\\
 \tilde{\cal D}_2^1&=& 0\,,\\
 {\cal E}_2^1 &=& {\cal D}_2^1\,\qquad
   \tilde{\cal E}_2^1= 0\,, \qquad \hat{\cal E}_2^1=\hat{\cal D}_2^1\,,\\
 \tilde{\cal F}_2^1&=& 0\,,\\
 {\cal G}_2^1 &=& {\cal F}_2^1\,\qquad
   \tilde{\cal G}_2^1= 0\,, \qquad \hat{\cal G}_2^1=\hat{\cal F}_2^1\,.
  \end{eqnarray}

\section{Quark initiated antennae}
\label{sec:qi}
The quark-initiated initial-final
antenna functions are obtained from the final-final antenna functions by
crossing a quark into the initial state. Their unintegrated analytical
expressions are obtained by a pure kinematical crossing from the
final-final expressions listed in~\cite{ourant}, with
no extra symmetry factors or overall signs applied, and
no further decompositions into sub-antennae. This procedure differs
from the approach used in our previous work on initial-final antenna functions
at NLO~\cite{hadant}. In the application of the antenna subtraction
formalism to the calculation of NNLO corrections to
$e^+e^-\to 3j$~\cite{our3j}, it
turned out that several decompositions can be circumvented by an appropriate
symmetrization of the real radiation matrix elements in the process under
consideration, and that the remaining required decompositions can be
derived on a case-by-case basis, often not along the lines
of~\cite{ourant,hadant}. A systematic decomposition would start from
a specific process under consideration, and is beyond the scope of the
present study.

The virtual one-loop antenna functions are obtained from the
final-final one-loop antenna functions~\cite{ourant}, where
some attention has to be paid to the correct analytic continuation
of the polylogarithmic functions. In the initial-final kinematics,
the relevant final state phase space consists of three different Riemann
sheets~\cite{allMC,distensor}, which have to be patched together
correctly.

All quark-initiated initial-final antenna functions are listed in
Table~\ref{tab:one}. We summarize their integrated forms in the following.

\subsection{Quark-quark antennae}
The next-to-leading order initial-final antenna functions were already
integrated in~\cite{hadant} through to finite terms
of order $\e^0$. Applying these antenna functions in an NNLO
calculation, one also needs their subleading terms through to $\e^2$.
To this order, the quark-initiated quark-antiquark antenna function is:
\begin{align}
\lefteqn{\mathcal{A}^{0}_{q,gq}=}\nl
+&\frac{1}{\eps^2}\Bigl[\delta(1-z)\Bigr]\nl
+&\frac{1}{\eps}\Bigl[\frac{1}{2}+\frac{3}{4} \delta(1-z)-\dd_{0}(z)+\frac{z}{2}\Bigr]\nl
+&\Bigl[\frac{3}{2}+\delta(1-z) \bigl(\frac{7}{4}-\frac{\pi ^2}{4}\bigr)-\frac{3}{4} \dd_{0}(z)+\dd_{1}(z)+\frac{1}{2} \h(0;z)-\frac{1}{1-z}\h(0;z)+\frac{1}{2} \h(1;z)\nl
&\quad+z \bigl(-\frac{1}{2}+\frac{1}{2} \h(0;z)+\frac{1}{2} \h(1;z)\bigr)\Bigr]\nl
+&\eps\Bigl[3-\frac{\pi ^2}{8}+\delta(1-z) \bigl(\frac{7}{2}-\frac{3 \pi ^2}{16}-\frac{7 \zeta_{3}}{3}\bigr)+\bigl(-\frac{7}{4}+\frac{\pi ^2}{4}\bigr) \dd_{0}(z)+\frac{3}{4} \dd_{1}(z)-\frac{1}{2} \dd_{2}(z)
+\frac{3}{2} \h(0;z)\nl
&\quad+\frac{3}{2} \h(1;z)+\frac{1}{2} \h(0,0;z)+\frac{1}{2} \h(0,1;z)+\frac{1}{2} \h(1,0;z)+\frac{1}{2} \h(1,1;z)+\frac{1}{1-z}\bigl(-\frac{3}{4} \h(0;z)\nl
&\quad-\h(0,0;z)-\h(0,1;z)-\h(1,0;z)\bigr)+z \bigl(-\frac{\pi ^2}{8}-\frac{1}{2} \h(0;z)-\frac{1}{2} \h(1;z)+\frac{1}{2} \h(0,0;z)\nl
&\quad+\frac{1}{2} \h(0,1;z)+\frac{1}{2} \h(1,0;z)+\frac{1}{2} \h(1,1;z)\bigr)\Bigr]\nl
+&\eps^{2}\Bigl[6-\frac{3 \pi ^2}{8}-\frac{7 \zeta_{3}}{6}+\delta(1-z) \bigl(7-\frac{7 \pi ^2}{16}-\frac{\pi ^4}{96}-\frac{7 \zeta_{3}}{4}\bigr)+\dd_{0}(z) \bigl(-\frac{7}{2}+\frac{3 \pi ^2}{16}
+\frac{7\zeta_{3}}{3}\bigr)\nl
&\quad+\bigl(\frac{7}{4}-\frac{\pi ^2}{4}\bigr) \dd_{1}(z)-\frac{3}{8} \dd_{2}(z)+\frac{1}{6} \dd_{3}(z)+3 \h(0;z)-\frac{1}{8} \pi ^2 \h(0;z)+3 \h(1;z)-\frac{1}{8} \pi ^2 \h(1;z)\nl
&\quad+\frac{3}{2} \h(0,0;z)+\frac{3}{2} \h(0,1;z)+\frac{3}{2} \h(1,0;z)+\frac{3}{2} \h(1,1;z)+\frac{1}{2} \h(0,0,0;z)+\frac{1}{2} \h(0,0,1;z)\nl
&\quad+\frac{1}{2} \h(0,1,0;z)+\frac{1}{2} \h(0,1,1;z)+\frac{1}{2} \h(1,0,0;z)+\frac{1}{2} \h(1,0,1;z)+\frac{1}{2} \h(1,1,0;z)\nl
&\quad+\frac{1}{2} \h(1,1,1;z)+\frac{1}{1-z}\bigl(-\frac{7}{4} \h(0;z)+\frac{1}{4} \pi ^2 \h(0;z)-\frac{3}{4} \h(0,0;z)-\frac{3}{4} \h(0,1;z)\nl
&\quad-\frac{3}{4} \h(1,0;z)-\h(0,0,0;z)-\h(0,0,1;z)-\h(0,1,0;z)-\h(0,1,1;z)-\h(1,0,0;z)\nl
&\quad-\h(1,0,1;z)-\h(1,1,0;z)\bigr)+z \bigl(\frac{\pi ^2}{8}-\frac{1}{8} \pi ^2 \h(0;z)-\frac{1}{8} \pi ^2 \h(1;z)-\frac{1}{2} \h(0,0;z)\nl
&\quad-\frac{1}{2} \h(0,1;z)-\frac{1}{2} \h(1,0;z)-\frac{1}{2} \h(1,1;z)+\frac{1}{2} \h(0,0,0;z)+\frac{1}{2} \h(0,0,1;z)+\frac{1}{2} \h(0,1,0;z)\nl
&\quad+\frac{1}{2} \h(0,1,1;z)+\frac{1}{2} \h(1,0,0;z)+\frac{1}{2} \h(1,0,1;z)+\frac{1}{2} \h(1,1,0;z)+\frac{1}{2} \h(1,1,1;z)-\frac{7 \zeta_{3}}{6}\bigr)\Bigr]\nl
&+\mathcal{O}\left(\epsilon^{3}\right)\,.
\end{align}
At NNLO, the quark-initiated initial-final
quark-antiquark double real radiation antenna are the
crossings of the corresponding final-final quark-antiquark antenna functions.
Depending on the
unresolved particles, they are of the $A_4^0$-type (two unresolved gluons),
$B_4^0$-type (unresolved secondary quark pair of different flavour) and
$C_4^0$-type (pure interference contribution for secondary quark pair
of identical flavour).

The integrated form of the initial-final $(q,ggq)$ antenna function reads
at leading colour:
\begin{align}
\lefteqn{\mathcal{A}^{0}_{q,ggq}=}\nl
+&\frac{1}{\eps^4}\Bigl[3 \delta(1-z)\Bigr]\nl
+&\frac{1}{\eps^3}\Bigl[3+\frac{19}{3} \delta(1-z)-6 \dd_{0}(z)+3 z\Bigr]\nl
+&\frac{1}{\eps^2}\Bigl[\frac{40}{3}+\delta(1-z) \bigl(\frac{1429}{72}-\frac{7 \pi^2}{3}\bigr)-\frac{38}{3} \dd_{0}(z)+12 \dd_{1}(z)+\frac{13}{2}\h(0;z)+6 \h(1;z)-\frac{12}{1-z} \h(0;z)\nl
&\quad+z \bigl(\frac{4}{3}+\frac{13}{2} \h(0;z)+6 \h(1;z)\bigr)\Bigr]\nl
+&\frac{1}{\eps}\Bigl[\frac{865}{18}-\frac{7 \pi^2}{3}+\delta(1-z) \bigl(\frac{23959}{432}-\frac{181 \pi^2}{36}-25 \zeta_{3}\bigr)+\bigl(-\frac{1429}{36}+\frac{14 \pi^2}{3}\bigr) \dd_{0}(z)
+\frac{76}{3} \dd_{1}(z)-12\dd_{2}(z)\nl
&\quad+\frac{61}{2} \h(0;z)+\frac{80}{3} \h(1;z)+\frac{29}{2} \h(0,0;z)+13 \h(0,1;z)+13 \h(1,0;z)+12 \h(1,1;z)\nl
&\quad+\frac{1}{1-z}\bigl(-\frac{\pi^2}{3}-\frac{55}{2}\h(0;z)-26 \h(0,0;z)-24 \h(0,1;z)-26 \h(1,0;z)\bigr)+z\bigl(\frac{38}{9}-\frac{7 \pi^2}{3}\nl
&\quad+\frac{9}{2} \h(0;z)+\frac{8}{3} \h(1;z)+\frac{29}{2} \h(0,0;z)+13 \h(0,1;z)+13\h(1,0;z)+12 \h(1,1;z)\bigr)\Bigr]\nl
+&\Bigl[\frac{3997}{27}-\frac{217 \pi^2}{18}-31 \zeta_{3}+\delta(1-z) \bigl(\frac{389623}{2592}-\frac{6857 \pi^2}{432}+\frac{163 \pi^4}{360}-\frac{1027\zeta_{3}}{18}\bigr)
+\dd_{0}(z) \bigl(-\frac{23959}{216}\nl
&\quad+\frac{181 \pi^2}{18}+50\zeta_{3}\bigr)+\bigl(\frac{1429}{18}-\frac{28 \pi^2}{3}\bigr) \dd_{1}(z)-\frac{76}{3} \dd_{2}(z)+8\dd_{3}(z)+\frac{653}{6} \h(0;z)-\frac{71}{12} \pi^2 \h(0;z)\nl
&\quad+\frac{865}{9} \h(1;z)-\frac{14}{3} \pi^2\h(1;z)+\frac{395}{6} \h(0,0;z)+61 \h(0,1;z)+\frac{163}{3} \h(1,0;z)\nl
&\quad+\frac{160}{3} \h(1,1;z)+\frac{61}{2}\h(0,0,0;z)+29 \h(0,0,1;z)+26 \h(0,1,0;z)+26 \h(0,1,1;z)\nl
&\quad+29 \h(1,0,0;z)+26 \h(1,0,1;z)+26\h(1,1,0;z)+24 \h(1,1,1;z)+z \bigl(\frac{220}{27}-\frac{25 \pi^2}{18}\nl
&\quad-31 \zeta_{3}+\frac{29}{3} \h(0;z)-\frac{71}{12} \pi^2 \h(0;z)+\frac{76}{9}\h(1;z)-\frac{14}{3} \pi^2 \h(1;z)+\frac{71}{6} \h(0,0;z)\nl
&\quad+9 \h(0,1;z)+\frac{19}{3} \h(1,0;z)+\frac{16}{3}\h(1,1;z)+\frac{61}{2} \h(0,0,0;z)+29 \h(0,0,1;z)\nl
&\quad+26 \h(0,1,0;z)+26 \h(0,1,1;z)+29 \h(1,0,0;z)+26\h(1,0,1;z)+26 \h(1,1,0;z)\nl
&\quad+24 \h(1,1,1;z)\bigr)+\frac{1}{1-z}\bigl(\frac{2\pi^2}{9}+10 \zeta_{3}-\frac{274}{3} \h(0;z)+\frac{32}{3} \pi^2 \h(0;z)-\frac{361}{6} \h(0,0;z)\nl
&\quad-55 \h(0,1;z)-\frac{161}{3}\h(1,0;z)-54 \h(0,0,0;z)-52 \h(0,0,1;z)-48 \h(0,1,0;z)\nl
&\quad-48 \h(0,1,1;z)-58 \h(1,0,0;z)-52\h(1,0,1;z)-52 \h(1,1,0;z)\bigr)\Bigr]\nl
&+\mathcal{O}\left(\epsilon\right)
\end{align}
and at subleading colour
\begin{align}
\lefteqn{\mathcal{\tilde{A}}^{0}_{q,ggq}=}\nl
+&\frac{1}{\eps^4}\Bigl[4 \delta(1-z)\Bigr]\nl
+&\frac{1}{\eps^3}\Bigl[4+6 \delta(1-z)-8 \dd_{0}(z)+4 z\Bigr]\nl
+&\frac{1}{\eps^2}\Bigl[13+\delta(1-z) \bigl(\frac{75}{4}-\frac{10 \pi^2}{3}\bigr)-12 \dd_{0}(z)+16 \dd_{1}(z)+9 \h(0;z)+8\h(1;z)-\frac{16}{1-z} \h(0;z)\nl
&\quad+z \bigl(1+9 \h(0;z)+8 \h(1;z)\bigr)\Bigr]\nl
+&\frac{1}{\eps}\Bigl[48-\frac{11 \pi^2}{3}+\delta(1-z) \bigl(\frac{417}{8}-5 \pi^2-\frac{116\zeta_{3}}{3}\bigr)+\bigl(-\frac{75}{2}+\frac{20 \pi^2}{3}\bigr) \dd_{0}(z)+24 \dd_{1}(z)-16 \dd_{2}(z)\nl
&\quad+30 \h(0;z)+26\h(1;z)+21 \h(0,0;z)+18 \h(0,1;z)+16 \h(1,0;z)+16 \h(1,1;z)\nl
&\quad+\frac{1}{1-z}\bigl(-24 \h(0;z)-36 \h(0,0;z)-32 \h(0,1;z)-32\h(1,0;z)\bigr)+z \bigl(4-\frac{11 \pi^2}{3}\nl
&\quad+2 \h(0;z)+2 \h(1;z)+21\h(0,0;z)+18 \h(0,1;z)+16 \h(1,0;z)+16 \h(1,1;z)\bigr)\Bigr]\nl
+&\Bigl[157-\frac{77 \pi^2}{6}-\frac{170 \zeta_{3}}{3}+\delta(1-z) \bigl(\frac{2275}{16}-\frac{125\pi^2}{8}+\frac{61 \pi^4}{90}-64 \zeta_{3}\bigr)+\dd_{0}(z) \bigl(-\frac{417}{4}+10 \pi^2\nl
&\quad+\frac{232 \zeta_{3}}{3}\bigr)+\bigl(75-\frac{40 \pi^2}{3}\bigr) \dd_{1}(z)-24 \dd_{2}(z)+\frac{32}{3} \dd_{3}(z)+103 \h(0;z)-\frac{59}{6}\pi^2 \h(0;z)+96 \h(1;z)\nl
&\quad-\frac{20}{3} \pi^2 \h(1;z)+72 \h(0,0;z)+60 \h(0,1;z)+50\h(1,0;z)+52 \h(1,1;z)\nl
&\quad+45 \h(0,0,0;z)+42 \h(0,0,1;z)+32 \h(0,1,0;z)+36 \h(0,1,1;z)+32\h(1,0,0;z)\nl
&\quad+32 \h(1,0,1;z)+32 \h(1,1,0;z)+32 \h(1,1,1;z)+\frac{1}{1-z}\bigl(+24\zeta_{3}-80\h(0;z)\nl
&\quad+\frac{52}{3} \pi^2 \h(0;z)-60 \h(0,0;z)-48 \h(0,1;z)-48 \h(1,0;z)-76 \h(0,0,0;z)\nl
&\quad-72\h(0,0,1;z)-56 \h(0,1,0;z)-64 \h(0,1,1;z)-64 \h(1,0,0;z)-64 \h(1,0,1;z)\nl
&\quad-64 \h(1,1,0;z)\bigr)+z \bigl(7+\frac{7 \pi^2}{6}-\frac{170 \zeta_{3}}{3}+5 \h(0;z)-\frac{59}{6} \pi^2\h(0;z)+8 \h(1;z)\nl
&\quad-\frac{20}{3} \pi^2 \h(1;z)+8 \h(0,0;z)+4 \h(0,1;z)+10 \h(1,0;z)+4\h(1,1;z)+45 \h(0,0,0;z)\nl
&\quad+42 \h(0,0,1;z)+32 \h(0,1,0;z)+36 \h(0,1,1;z)+32 \h(1,0,0;z)+32\h(1,0,1;z)\nl
&\quad+32 \h(1,1,0;z)+32 \h(1,1,1;z)\bigr)\Bigr]\nl
&+\mathcal{O}\left(\epsilon\right)\;.
\end{align}

Crossing a primary quark (coupled to the external current) into the initial
state, one obtains the $(q,q^{\prime}\bar{q}^{\prime}q)$
antenna function, whose integral yields
\begin{align}
\lefteqn{\mathcal{B}^{0}_{q,q^{\prime}\bar{q}^{\prime}q}=}\nl
+&\frac{1}{\eps^3}\Bigl[-\frac{1}{3}\delta(1-z)\Bigr]\nl
+&\frac{1}{\eps^2}\Bigl[-\frac{1}{3}-\frac{19}{18} \delta(1-z)+\frac{2}{3} \dd_{0}(z)-\frac{z}{3}\Bigr]\nl
+&\frac{1}{\eps}\Bigl[-\frac{17}{9}+\delta(1-z) \bigl(-\frac{373}{108}+\frac{5 \pi^2}{18}\bigr)+\frac{19}{9} \dd_{0}(z)-\frac{4}{3} \dd_{1}(z)-\h(0;z)-\frac{2}{3} \h(1;z)+\frac{2}{1-z}\h(0;z)\nl
&\quad+z \bigl(-\frac{5}{9}-\h(0;z)-\frac{2}{3} \h(1;z)\bigr)\Bigr]\nl
+&\Bigl[-\frac{241}{27}+\frac{7 \pi^2}{18}+\delta(1-z)\bigl(-\frac{7081}{648}+\frac{95 \pi^2}{108}+\frac{32 \zeta_{3}}{9}\bigr)+\bigl(\frac{373}{54}-\frac{5 \pi^2}{9}\bigr) \dd_{0}(z)
-\frac{38}{9} \dd_{1}(z)+\frac{4}{3}\dd_{2}(z)\nl
&\quad-\frac{19}{3} \h(0;z)-\frac{34}{9} \h(1;z)-\frac{7}{3} \h(0,0;z)-2 \h(0,1;z)-\frac{4}{3}\h(1,0;z)-\frac{4}{3} \h(1,1;z)\nl
&\quad+\frac{1}{1-z}\bigl(-\frac{2 \pi^2}{9}+\frac{22}{3} \h(0;z)+\frac{14}{3} \h(0,0;z)+4 \h(0,1;z)+\frac{8}{3}\h(1,0;z)\bigr)+z \bigl(-\frac{19}{27}\nl
&\quad+\frac{7 \pi^2}{18}-\frac{5}{3} \h(0;z)-\frac{10}{9} \h(1;z)-\frac{7}{3}\h(0,0;z)-2 \h(0,1;z)-\frac{4}{3} \h(1,0;z)-\frac{4}{3} \h(1,1;z)\bigr)\Bigr]\nl
&+\mathcal{O}\left(\epsilon\right)\,.
\end{align}

Correspondingly, crossing a secondary quark (not coupled to the
external current) into the initial
state, one obtains the $(q^{\prime},q \bar{q}q^{\prime})$
antenna function, whose integral yields
\begin{align}
\lefteqn{\mathcal{B}^{0}_{q^{\prime},q\bar{q}q^{\prime}}=}\nl
+&\frac{1}{\eps^2}\Bigl[1+2 \h(0;z)+\frac{4}{3 z}-\frac{4}{3} z^2+z \bigl(-1+2 \h(0;z)\bigr)\Bigr]\nl
+&\frac{1}{\eps}\Bigl[\frac{10}{3}-\frac{2 \pi^2}{3}+\h(0;z)+2\h(1;z)+6 \h(0,0;z)+4 \h(0,1;z)+\frac{1}{z}\bigl(\frac{4}{9}+\frac{8}{3} \h(1;z)\bigr)\nl
&\quad+z \bigl(-\frac{22}{3}-\frac{2\pi^2}{3}-7 \h(0;z)-2 \h(1;z)+6 \h(0,0;z)+4 \h(0,1;z)\bigr)\nl
&\quad+z^2 \bigl(\frac{32}{9}-\frac{16}{3} \h(0;z)-\frac{8}{3} \h(1;z)\bigr)\Bigr]\nl
+&\Bigl[\frac{25}{9}-\frac{\pi^2}{2}-4 \zeta_{3}+\frac{88}{3} \h(0;z)-\frac{7}{3} \pi^2 \h(0;z)+\frac{20}{3} \h(1;z)-8\h(-1,0;z)+\h(0,0;z)\nl
&\quad+2 \h(0,1;z)+4 \h(1,0;z)+z^2 \bigl(-\frac{208}{27}+2 \pi^2+\frac{128}{9}\h(0;z)+\frac{64}{9} \h(1;z)-\frac{8}{3} \h(-1,0;z)\nl
&\quad-\frac{40}{3} \h(0,0;z)-\frac{32}{3} \h(0,1;z)-\frac{16}{3}\h(1,0;z)-\frac{16}{3} \h(1,1;z)\bigr)+4 \h(1,1;z)+14 \h(0,0,0;z)\nl
&\quad+12 \h(0,0,1;z)+8\h(0,1,0;z)+8 \h(0,1,1;z)+\frac{1}{z}\bigl(\frac{532}{27}-\frac{2 \pi^2}{3}+\frac{8}{9} \h(1;z)\nl
&\quad-\frac{8}{3}\h(-1,0;z)+\frac{16}{3} \h(1,0;z)+\frac{16}{3} \h(1,1;z)\bigr)+z \bigl(-\frac{133}{9}+\frac{7 \pi^2}{6}-4 \zeta_{3}-4 \h(0;z)\nl
&\quad-\frac{7}{3} \pi^2 \h(0;z)-\frac{44}{3}\h(1;z)-8 \h(-1,0;z)-11 \h(0,0;z)-14 \h(0,1;z)-4 \h(1,0;z)\nl
&\quad-4 \h(1,1;z)+14\h(0,0,0;z)+12 \h(0,0,1;z)+8 \h(0,1,0;z)+8 \h(0,1,1;z)\bigr)\Bigr]\nl
&+\mathcal{O}\left(\epsilon\right)\,.
\end{align}

As explained in Section~\ref{sec:ant} above, there are three different
crossings for the identical flavour only antenna functions $C_4^0$, which
are symmetrized over the antiquarks, but not over the quarks.

Crossing either antiquark into the initial state yields the $(q,q\bar{q}q)$
antenna, whose integral is:
\begin{align}
\lefteqn{\mathcal{C}^{0}_{q,q\bar{q}q}=}\nl
+&\frac{1}{\eps}\Bigl[2-\frac{\pi^2}{12}+\frac{1}{2} \h(0;z)-\frac{1}{2} \h(0,0;z)-\frac{1}{2} \h(1,0;z)+\frac{1}{1-z}\bigl(\frac{\pi^2}{6}+\frac{3}{4}\h(0;z)+\h(0,0;z)\nl
&\quad+\h(1,0;z)\bigr)+z \bigl(-\frac{7}{4}-\frac{\pi^2}{12}+\frac{1}{2}\h(0;z)-\frac{1}{2} \h(0,0;z)-\frac{1}{2} \h(1,0;z)\bigr)\Bigr]\nl
+&\Bigl[\frac{1}{4}-\frac{5 \pi^2}{12}-\frac{3 \zeta_{3}}{2}-\frac{1}{6} \pi^2 \h(0;z)+4 \h(1;z)+2 \h(-1,0;z)-2 \h(0,0;z)+\h(0,1;z)\nl
&\quad-\frac{5}{2}\h(1,0;z)-2 \h(0,-1,0;z)-\frac{3}{2} \h(0,0,0;z)-\h(0,0,1;z)-\h(0,1,0;z)\nl
&\quad-\frac{7}{2}\h(1,0,0;z)-\h(1,0,1;z)-\h(1,1,0;z)+\frac{1}{1-z}\bigl(\frac{\pi^2}{4}+3 \zeta_{3}+3 \h(0;z)+\frac{1}{3} \pi^2 \h(0;z)\nl
&\quad+\frac{21}{4}\h(0,0;z)+\frac{3}{2} \h(0,1;z)+3 \h(1,0;z)+4 \h(0,-1,0;z)+3 \h(0,0,0;z)\nl
&\quad+2 \h(0,0,1;z)+2\h(0,1,0;z)+7 \h(1,0,0;z)+2 \h(1,0,1;z)+2 \h(1,1,0;z)\bigr)\nl
&\quad+z \bigl(\frac{7}{4}+\frac{\pi^2}{12}-\frac{27}{4} \h(0;z)-\frac{1}{6}\pi^2 \h(0;z)-\frac{7}{2} \h(1;z)+2 \h(-1,0;z)+\h(0,0;z)\nl
&\quad+\h(0,1;z)+\frac{1}{2} \h(1,0;z)-2\h(0,-1,0;z)-\frac{3}{2} \h(0,0,0;z)-\h(0,0,1;z)-\h(0,1,0;z)\nl
&\quad-\frac{7}{2} \h(1,0,0;z)-\h(1,0,1;z)-\h(1,1,0;z)-\frac{3\zeta_{3}}{2}\bigr)\Bigr]\nl
&+\mathcal{O}\left(\epsilon\right)\,.
\end{align}

Crossing the quark which couples to the external current to the initial state
yields the $(\bar{q},\bar{q}q\bar{q})$ antenna function, which integrates to:
\begin{align}
\lefteqn{\mathcal{C}^{0}_{\bar{q},\bar{q}q\bar{q}}=}\nl
+&\frac{1}{\eps}\Bigl[\delta(1-z) \bigl(-\frac{13}{8}+\frac{\pi^2}{4}-\zeta_{3}\bigr)\Bigr]\nl
+&\Bigl[-5+2 \pi^2+\delta(1-z)
\bigl(-\frac{175}{16}+\frac{\pi^2}{2}+\frac{17 \pi^4}{180}-\frac{107 \zeta_{3}}{2}\bigr)+\dd_{0}(z) \bigl(\frac{13}{4}-\frac{\pi^2}{2}+2 \zeta_{3}\bigr)-5 \h(0;z)\nl
&\quad-6 \h(-1,0;z)+15\h(0,0;z)+9 \h(1,0;z)-\frac{2}{z^2} \h(-1,0;z)+\frac{1}{z}\bigl(-2+2 \h(0;z)\nl
&\quad-6 \h(-1,0;z)\bigr)+z (2 \pi^2-2 \h(-1,0;z)+9 \h(0,0;z)+7 \h(1,0;z))+\frac{1}{1-z}\bigl(-\frac{3 \pi^2}{2}\nl
&\quad-6\zeta_{3}+10 \h(0;z)+\frac{2}{3}\pi^2 \h(1;z)-15 \h(0,0;z)-9 \h(1,0;z)+4 \h(0,-1,0;z)\nl
&\quad+8 \h(1,0,0;z)+4 \h(1,1,0;z)\bigr)\Bigr]\nl
&+\mathcal{O}\left(\epsilon\right)\,.
\end{align}

Finally, crossing the quark not coupled to the external current (and thus
participating in the collinear splitting), one obtains the
$(\bar{q},q\bar{q}\bar{q})$ antenna function, integrating to:
\begin{align}
\lefteqn{\mathcal{C}^{0}_{\bar{q},q\bar{q}\bar{q}}=}\nl
+&\frac{1}{\eps}\Bigl[-2-\frac{\pi^2}{6}-\h(0;z)-2 \h(-1,0;z)+\h(0,0;z)+\frac{1}{1+z}\bigl(\frac{\pi^2}{3}+4 \h(-1,0;z)-2 \h(0,0;z)\bigr)\nl
&\quad+z\bigl(2+\frac{\pi^2}{6}-\h(0;z)+2 \h(-1,0;z)-\h(0,0;z)\bigr)\Bigr]\nl
+&\Bigl[\frac{15}{2}+\frac{\pi^2}{6}-5 \zeta_{3}+\pi^2 \h(-1;z)-\frac{1}{2} \h(0;z)-\frac{1}{3} \pi^2 \h(0;z)-4\h(1;z)-4 \h(-1,0;z)\nl
&\quad-2 \h(0,0;z)-2\h(0,1;z)+4 \h(-1,-1,0;z)-8 \h(-1,0,0;z)-4 \h(-1,0,1;z)\nl
&\quad-4 \h(0,-1,0;z)+3 \h(0,0,0;z)+2\h(0,0,1;z)-\frac{2}{z^2} \h(-1,0;z)+\frac{1}{z}\bigl(-2+2 \h(0;z)\nl
&\quad+6 \h(-1,0;z)\bigr)+\frac{1}{1+z}\bigl(+14 \zeta_{3}-\frac{8}{3}\pi^2 \h(-1;z)-4 \h(0;z)+\frac{2}{3} \pi^2 \h(0;z)\nl
&\quad-16 \h(-1,-1,0;z)+20 \h(-1,0,0;z)+8\h(-1,0,1;z)+12 \h(0,-1,0;z)\nl
&\quad-6 \h(0,0,0;z)-4 \h(0,0,1;z)\bigr)+z \bigl(-\frac{11}{2}+\frac{\pi^2}{6}+5 \zeta_{3}-\pi^2 \h(-1;z)+\frac{19}{2} \h(0;z)\nl
&\quad+\frac{1}{3}\pi^2 \h(0;z)+4 \h(1;z)-4 \h(-1,0;z)-2 \h(0,0;z)-2 \h(0,1;z)-4 \h(-1,-1,0;z)\nl
&\quad+8\h(-1,0,0;z)+4 \h(-1,0,1;z)+4 \h(0,-1,0;z)-3 \h(0,0,0;z)-2 \h(0,0,1;z)\bigr)\Bigr]\nl
&+\mathcal{O}\left(\epsilon\right)\,.
\end{align}

The integrated one-loop quark-antiquark antenna functions
at leading and subleading colour read:
\begin{align}
\lefteqn{\mathcal{A}^{1}_{q,gq}=}\nl
+&\frac{1}{\eps^4}\Bigl[-\frac{1}{4}\delta(1-z)\Bigr]\nl
+&\frac{1}{\eps^3}\Bigl[-\frac{1}{4}-\frac{53}{24} \delta(1-z)-\frac{z}{4}+\frac{1}{2} \dd_{0}(z)\Bigr]\nl
+&\frac{1}{\eps^2}\Bigl[-\frac{13}{6}+\delta(1-z) \bigl(-\frac{43}{16}+\frac{5 \pi^2}{24}\bigr)+\frac{31}{12} \dd_{0}(z)-\dd_{1}(z)-\frac{3}{4}
\h(0;z)-\frac{1}{2}\h(1;z)+\frac{3 \h(0;z)}{2 (1-z)}\nl
&\quad-z \bigl(\frac{2}{3}+\frac{3}{4} \h(0;z)+\frac{1}{2} \h(1;z)\bigr)\Bigr]\nl
+&\frac{1}{\eps}\Bigl[-\frac{27}{4}+\frac{\pi^2}{6}+\delta(1-z) \bigl(-\frac{167}{24}+\frac{37 \pi^2}{48}+\frac{7 \zeta_{3}}{6}\bigr)+\bigl(4-\frac{5\pi^2}{12}\bigr) \dd_{0}(z)
-\frac{10}{3} \dd_{1}(z)+\dd_{2}(z)-\frac{14}{3} \h(0;z)\nl
&\quad-\frac{41}{12} \h(1;z)-2\h(0,0;z)-\frac{5}{4} \h(0,1;z)-\frac{3}{2} \h(1,0;z)-\h(1,1;z)+\frac{1}{1-z}\bigl(\frac{\pi^2}{12}\nl
&\quad+\frac{49}{12} \h(0;z)+4\h(0,0;z)+\frac{5}{2} \h(0,1;z)+3 \h(1,0;z)\bigr)+z \bigl(\frac{25}{24}+\frac{\pi^2}{6}-\frac{1}{6} \h(0;z)\nl
&\quad-\frac{5}{12}\h(1;z)-2 \h(0,0;z)-\frac{5}{4} \h(0,1;z)-\frac{3}{2} \h(1,0;z)-\h(1,1;z)\bigr)\Bigr]\nl
+&\Bigl[-\frac{67}{4}+\frac{67 \pi^2}{48}+\frac{29 \zeta_{3}}{12}+\delta(1-z) \bigl(-\frac{773}{48}+\frac{23 \pi^2}{16}-\frac{67\pi^4}{1440}+\frac{217 \zeta_{3}}{36}\bigr)
+\bigl(\frac{257}{24}-\frac{13 \pi^2}{12}\nl
&\quad-\frac{7 \zeta_{3}}{3}\bigr) \dd_{0}(z)+\bigl(-\frac{53}{8}+\frac{5\pi^2}{6}\bigr) \dd_{1}(z)+\frac{29}{12} \dd_{2}(z)-\frac{2}{3} \dd_{3}(z)-\frac{117}{8} \h(0;z)
+\frac{5}{8} \pi^2\h(0;z)-\frac{89}{8} \h(1;z)\nl
&\quad+\frac{1}{3} \pi^2 \h(1;z)-\frac{119}{12} \h(0,0;z)-\frac{43}{6}\h(0,1;z)-\frac{89}{12} \h(1,0;z)-\frac{71}{12} \h(1,1;z)\nl
&\quad-\frac{9}{2} \h(0,0,0;z)-3 \h(0,0,1;z)-3\h(0,1,0;z)-\frac{9}{4} \h(0,1,1;z)-4 \h(1,0,0;z)\nl
&\quad-\frac{5}{2} \h(1,0,1;z)-3 \h(1,1,0;z)+\frac{1}{1-z}\bigl(\frac{\pi^2}{8}-\frac{5\zeta_{3}}{2}+\frac{69}{8} \h(0;z)-\frac{5}{4} \pi^2 \h(0;z)\nl
&\quad+\frac{1}{6} \pi^2 \h(1;z)+\frac{47}{6}\h(0,0;z)+\frac{67}{12} \h(0,1;z)+\frac{19}{3} \h(1,0;z)+9 \h(0,0,0;z)\nl
&\quad+6 \h(0,0,1;z)+6\h(0,1,0;z)+\frac{9}{2} \h(0,1,1;z)+8 \h(1,0,0;z)+5 \h(1,0,1;z)\nl
&\quad+6 \h(1,1,0;z)\bigr)+z\bigl(\frac{1}{8}-\frac{5 \pi^2}{48}+\frac{29 \zeta_{3}}{12}+\frac{49}{24} \h(0;z)+\frac{5}{8} \pi^2 \h(0;z)+\frac{23}{12}\h(1;z)\nl
&\quad+\frac{1}{3} \pi^2 \h(1;z)+\frac{7}{12} \h(0,0;z)+\frac{1}{3} \h(0,1;z)+\frac{1}{12}\h(1,0;z)+\frac{1}{12} \h(1,1;z)\nl
&\quad-\frac{9}{2} \h(0,0,0;z)-3 \h(0,0,1;z)-3 \h(0,1,0;z)-\frac{9}{4}\h(0,1,1;z)-4 \h(1,0,0;z)\nl
&\quad-\frac{5}{2} \h(1,0,1;z)-3 \h(1,1,0;z)-2 \h(1,1,1;z)\bigr)-2\h(1,1,1;z)\Bigr]\nl
&+\mathcal{O}\left(\epsilon\right)\,,
\end{align}
\begin{align}
\lefteqn{\mathcal{\tilde{A}}^{1}_{q,gq}=}\nl
+&\frac{1}{\eps^2}\Bigl[\delta(1-z) \bigl(-\frac{5}{16}+\frac{\pi^2}{12}\bigr)-\frac{1}{4} \h(0;z)+\frac{\h(0;z)}{2 (1-z)}-\frac{1}{4}z \h(0;z)\Bigr]\nl
+&\frac{1}{\eps}\Bigl[-1+\frac{\pi^2}{8}+\delta(1-z) \bigl(-\frac{5}{4}+\frac{5 \zeta_{3}}{2}\bigr)+\bigl(\frac{5}{8}-\frac{\pi^2}{6}\bigr)\dd_{0}(z)-\frac{5}{4} \h(0;z)-\h(0,0;z)-\frac{1}{4}\h(0,1;z)\nl
&\quad+\frac{1}{1-z}\bigl(-\frac{\pi^2}{12}+\frac{3}{4} \h(0;z)+2 \h(0,0;z)+\frac{1}{2} \h(0,1;z)\bigr)+z \bigl(-\frac{1}{8}+\frac{\pi^2}{8}+\frac{1}{4} \h(0;z)\nl
&\quad-\h(0,0;z)-\frac{1}{4}\h(0,1;z)\bigr)\Bigr]\nl
+&\Bigl[-6-\frac{\pi^2}{24}+\frac{9 \zeta_{3}}{4}+\delta(1-z) \bigl(-\frac{61}{16}+\frac{25 \pi^2}{96}-\frac{43 \pi^4}{1440}+16\zeta_{3}\bigr)+\bigl(\frac{5}{2}
-5 \zeta_{3}\bigr) \dd_{0}(z)+\bigl(-\frac{5}{4}+\frac{\pi^2}{3}\bigr) \dd_{1}(z)\nl
&\quad-\frac{35}{8}\h(0;z)+\frac{13}{24} \pi^2 \h(0;z)-\frac{7}{8} \h(1;z)+\frac{1}{6} \pi^2 \h(1;z)-7 \h(0,0;z)-\frac{5}{4}\h(0,1;z)\nl
&\quad-2 \h(1,0;z)-\frac{5}{2} \h(0,0,0;z)-\h(0,0,1;z)-\frac{1}{4} \h(0,1,1;z)+\frac{1}{2}\h(1,0,0;z)\nl
&\quad+\frac{1}{2}\h(1,1,0;z)+\frac{1}{1-z}\bigl(\frac{\pi^2}{8}+\frac{3 \zeta_{3}}{2}+\frac{7}{4} \h(0;z)-\frac{13}{12} \pi^2\h(0;z)-\frac{1}{3} \pi^2 \h(1;z)\nl
&\quad+6 \h(0,0;z)+\frac{3}{4} \h(0,1;z)+\frac{3}{2} \h(1,0;z)+5\h(0,0,0;z)+2 \h(0,0,1;z)+\frac{1}{2} \h(0,1,1;z)\nl
&\quad-2 \h(1,0,0;z)-2 \h(1,1,0;z)\bigr)+z\bigl(-\frac{1}{8}-\frac{23 \pi^2}{24}+\frac{9 \zeta_{3}}{4}+\frac{1}{8} \h(0;z)+\frac{13}{24} \pi^2 \h(0;z)\nl
&\quad+\frac{1}{4}\h(1;z)+\frac{1}{6} \pi^2 \h(1;z)-\frac{3}{2} \h(0,0;z)+\frac{1}{4} \h(0,1;z)-\frac{5}{2} \h(1,0;z)-\frac{5}{2}\h(0,0,0;z)\nl
&\quad-\h(0,0,1;z)-\frac{1}{4} \h(0,1,1;z)+\frac{1}{2} \h(1,0,0;z)+\frac{1}{2} \h(1,1,0;z)\bigr)\Bigr]\nl
&+\mathcal{O}\left(\epsilon\right)\;.
\end{align}

The quark loop contribution is:
\begin{align}
\lefteqn{\mathcal{\hat{A}}^{1}_{q,gq}=}\nl
+&\frac{1}{\eps^3}\Bigl[\frac{1}{3}\delta(1-z)\Bigr]\nl
+&\frac{1}{\eps^2}\Bigl[\frac{1}{6}+\frac{1}{4}\delta(1-z)-\frac{1}{3} \dd_{0}(z)+\frac{z}{6}\Bigr]\nl
+&\frac{1}{\eps}\Bigl[\frac{1}{2}+\delta(1-z) \bigl(\frac{7}{12}-\frac{\pi^2}{12}\bigr)-\frac{1}{4} \dd_{0}(z)+\frac{1}{3} \dd_{1}(z)+\frac{1}{6}\h(0;z)+\frac{1}{6}\h(1;z)-\frac{1}{3 (1-z)}\h(0;z)\nl
&\quad+z \bigl(-\frac{1}{6}+\frac{1}{6} \h(0;z)+\frac{1}{6} \h(1;z)\bigr)\Bigr]\nl
+&\Bigl[1-\frac{\pi^2}{24}+\delta(1-z) \bigl(\frac{7}{6}-\frac{\pi^2}{16}-\frac{7 \zeta_{3}}{9}\bigr)+\bigl(-\frac{7}{12}+\frac{\pi^2}{12}\bigr)\dd_{0}(z)+\frac{1}{4} \dd_{1}(z)-\frac{1}{6} \dd_{2}(z)
+\frac{1}{2} \h(0;z)\nl
&\quad+\frac{1}{2} \h(1;z)+\frac{1}{6}\h(0,0;z)+\frac{1}{6} \h(0,1;z)+\frac{1}{6} \h(1,0;z)+\frac{1}{6} \h(1,1;z)+\frac{1}{1-z}\bigl(-\frac{1}{4} \h(0;z)\nl
&\quad-\frac{1}{3} \h(0,0;z)-\frac{1}{3} \h(0,1;z)-\frac{1}{3}\h(1,0;z)\bigr)+z \bigl(-\frac{\pi^2}{24}-\frac{1}{6} \h(0;z)-\frac{1}{6} \h(1;z)+\frac{1}{6}\h(0,0;z)\nl
&\quad+\frac{1}{6} \h(0,1;z)+\frac{1}{6} \h(1,0;z)+\frac{1}{6} \h(1,1;z)\bigr)\Bigr]\nl
&+\mathcal{O}\left(\epsilon\right)\,.
\end{align}

\subsection{Quark-gluon antennae}
The integrated NLO quark-gluon initial-final antenna
with an initial state quark reads  through to $\e^2$
for the gluon radiation:
\begin{align}
\lefteqn{\mathcal{D}^{0}_{q,gg}=}\nl
+&\frac{1}{\eps^2}\Bigl[2\delta(1-z)\Bigr]\nl
+&\frac{1}{\eps}\Bigl[1+\frac{11}{6} \delta(1-z)+z-2 \dd_{0}(z)\Bigr]\nl
+&\Bigl[1+\delta(1-z) \bigl(\frac{67}{18}-\frac{\pi ^2}{2}\bigr)-\frac{1}{3 z}-\frac{11}{6} \dd_{0}(z)+2 \dd_{1}(z)+\h(0;z)+\h(1;z)-\frac{2}{1-z} \h(0;z)\nl
&\quad+z (-1+\h(0;z)+\h(1;z))\Bigr]\nl
+&\eps\Bigl[1-\frac{\pi ^2}{4}+\delta(1-z) \bigl(\frac{202}{27}-\frac{11 \pi ^2}{24}-\frac{14 \zeta_{3}}{3}\bigr)+\bigl(-\frac{67}{18}+\frac{\pi ^2}{2}\bigr) \dd_{0}(z)+\frac{11}{6} \dd_{1}(z)-\dd_{2}(z)
+\h(0;z)\nl
&\quad+\h(1;z)+\h(0,0;z)+\h(0,1;z)+\h(1,0;z)+\h(1,1;z)-\frac{1}{z}\bigl(\frac{2}{9}+\frac{1}{3} \h(0;z)+\frac{1}{3} \h(1;z)\bigr)\nl
&\quad+\frac{1}{1-z}\bigl(-\frac{11}{6} \h(0;z)-2 \h(0,0;z)-2 \h(0,1;z)-2 \h(1,0;z)\bigr)+z \bigl(-\frac{\pi ^2}{4}-\h(0;z)\nl
&\quad-\h(1;z)+\h(0,0;z)+\h(0,1;z)+\h(1,0;z)+\h(1,1;z)\bigr)\Bigr]\nl
+&\eps^{2}\Bigl[3-\frac{\pi ^2}{4}-\frac{7 \zeta_{3}}{3}+\delta(1-z) \bigl(\frac{1214}{81}-\frac{67 \pi ^2}{72}-\frac{\pi ^4}{48}-\frac{77 \zeta_{3}}{18}\bigr)+\dd_{0}(z) \bigl(-\frac{202}{27}+\frac{11 \pi ^2}{24}+\frac{14 \zeta_{3}}{3}\bigr)\nl
&\quad+\bigl(\frac{67}{18}-\frac{\pi ^2}{2}\bigr) \dd_{1}(z)-\frac{11}{12} \dd_{2}(z)+\frac{1}{3} \dd_{3}(z)+\h(0;z)-\frac{1}{4} \pi ^2 \h(0;z)+\h(1;z)-\frac{1}{4} \pi ^2 \h(1;z)\nl
&\quad+\h(0,0;z)+\h(0,1;z)+\h(1,0;z)+\h(1,1;z)+\h(0,0,0;z)+\h(0,0,1;z)\nl
&\quad+\h(0,1,0;z)+\h(0,1,1;z)+\h(1,0,0;z)+\h(1,0,1;z)+\h(1,1,0;z)+\h(1,1,1;z)\nl
&\quad+\frac{1}{z}\bigl(-\frac{13}{27}+\frac{\pi ^2}{12}-\frac{2}{9} \h(0;z)-\frac{2}{9} \h(1;z)-\frac{1}{3} \h(0,0;z)-\frac{1}{3} \h(0,1;z)-\frac{1}{3} \h(1,0;z)\nl
&\quad-\frac{1}{3} \h(1,1;z)\bigr)+\frac{1}{1-z}\bigl(-\frac{67}{18} \h(0;z)+\frac{1}{2} \pi ^2 \h(0;z)-\frac{11}{6} \h(0,0;z)-\frac{11}{6} \h(0,1;z)\nl
&\quad-\frac{11}{6} \h(1,0;z)-2 \h(0,0,0;z)-2 \h(0,0,1;z)-2 \h(0,1,0;z)-2 \h(0,1,1;z)\nl
&\quad-2 \h(1,0,0;z)-2 \h(1,0,1;z)-2 \h(1,1,0;z)\bigr)+z \bigl(\frac{\pi ^2}{4}-\frac{7\zeta_{3}}{3}-\frac{1}{4} \pi ^2 \h(0;z)\nl
&\quad-\frac{1}{4} \pi ^2 \h(1;z)-\h(0,0;z)-\h(0,1;z)-\h(1,0;z)-\h(1,1;z)+\h(0,0,0;z)\nl
&\quad+\h(0,0,1;z)+\h(0,1,0;z)+\h(0,1,1;z)+\h(1,0,0;z)+\h(1,0,1;z)+\h(1,1,0;z)\nl
&\quad+\h(1,1,1;z)\bigr)\Bigr]\nl
&+\mathcal{O}\left(\epsilon^{3}\right)\,,
\end{align}
while we have to distinguish two cases for the secondary quark radiation.
Crossing the primary quark into the initial state, one obtains
\begin{align}
\lefteqn{\mathcal{E}^{0}_{q,q^{\prime}\bar{q}^{\prime}}=}\nl
-&\frac{1}{\eps}\Bigl[\frac{1}{3}\delta(1-z)\Bigr]\nl
-&\Bigl[\frac{5}{9} \delta(1-z)+\frac{1}{6 z}-\frac{1}{3} \dd_{0}(z)\Bigr]\nl
+&\eps\Bigl[\delta(1-z) \bigl(-\frac{28}{27}+\frac{\pi ^2}{12}\bigr)+\frac{5}{9} \dd_{0}(z)-\frac{1}{3} \dd_{1}(z)+\frac{1}{3(1-z)}\h(0;z)\nl
&\quad-\frac{1}{z}\bigl(\frac{4}{9}+\frac{1}{6} \h(0;z)+\frac{1}{6} \h(1;z)\bigr)\Bigr]\nl
+&\eps^{2}\Bigl[+\delta(1-z) \bigl(-\frac{164}{81}+\frac{5 \pi ^2}{36}+\frac{7 \zeta_{3}}{9}\bigr)+\bigl(\frac{28}{27}-\frac{\pi ^2}{12}\bigr) \dd_{0}(z)-\frac{5}{9} \dd_{1}(z)+\frac{1}{6} \dd_{2}(z)
+\frac{1}{1-z}\bigl(\frac{5}{9} \h(0;z)\nl
&\quad+\frac{1}{3} \h(0,0;z)+\frac{1}{3} \h(0,1;z)+\frac{1}{3} \h(1,0;z)\bigr)+\frac{1}{z}\bigl(-\frac{26}{27}+\frac{\pi ^2}{24}-\frac{4}{9} \h(0;z)-\frac{4}{9} \h(1;z)\nl
&\quad-\frac{1}{6} \h(0,0;z)-\frac{1}{6} \h(0,1;z)-\frac{1}{6} \h(1,0;z)-\frac{1}{6} \h(1,1;z)\bigr)\Bigr]\nl
&+\mathcal{O}\left(\epsilon^{3}\right)\,,
\end{align}
while crossing the secondary quark results in
\begin{align}
\lefteqn{\mathcal{E}^{0}_{q^{\prime},q^{\prime}q}=}\nl
+&\frac{1}{\eps}\Bigl[-1+\frac{1}{z}+\frac{z}{2}\Bigr]\nl
+&\Bigl[-\frac{3}{2}-\h(0;z)-\h(1;z)+\frac{1}{z}\bigl(2+\h(0;z)+\h(1;z)\bigr)+z \bigl(\frac{1}{2} \h(0;z)+\frac{1}{2} \h(1;z)\bigr)\Bigr]\nl
+&\eps\Bigl[-3+\frac{\pi ^2}{4}-\frac{3}{2} \h(0;z)-\frac{3}{2} \h(1;z)-\h(0,0;z)-\h(0,1;z)-\h(1,0;z)-\h(1,1;z)\nl
&\quad+\frac{1}{z}\bigl(4-\frac{\pi ^2}{4}+2 \h(0;z)+2 \h(1;z)+\h(0,0;z)+\h(0,1;z)+\h(1,0;z)+\h(1,1;z)\bigr)\nl
&\quad+z \bigl(-\frac{\pi ^2}{8}+\frac{1}{2} \h(0,0;z)+\frac{1}{2} \h(0,1;z)+\frac{1}{2} \h(1,0;z)+\frac{1}{2} \h(1,1;z)\bigr)\Bigr]\nl
+&\eps^{2}\Bigl[-6+\frac{3 \pi ^2}{8}+\frac{7 \zeta_{3}}{3}-3 \h(0;z)+\frac{1}{4} \pi ^2 \h(0;z)-3 \h(1;z)+\frac{1}{4} \pi ^2 \h(1;z)-\frac{3}{2} \h(0,0;z)\nl
&\quad-\frac{3}{2} \h(0,1;z)-\frac{3}{2} \h(1,0;z)-\frac{3}{2} \h(1,1;z)-\h(0,0,0;z)-\h(0,0,1;z)-\h(0,1,0;z)\nl
&\quad-\h(0,1,1;z)-\h(1,0,0;z)-\h(1,0,1;z)-\h(1,1,0;z)-\h(1,1,1;z)+\frac{1}{z}\bigl(8-\frac{\pi^2}{2}\nl
&\quad-\frac{7\zeta_{3}}{3}+4 \h(0;z)-\frac{1}{4} \pi ^2 \h(0;z)+4 \h(1;z)-\frac{1}{4} \pi ^2 \h(1;z)+2 \h(0,0;z)+2 \h(0,1;z)\nl
&\quad+2 \h(1,0;z)+2 \h(1,1;z)+\h(0,0,0;z)+\h(0,0,1;z)+\h(0,1,0;z)+\h(0,1,1;z)\nl
&\quad+\h(1,0,0;z)+\h(1,0,1;z)+\h(1,1,0;z)+\h(1,1,1;z)\bigr)+z \bigl(-\frac{7 \zeta_{3}}{6}-\frac{1}{8} \pi ^2 \h(0;z)\nl
&\quad-\frac{1}{8} \pi ^2 \h(1;z)+\frac{1}{2} \h(0,0,0;z)+\frac{1}{2} \h(0,0,1;z)+\frac{1}{2} \h(0,1,0;z)+\frac{1}{2} \h(0,1,1;z)\nl
&\quad+\frac{1}{2} \h(1,0,0;z)+\frac{1}{2} \h(1,0,1;z)+\frac{1}{2} \h(1,1,0;z)+\frac{1}{2} \h(1,1,1;z)\bigr)\Bigr]\nl
&+\mathcal{O}\left(\epsilon^{3}\right)\,.
\end{align}

The NNLO double real radiation quark-gluon antenna functions
are of the $D_4^0$ type (quark-gluon-gluon-gluon) and of $E_4^0$-type
(quark-quark-antiquark-gluon). While the former has only one
possible quark-initiated crossing, we have to distinguish several
different cases for the latter.

The integrated form of the initial-final $(q,ggg)$ antenna function reads:
\begin{align}
\lefteqn{\mathcal{D}^{0}_{q,ggg}=}\nl
+&\frac{1}{\eps^4}\Bigl[10 \delta(1-z)\Bigr]\nl
+&\frac{1}{\eps^3}\Bigl[10+22 \delta(1-z)-20 \dd_{0}(z)+10 z\Bigr]\nl
+&\frac{1}{\eps^2}\Bigl[17+\delta(1-z) \bigl(\frac{590}{9}-8 \pi^2\bigr)-44 \dd_{0}(z)+40 \dd_{1}(z)+22\h(0;z)+20\h(1;z)-\frac{40\h(0;z)}{1-z}-\frac{4}{z}\nl
&\quad+z\bigl(5+22\h(0;z)+20\h(1;z)\bigr)\Bigr]\nl
+&\frac{1}{\eps}\Bigl[\frac{107}{3}-\frac{25 \pi^2}{3}+\delta(1-z) \bigl(\frac{4868}{27}-\frac{154 \pi^2}{9}-\frac{272 \zeta_{3}}{3}\bigr)+\bigl(-\frac{1180}{9}+16 \pi^2\bigr) \dd_{0}(z)+88 \dd_{1}(z)-40 \dd_{2}(z)\nl
&\quad+\frac{113}{3}\h(0;z)+34 \h(1;z)+50 \h(0,0;z)+44 \h(0,1;z)+42 \h(1,0;z)+40\h(1,1;z)\nl
&\quad+\frac{1}{1-z}\bigl(-\frac{2\pi^2}{3}-\frac{277}{3} \h(0;z)-88 \h(0,0;z)-80 \h(0,1;z)-84 \h(1,0;z)\bigr)+\frac{1}{z}\bigl(-\frac{112}{9}\nl
&\quad-8 \h(0;z)-8 \h(1;z)\bigr)+z \bigl(22-\frac{25 \pi^2}{3}+\frac{41}{3} \h(0;z)+10 \h(1;z)+50 \h(0,0;z)\nl
&\quad+44 \h(0,1;z)+42\h(1,0;z)+40 \h(1,1;z)\bigr)\Bigr]\nl
+&\Bigl[\frac{565}{9}-\frac{145 \pi^2}{18}-\frac{308\zeta_{3}}{3}+\delta(1-z)\bigl(\frac{25811}{54}-\frac{1408 \pi^2}{27}+\frac{319 \pi^4}{180}-\frac{934 \zeta_{3}}{3}\bigr)
+\dd_{0}(z) \bigl(-\frac{9736}{27}\nl
&\quad+\frac{308 \pi^2}{9}+\frac{544 \zeta_{3}}{3}\bigr)+\bigl(\frac{2360}{9}-32 \pi^2\bigr) \dd_{1}(z)-88 \dd_{2}(z)+\frac{80}{3} \dd_{3}(z)-\frac{4}{3}\pi^2 \h(-1;z)+\frac{616}{9} \h(0;z)\nl
&\quad-\frac{65}{3} \pi^2 \h(0;z)+\frac{214}{3} \h(1;z)-\frac{50}{3}\pi^2 \h(1;z)-8 \h(-1,0;z)+121 \h(0,0;z)\nl
&\quad+\frac{226}{3} \h(0,1;z)+96 \h(1,0;z)+68\h(1,1;z)-16 \h(-1,-1,0;z)+8 \h(-1,0,0;z)\nl
&\quad+4 \h(0,-1,0;z)+106 \h(0,0,0;z)+100 \h(0,0,1;z)+84\h(0,1,0;z)+88 \h(0,1,1;z)\nl
&\quad+82 \h(1,0,0;z)+84 \h(1,0,1;z)+80 \h(1,1,0;z)+80 \h(1,1,1;z)+\frac{1}{z}\bigl(-\frac{2087}{54}+\frac{34\pi^2}{9}\nl
&\quad-4 \zeta_{3}-\frac{2}{3} \pi^2 \h(-1;z)-\frac{224}{9} \h(0;z)-\frac{224}{9} \h(1;z)+\frac{2}{3} \pi^2\h(1;z)-4 \h(-1,0;z)\nl
&\quad-16 \h(0,0;z)-16 \h(0,1;z)-\frac{44}{3} \h(1,0;z)-16 \h(1,1;z)-8\h(-1,-1,0;z)\nl
&\quad+4 \h(-1,0,0;z)+8 \h(0,-1,0;z)+8 \h(1,0,0;z)+4 \h(1,1,0;z)\bigr)+\frac{1}{1-z}\bigl(-\frac{11 \pi^2}{3}\nl
&\quad+32 \zeta_{3}-\frac{2368}{9} \h(0;z)+\frac{116}{3} \pi^2 \h(0;z)+\frac{4}{3}\pi^2 \h(1;z)-243 \h(0,0;z)-\frac{554}{3} \h(0,1;z)\nl
&\quad-\frac{620}{3} \h(1,0;z)+8 \h(0,-1,0;z)-184\h(0,0,0;z)-176 \h(0,0,1;z)-152 \h(0,1,0;z)\nl
&\quad-160 \h(0,1,1;z)-164 \h(1,0,0;z)-168 \h(1,0,1;z)-160\h(1,1,0;z)\bigr)+z \bigl(\frac{443}{6}\nl
&\quad-\frac{25\pi^2}{18}-\frac{320 \zeta_{3}}{3}-\frac{2}{3} \pi^2 \h(-1;z)+\frac{391}{9} \h(0;z)-\frac{65}{3} \pi^2 \h(0;z)+44 \h(1;z)\nl
&\quad-\frac{50}{3}\pi^2 \h(1;z)-4 \h(-1,0;z)+47 \h(0,0;z)+\frac{82}{3} \h(0,1;z)+34 \h(1,0;z)\nl
&\quad+20 \h(1,1;z)-8\h(-1,-1,0;z)+4 \h(-1,0,0;z)+106 \h(0,0,0;z)+100 \h(0,0,1;z)\nl
&\quad+84 \h(0,1,0;z)+88 \h(0,1,1;z)+82\h(1,0,0;z)+84 \h(1,0,1;z)+80 \h(1,1,0;z)\nl
&\quad+80 \h(1,1,1;z)\bigr)\Bigr]\nl
&+\mathcal{O}\left(\epsilon\right)\,.
\end{align}

For the $E_4^0$-type antenna functions, we can either cross the primary
quark or the secondary quark into the initial state. In both cases,
we have leading and subleading colour contributions.

Crossing the primary quark, one obtains the $(q,q^{\prime}\bar{q}^{\prime}g)$
antenna functions, whose integral reads at leading colour
\begin{align}
\lefteqn{\mathcal{E}^{0}_{q,q^{\prime}\bar{q}^{\prime}g}=}\nl
+&\frac{1}{\eps^3}\Bigl[-\frac{5)}{3} \delta(1-z\Bigr]\nl
+&\frac{1}{\eps^2}\Bigl[-1-\frac{47}{9} \delta(1-z)+\frac{10}{3} \dd_{0}(z)-\frac{2}{3 z}-z\Bigr]\nl
+&\frac{1}{\eps}\Bigl[-\frac{4}{3}+\delta(1-z) \bigl(-\frac{557}{36}+\frac{23 \pi^2}{18}\bigr)+\frac{94}{9} \dd_{0}(z)-\frac{20}{3} \dd_{1}(z)-\frac{7}{3}\h(0;z)-2 \h(1;z)+\frac{22 \h(0;z)}{3 (1-z)}\nl
&\quad+\frac{1}{z}\bigl(-\frac{26}{9}-\frac{4}{3}\h(0;z)-\frac{4}{3} \h(1;z)\bigr)+z \bigl(-\frac{3}{2}-\frac{7}{3} \h(0;z)-2 \h(1;z)\bigr)\Bigr]\nl
+&\Bigl[\frac{10}{9}+\frac{\pi^2}{2}+\delta(1-z) \bigl(-\frac{28613}{648}+\frac{25 \pi^2}{6}+\frac{118 \zeta_{3}}{9}\bigr)+\bigl(\frac{557}{18}-\frac{23 \pi^2}{9}\bigr) \dd_{0}(z)
-\frac{188}{9} \dd_{1}(z)+\frac{20}{3}\dd_{2}(z)\nl
&\quad-\frac{38}{9} \h(0;z)-\frac{8}{3} \h(1;z)-\frac{19}{3} \h(0,0;z)-\frac{14}{3} \h(0,1;z)-\frac{16}{3}\h(1,0;z)-4 \h(1,1;z)\nl
&\quad+\frac{1}{1-z}\bigl(22 \h(0;z)+\frac{50}{3} \h(0,0;z)+\frac{44}{3} \h(0,1;z)+\frac{44}{3} \h(1,0;z)\bigr)+\frac{1}{z}\bigl(-\frac{103}{9}+\frac{5 \pi^2}{9}\nl
&\quad-\frac{52}{9}\h(0;z)-\frac{52}{9} \h(1;z)-\frac{8}{3} \h(0,0;z)-\frac{8}{3} \h(0,1;z)-\frac{8}{3} \h(1,0;z)-\frac{8}{3}\h(1,1;z)\bigr)\nl
&\quad+z\bigl(-\frac{41}{12}+\frac{5 \pi^2}{6}-\frac{32}{9} \h(0;z)-3 \h(1;z)-\frac{16}{3} \h(0,0;z)-\frac{14}{3}\h(0,1;z)-\frac{13}{3} \h(1,0;z)\nl
&\quad-4 \h(1,1;z)\bigr)\Bigr]\nl
&+\mathcal{O}\left(\epsilon\right)
\end{align}
and at subleading colour
\begin{align}
\lefteqn{\mathcal{\tilde{E}}^{0}_{q,q^{\prime}\bar{q}^{\prime}g}=}\nl
+&\frac{1}{\eps^3}\Bigl[-\frac{2}{3} \delta(1-z)\Bigr]\nl
+&\frac{1}{\eps^2}\Bigl[-\frac{19}{9} \delta(1-z)+\frac{4}{3} \dd_{0}(z)-\frac{2}{3 z}\Bigr]\nl
+&\frac{1}{\eps}\Bigl[\delta(1-z) \bigl(-\frac{373}{54}+\frac{5 \pi^2}{9}\bigr)+\frac{38}{9} \dd_{0}(z)-\frac{8}{3} \dd_{1}(z)+\frac{8 \h(0;z)}{3(1-z)}
+\frac{1}{z}\bigl(-\frac{25}{9}-\frac{4}{3} \h(0;z)-\frac{4}{3} \h(1;z)\bigr)\Bigr]\nl
+&\Bigl[+\delta(1-z) \bigl(-\frac{6973}{324}+\frac{95 \pi^2}{54}+\frac{64 \zeta_{3}}{9}\bigr)+\bigl(\frac{373}{27}-\frac{10 \pi^2}{9}\bigr) \dd_{0}(z)-\frac{76}{9} \dd_{1}(z)+\frac{8}{3} \dd_{2}(z)
+\frac{1}{1-z}\bigl(\frac{76}{9}\h(0;z)\nl
&\quad+\frac{16}{3} \h(0,0;z)+\frac{16}{3} \h(0,1;z)+\frac{16}{3} \h(1,0;z)\bigr)+\frac{1}{z}\bigl(-\frac{523}{54}+\frac{5\pi^2}{9}\nl
&\quad-\frac{50}{9} \h(0;z)-\frac{50}{9} \h(1;z)-\frac{8}{3} \h(0,0;z)-\frac{8}{3} \h(0,1;z)-\frac{8}{3}\h(1,0;z)-\frac{8}{3} \h(1,1;z)\bigr)\Bigr]\nl
&+\mathcal{O}\left(\epsilon\right)\,.
\end{align}

Crossing the secondary quark, one obtains the $(q^{\prime},q^{\prime}qg)$
antenna functions, whose integral reads at leading colour
\begin{align}
\lefteqn{\mathcal{E}^{0}_{q^{\prime},q^{\prime}qg}=}\nl
+&\frac{1}{\eps^3}\Bigl[-4+\frac{4}{z}+2 z\Bigr]\nl
+&\frac{1}{\eps^2}\Bigl[-15-\frac{4 z^2}{3}-4 \h(0;z)-8 \h(1;z)+\frac{1}{z}\bigl(\frac{55}{3}+8\h(0;z)+8 \h(1;z)\bigr)\nl
&\quad+z \bigl(\frac{3}{2}+\frac{13}{2} \h(0;z)+4 \h(1;z)\bigr)\Bigr]\nl
+&\frac{1}{\eps}\Bigl[-\frac{235}{6}+\frac{7 \pi^2}{3}-39 \h(0;z)-30\h(1;z)+4 \h(-1,0;z)-8 \h(0,0;z)-8 \h(0,1;z)\nl
&\quad-16 \h(1,0;z)-16 \h(1,1;z)+\frac{1}{z}\bigl(\frac{1537}{36}-3 \pi^2+28 \h(0;z)+\frac{110}{3}\h(1;z)+4 \h(-1,0;z)\nl
&\quad+16 \h(0,0;z)+16 \h(0,1;z)+16 \h(1,0;z)+16 \h(1,1;z)\bigr)+z\bigl(\frac{47}{12}-\frac{7 \pi^2}{3}-\frac{5}{2} \h(0;z)\nl
&\quad+3 \h(1;z)+2 \h(-1,0;z)+\frac{31}{2} \h(0,0;z)+13\h(0,1;z)+8 \h(1,0;z)+8 \h(1,1;z)\bigr)\nl
&\quad+z^2 \bigl(\frac{32}{9}-\frac{16}{3} \h(0;z)-\frac{8}{3} \h(1;z)\bigr)\Bigr]\nl
+&\Bigl[-\frac{2695}{18}+\frac{115 \pi^2}{6}+\frac{170\zeta_{3}}{3}-2 \pi^2 \h(-1;z)-\frac{337}{6} \h(0;z)+\frac{14}{3} \pi^2\h(0;z)-\frac{235}{3} \h(1;z)\nl
&\quad+\frac{20}{3} \pi^2 \h(1;z)-89 \h(0,0;z)-78 \h(0,1;z)-52\h(1,0;z)-60 \h(1,1;z)\nl
&\quad-8\h(-1,-1,0;z)+16 \h(-1,0,0;z)+8 \h(-1,0,1;z)+8 \h(0,-1,0;z)-16 \h(0,0,0;z)\nl
&\quad-16 \h(0,0,1;z)-12\h(0,1,0;z)-16 \h(0,1,1;z)-32 \h(1,0,0;z)-32 \h(1,0,1;z)\nl
&\quad-32 \h(1,1,0;z)-32 \h(1,1,1;z)+\frac{1}{z}\bigl(\frac{37925}{216}-\frac{91\pi^2}{6}-\frac{134 \zeta_{3}}{3}-2 \pi^2 \h(-1;z)\nl
&\quad+\frac{193}{2} \h(0;z)-8 \pi^2 \h(0;z)+\frac{1537}{18} \h(1;z)-\frac{20}{3}\pi^2 \h(1;z)+\frac{4}{3} \h(-1,0;z)+56 \h(0,0;z)\nl
&\quad+56 \h(0,1;z)+\frac{205}{3} \h(1,0;z)+\frac{220}{3}\h(1,1;z)-8 \h(-1,-1,0;z)+16 \h(-1,0,0;z)\nl
&\quad+8 \h(-1,0,1;z)+8 \h(0,-1,0;z)+32 \h(0,0,0;z)+32\h(0,0,1;z)+28 \h(0,1,0;z)\nl
&\quad+32 \h(0,1,1;z)+32 \h(1,0,0;z)+32 \h(1,0,1;z)+32 \h(1,1,0;z)+32\h(1,1,1;z)\bigr)\nl
&\quad+z \bigl(\frac{1069}{72}-\frac{3 \pi^2}{4}-\frac{82 \zeta_{3}}{3}-\pi^2 \h(-1;z)+22 \h(0;z)-\frac{83}{12}\pi^2 \h(0;z)+\frac{47}{6} \h(1;z)\nl
&\quad-\frac{10}{3} \pi^2 \h(1;z)-4 \h(-1,0;z)-\frac{23}{2}\h(0,0;z)-5 \h(0,1;z)+3 \h(1,0;z)+6 \h(1,1;z)\nl
&\quad-4 \h(-1,-1,0;z)+8 \h(-1,0,0;z)+4\h(-1,0,1;z)+4 \h(0,-1,0;z)+\frac{67}{2} \h(0,0,0;z)\nl
&\quad+31 \h(0,0,1;z)+24 \h(0,1,0;z)+26\h(0,1,1;z)+16 \h(1,0,0;z)+16 \h(1,0,1;z)\nl
&\quad+16 \h(1,1,0;z)+16 \h(1,1,1;z)\bigr)+z^2 \bigl(-\frac{208}{27}+2 \pi^2+\frac{128}{9} \h(0;z)+\frac{64}{9} \h(1;z)\nl
&\quad-\frac{8}{3} \h(-1,0;z)-\frac{40}{3}\h(0,0;z)-\frac{32}{3} \h(0,1;z)-\frac{16}{3} \h(1,0;z)-\frac{16}{3} \h(1,1;z)\bigr)\Bigr]\nl
&+\mathcal{O}\left(\epsilon\right)
\end{align}
and at subleading colour
\begin{align}
\lefteqn{\mathcal{\tilde{E}}^{0}_{q^{\prime},q^{\prime}qg}=}\nl
+&\frac{1}{\eps^3}\Bigl[-2+\frac{2}{z}+z\Bigr]\nl
+&\frac{1}{\eps^2}\Bigl[-8-5 \h(0;z)-4 \h(1;z)+\frac{1}{z}\bigl(9+4\h(0;z)+4 \h(1;z)\bigr)+z \bigl(\frac{7}{4}+\frac{5}{2} \h(0;z)+2 \h(1;z)\bigr)\Bigr]\nl
+&\frac{1}{\eps}\Bigl[-\frac{127}{4}+2 \pi^2-16 \h(0;z)-16 \h(1;z)-11 \h(0,0;z)-10 \h(0,1;z)-8 \h(1,0;z)\nl
&\quad-8\h(1,1;z)+\frac{1}{z}\bigl(\frac{75}{2}-\frac{5\pi^2}{3}+18 \h(0;z)+18\h(1;z)+8 \h(0,0;z)+8 \h(0,1;z)\nl
&\quad+8 \h(1,0;z)+8 \h(1,1;z)\bigr)+z \bigl(\frac{7}{2}-\pi^2+\frac{19}{4} \h(0;z)+\frac{7}{2} \h(1;z)+\frac{11}{2} \h(0,0;z)\nl
&\quad+5\h(0,1;z)+4 \h(1,0;z)+4 \h(1,1;z)\bigr)\Bigr]\nl
+&\Bigl[-\frac{499}{4}+\frac{20 \pi^2}{3}+\frac{70\zeta_{3}}{3}-\frac{259}{4} \h(0;z)+\frac{9}{2} \pi^2 \h(0;z)-\frac{127}{2} \h(1;z)+\frac{10}{3}\pi^2 \h(1;z)\nl
&\quad-32 \h(0,0;z)-32 \h(0,1;z)-32 \h(1,0;z)-32 \h(1,1;z)-23 \h(0,0,0;z)\nl
&\quad-22\h(0,0,1;z)-20 \h(0,1,0;z)-20 \h(0,1,1;z)-16 \h(1,0,0;z)-16 \h(1,0,1;z)\nl
&\quad-16 \h(1,1,0;z)-16\h(1,1,1;z)+\frac{1}{z}\bigl(\frac{565}{4}-\frac{15 \pi^2}{2}+75 \h(0;z)-\frac{10}{3} \pi^2 \h(0;z)\nl
&\quad+75\h(1;z)-\frac{10}{3} \pi^2 \h(1;z)+36 \h(0,0;z)+36 \h(0,1;z)+36 \h(1,0;z)+36 \h(1,1;z)\nl
&\quad+16\h(0,0,0;z)+16 \h(0,0,1;z)+16 \h(0,1,0;z)+16 \h(0,1,1;z)+16 \h(1,0,0;z)\nl
&\quad+16 \h(1,0,1;z)+16\h(1,1,0;z)+16 \h(1,1,1;z)-\frac{64 \zeta_{3}}{3}\bigr)+z \bigl(13-\frac{15 \pi^2}{8}+7 \h(0;z)\nl
&\quad-\frac{9}{4}\pi^2 \h(0;z)+7 \h(1;z)-\frac{5}{3} \pi^2 \h(1;z)+\frac{43}{4} \h(0,0;z)+\frac{19}{2} \h(0,1;z)+7\h(1,0;z)\nl
&\quad+7 \h(1,1;z)+\frac{23}{2} \h(0,0,0;z)+11 \h(0,0,1;z)+10 \h(0,1,0;z)+10 \h(0,1,1;z)\nl
&\quad+8\h(1,0,0;z)+8 \h(1,0,1;z)+8 \h(1,1,0;z)+8 \h(1,1,1;z)-\frac{35 \zeta_{3}}{3}\bigr)\Bigr]\nl
&+\mathcal{O}\left(\epsilon\right)\,.
\end{align}

The one-loop virtual antenna functions at NNLO are obtained by
crossing the results from~\cite{ourant}. For the $(q,gg)$-case,
we have the leading colour and the quark loop correction,
which integrate to
\begin{align}
\lefteqn{\mathcal{D}^{1}_{q,gg}=}\nl
+&\frac{1}{\eps^4}\Bigl[-\frac{1}{2}\delta(1-z)\Bigr]\nl
+&\frac{1}{\eps^3}\Bigl[-\frac{1}{2}-\frac{55}{12} \delta(1-z)+\dd_{0}(z)-\frac{z}{2}\Bigr]\nl
+&\frac{1}{\eps^2}\Bigl[-\frac{7}{3}+\delta(1-z) \bigl(-\frac{85}{12}+\frac{7 \pi^2}{12}\bigr)+\frac{11}{2} \dd_{0}(z)-2 \dd_{1}(z)-2\h(0;z)-\h(1;z)
+\frac{4 \h(0;z)}{1-z}+\frac{1}{3 z}\nl
&\quad+z \bigl(-\frac{4}{3}-2 \h(0;z)-\h(1;z)\bigr)\Bigr]\nl
+&\frac{1}{\eps}\Bigl[-\frac{17}{6}+\frac{7 \pi^2}{12}+\delta(1-z) \bigl(-\frac{1967}{108}+\frac{361 \pi^2}{72}+\frac{22 \zeta_{3}}{3}\bigr)+\bigl(\frac{389}{36}
-\frac{7\pi^2}{6}\bigr) \dd_{0}(z)-\frac{22}{3} \dd_{1}(z)+2 \dd_{2}(z)\nl
&\quad-\frac{23}{6} \h(0;z)-\frac{17}{6} \h(1;z)-6 \h(0,0;z)-3 \h(0,1;z)-3 \h(1,0;z)-2\h(1,1;z)\nl
&\quad+\frac{1}{1-z}\bigl(11\h(0;z)+12 \h(0,0;z)+6 \h(0,1;z)+6 \h(1,0;z)\bigr)+\frac{1}{z}\bigl(\frac{14}{9}+\frac{4}{3}\h(0;z)\nl
&\quad+\frac{2}{3} \h(1;z)\bigr)+z \bigl(\frac{11}{6}+\frac{7 \pi^2}{12}+\frac{1}{6}\h(0;z)-\frac{5}{6} \h(1;z)-6 \h(0,0;z)-3 \h(0,1;z)\nl
&\quad-3 \h(1,0;z)-2 \h(1,1;z)\bigr)\Bigr]\nl
+&\Bigl[-\frac{17}{3}+\frac{67 \pi^2}{72}+\frac{25 \zeta_{3}}{3}+\delta(1-z) \bigl(-\frac{14453}{324}+\frac{3023 \pi^2}{432}-\frac{11\pi^4}{72}+\frac{269 \zeta_{3}}{6}\bigr)
+\bigl(\frac{3197}{108}-\frac{52 \pi^2}{9}\nl
&\quad-\frac{44 \zeta_{3}}{3}\bigr) \dd_{0}(z)+\bigl(-\frac{73}{4}+\frac{7\pi^2}{3}\bigr) \dd_{1}(z)+\frac{11}{2} \dd_{2}(z)-\frac{4}{3} \dd_{3}(z)-\frac{9}{2} \h(0;z)+\frac{7}{3} \pi^2
\h(0;z)-\frac{23}{6} \h(1;z)\nl
&\quad+\frac{4}{3} \pi^2 \h(1;z)-\frac{79}{6} \h(0,0;z)-\frac{29}{6} \h(0,1;z)-\frac{61}{6}\h(1,0;z)-\frac{23}{6} \h(1,1;z)\nl
&\quad-14\h(0,0,0;z)-8 \h(0,0,1;z)-6 \h(0,1,0;z)-5 \h(0,1,1;z)-6 \h(1,0,0;z)\nl
&\quad-5 \h(1,0,1;z)-4\h(1,1,0;z)-4 \h(1,1,1;z)+\frac{1}{1-z}\bigl(\frac{11 \pi^2}{18}-2 \zeta_{3}+\frac{925}{36} \h(0;z)\nl
&\quad-\frac{14}{3} \pi^2 \h(0;z)-\frac{1}{3}\pi^2 \h(1;z)+\frac{88}{3} \h(0,0;z)+\frac{44}{3} \h(0,1;z)+\frac{55}{3} \h(1,0;z)\nl
&\quad+28 \h(0,0,0;z)+16\h(0,0,1;z)+12 \h(0,1,0;z)+10 \h(0,1,1;z)+12 \h(1,0,0;z)\nl
&\quad+10 \h(1,0,1;z)+8 \h(1,1,0;z)\bigr)+\frac{1}{z}\bigl(\frac{145}{54}-\frac{11 \pi^2}{18}+\frac{53}{18} \h(0;z)+\frac{5}{2}\h(1;z)\nl
&\quad+\frac{10}{3} \h(0,0;z)+2 \h(0,1;z)+\frac{4}{3} \h(1,0;z)+\frac{4}{3} \h(1,1;z)\bigr)+z\bigl(\frac{79 \pi^2}{72}+\frac{25 \zeta_{3}}{3}+\frac{11}{6} \h(0;z)\nl
&\quad+\frac{7}{3} \pi^2 \h(0;z)+\frac{11}{6}\h(1;z)+\frac{4}{3} \pi^2 \h(1;z)+\frac{17}{6} \h(0,0;z)+\frac{7}{6} \h(0,1;z)-\frac{1}{6} \h(1,0;z)\nl
&\quad+\frac{1}{6}\h(1,1;z)-14 \h(0,0,0;z)-8 \h(0,0,1;z)-6 \h(0,1,0;z)-5 \h(0,1,1;z)\nl
&\quad-6 \h(1,0,0;z)-5\h(1,0,1;z)-4 \h(1,1,0;z)-4 \h(1,1,1;z)\bigr)\Bigr]\nl
&+\mathcal{O}\left(\epsilon\right)
\end{align}
and
\begin{align}
\lefteqn{\mathcal{\hat{D}}^{1}_{q,gg}=}\nl
+&\frac{1}{\eps^3}\Bigl[\frac{2}{3} \delta(1-z)\Bigr]\nl
+&\frac{1}{\eps^2}\Bigl[\frac{1}{3}+\frac{11}{18} \delta(1-z)-\frac{2}{3} \dd_{0}(z)+\frac{z}{3}\Bigr]\nl
+&\frac{1}{\eps}\Bigl[\frac{1}{3}+\delta(1-z) \bigl(\frac{143}{108}-\frac{\pi^2}{2}\bigr)-\frac{11}{18} \dd_{0}(z)+\frac{2}{3} \dd_{1}(z)+\frac{1}{3}\h(0;z)+\frac{1}{3}\h(1;z)
-\frac{2 \h(0;z)}{3 (1-z)}-\frac{1}{9 z}\nl
&\quad+z \bigl(-\frac{1}{3}+\frac{1}{3} \h(0;z)+\frac{1}{3} \h(1;z)\bigr)\Bigr]\nl
+&\Bigl[\frac{1}{2}-\frac{\pi^2}{4}+\delta(1-z) \bigl(\frac{979}{324}-\frac{11 \pi^2}{24}-\frac{14 \zeta_{3}}{9}\bigr)+\bigl(-\frac{38}{27}+\frac{\pi^2}{2}\bigr)\dd_{0}(z)+\frac{11}{18} \dd_{1}(z)-\frac{1}{3} \dd_{2}(z)+\frac{1}{3} \h(0;z)\nl
&\quad+\frac{1}{3} \h(1;z)+\frac{1}{3} \h(0,0;z)+\frac{1}{3} \h(0,1;z)+\frac{1}{3} \h(1,0;z)+\frac{1}{3} \h(1,1;z)+\frac{1}{1-z}\bigl(-\frac{11}{18}\h(0;z)\nl
&\quad-\frac{2}{3} \h(0,0;z)-\frac{2}{3} \h(0,1;z)-\frac{2}{3} \h(1,0;z)\bigr)+\frac{1}{z}\bigl(-\frac{2}{27}-\frac{1}{9} \h(0;z)-\frac{1}{9}\h(1;z)\bigr)+z\bigl(
-\frac{\pi^2}{4}\nl
&\quad-\frac{1}{3} \h(0;z)-\frac{1}{3} \h(1;z)+\frac{1}{3} \h(0,0;z)+\frac{1}{3} \h(0,1;z)+\frac{1}{3}\h(1,0;z)+\frac{1}{3} \h(1,1;z)\bigr)\Bigr]\nl
&+\mathcal{O}\left(\epsilon\right)\,.
\end{align}

For the $(q,q^\prime\bar q^\prime)$-case (primary quark
in the initial state), there are leading and
subleading colour as well as  quark loop contributions. Their integrals
are:
\begin{align}
\lefteqn{\mathcal{E}^{1}_{q,q^{\prime}\bar{q}^{\prime}}=}\nl
+&\frac{1}{\eps^2}\Bigl[\frac{11}{18} \delta(1-z)\Bigr]\nl
+&\frac{1}{\eps}\Bigl[\frac{119}{108} \delta(1-z)-\frac{11}{18} \dd_{0}(z)-\frac{2 \h(0;z)}{3(1-z)}+\frac{1}{z}\bigl(\frac{2}{9}+\frac{1}{3} \h(0;z)\bigr)\Bigr]\nl
+&\Bigl[-\frac{1}{6}+\delta(1-z) \bigl(\frac{787}{324}-\frac{17 \pi^2}{24}\bigr)+\frac{2 \pi^2}{9}-\frac{32}{27} \dd_{0}(z)+\frac{11}{18}\dd_{1}(z)-\frac{1}{6} \h(0;z)
+\frac{2}{3} \h(0,0;z)\nl
&\quad+\frac{2}{3} \h(1,0;z)+\frac{1}{1-z}\bigl(-\frac{\pi^2}{9}-\frac{31}{18} \h(0;z)-\frac{8}{3}\h(0,0;z)-\frac{2}{3} \h(0,1;z)-\frac{4}{3} \h(1,0;z)\bigr)\nl
&\quad+\frac{1}{z}\bigl(\frac{119}{216}-\frac{\pi^2}{18}+\frac{37}{36} \h(0;z)-\frac{1}{9} \h(1;z)+\h(0,0;z)+\frac{1}{3}\h(0,1;z)+\frac{1}{3} \h(1,0;z)\bigr)\nl
&\quad+z \bigl(\frac{\pi^2}{18}+\frac{1}{6} \h(0,0;z)+\frac{1}{6}\h(1,0;z)\bigr)\Bigr]\nl
&+\mathcal{O}\left(\epsilon\right)\,,
\end{align}
\begin{align}
\lefteqn{\mathcal{\tilde{E}}^{1}_{q,q^{\prime}\bar{q}^{\prime}}=}\nl
+&\frac{1}{\eps^3}\Bigl[\frac{1}{6}\delta(1-z)\Bigr]\nl
+&\frac{1}{\eps^2}\Bigl[\frac{19}{36} \delta(1-z)-\frac{1}{3} \dd_{0}(z)+\frac{1}{6 z}\Bigr]\nl
+&\frac{1}{\eps}\Bigl[\delta(1-z) \bigl(\frac{173}{108}-\frac{5 \pi^2}{36}\bigr)-\frac{19}{18} \dd_{0}(z)+\frac{2}{3} \dd_{1}(z)-\frac{2 \h(0;z)}{3(1-z)}
+\frac{1}{z}\bigl(\frac{25}{36}+\frac{1}{3} \h(0;z)+\frac{1}{3} \h(1;z)\bigr)\Bigr]\nl
+&\Bigl[\delta(1-z) \bigl(\frac{343}{81}-\frac{95 \pi^2}{216}-\frac{7 \zeta_{3}}{9}\bigr)+\bigl(-\frac{173}{54}+\frac{5 \pi^2}{18}\bigr)\dd_{0}(z)+\frac{19}{9} \dd_{1}(z)
-\frac{2}{3} \dd_{2}(z)+\frac{1}{1-z}\bigl(-\frac{19}{9} \h(0;z)\nl
&\quad-\frac{4}{3} \h(0,0;z)-\frac{4}{3}\h(0,1;z)-\frac{4}{3} \h(1,0;z)\bigr)+\frac{1}{z}\bigl(\frac{62}{27}-\frac{5 \pi^2}{36}+\frac{25}{18} \h(0;z)+\frac{25}{18}\h(1;z)\nl
&\quad+\frac{2}{3} \h(0,0;z)+\frac{2}{3} \h(0,1;z)+\frac{2}{3} \h(1,0;z)+\frac{2}{3} \h(1,1;z)\bigr)\Bigr]\nl
&+\mathcal{O}\left(\epsilon\right)
\end{align}
and
\begin{align}
\lefteqn{\mathcal{\hat{E}}^{1}_{q,q^{\prime}\bar{q}^{\prime}}=}\nl
+&\frac{1}{\eps}\Bigl[\frac{5}{27} \delta(1-z)-\frac{1}{9} \dd_{0}(z)+\frac{1}{18 z}\Bigr]\nl
+&\Bigl[\delta(1-z) \bigl(\frac{53}{81}-\frac{\pi^2}{108}\bigr)-\frac{5}{9} \dd_{0}(z)+\frac{1}{3} \dd_{1}(z)-\frac{\h(0;z)}{3(1-z)}+\frac{1}{z}\bigl(\frac{1}{3}+\frac{1}{6} \h(0;z)
+\frac{1}{6} \h(1;z)\bigr)\Bigr]\nl
&+\mathcal{O}\left(\epsilon\right)\,.
\end{align}

Likewise, the $(q^\prime, q^\prime q)$-case (secondary quark
in the initial state), there are leading and
subleading colour as well as  quark loop contributions, which integrate to:
\begin{align}
\lefteqn{\mathcal{E}^{1}_{q^{\prime},q^{\prime}q}=}\nl
+&\frac{1}{\eps^2}\Bigl[\frac{3}{4}+\h(0;z)+\frac{1}{2} \h(1;z)+\frac{1}{z}\bigl(-\frac{3}{4}-\h(0;z)-\frac{1}{2} \h(1;z)\bigr)-z \bigl(\frac{3}{8}+\frac{1}{2}
\h(0;z)+\frac{1}{4} \h(1;z)\bigr)\Bigr]\nl
+&\frac{1}{\eps}\Bigl[-\frac{41}{36}-\frac{\pi^2}{3}+\frac{5}{3} \h(0;z)+\frac{7}{4} \h(1;z)+4 \h(0,0;z)+2 \h(0,1;z)+2\h(1,0;z)+\frac{3}{2} \h(1,1;z)\nl
&\quad+\frac{1}{z}\bigl(\frac{29}{36}+\frac{\pi^2}{3}-\frac{8}{3} \h(0;z)-\frac{9}{4} \h(1;z)-4 \h(0,0;z)-2\h(0,1;z)-2 \h(1,0;z)\nl
&\quad-\frac{3}{2} \h(1,1;z)\bigr)+z \bigl(\frac{10}{9}+\frac{\pi^2}{6}+\frac{1}{6} \h(0;z)-\frac{3}{8}\h(1;z)-2 \h(0,0;z)-\h(0,1;z)\nl
&\quad-\h(1,0;z)-\frac{3}{4} \h(1,1;z)\bigr)\Bigr]\nl
+&\Bigl[-\frac{347}{54}-\frac{43 \pi^2}{12}-5 \zeta_{3}-\frac{1}{36} \h(0;z)-\frac{11}{6} \pi^2 \h(0;z)+\frac{49}{36}\h(1;z)-\frac{3}{4} \pi^2 \h(1;z)+\frac{3}{2} \h(0,0;z)\nl
&\quad+\frac{11}{3} \h(0,1;z)-\frac{1}{3} \h(1,0;z)+\frac{15}{4}\h(1,1;z)+10 \h(0,0,0;z)+6 \h(0,0,1;z)\nl
&\quad+4 \h(0,1,0;z)+4 \h(0,1,1;z)+5 \h(1,0,0;z)+4\h(1,0,1;z)+4 \h(1,1,0;z)\nl
&\quad+\frac{7}{2} \h(1,1,1;z)+\frac{1}{z}\bigl(\frac{871}{108}+\frac{41 \pi^2}{12}+5 \zeta_{3}-\frac{31}{18} \h(0;z)+\frac{11}{6}\pi^2 \h(0;z)-\frac{115}{36} \h(1;z)\nl
&\quad+\frac{3}{4} \pi^2 \h(1;z)-\frac{15}{2} \h(0,0;z)-\frac{17}{3}\h(0,1;z)-\frac{19}{6} \h(1,0;z)-\frac{21}{4} \h(1,1;z)\nl
&\quad-10 \h(0,0,0;z)-6 \h(0,0,1;z)-4\h(0,1,0;z)-4 \h(0,1,1;z)-5 \h(1,0,0;z)\nl
&\quad-4 \h(1,0,1;z)-4 \h(1,1,0;z)-\frac{7}{2} \h(1,1,1;z)\bigr)+z\bigl(\frac{121}{54}+\frac{9 \pi^2}{8}+\frac{5 \zeta_{3}}{2}+\frac{20}{9} \h(0;z)\nl
&\quad+\frac{11}{12} \pi^2 \h(0;z)+\frac{10}{9}\h(1;z)+\frac{3}{8} \pi^2 \h(1;z)+\frac{9}{4} \h(0,0;z)+\frac{1}{6} \h(0,1;z)+\frac{7}{6} \h(1,0;z)\nl
&\quad-\frac{3}{8}\h(1,1;z)-5 \h(0,0,0;z)-3 \h(0,0,1;z)-2 \h(0,1,0;z)-2 \h(0,1,1;z)\nl
&\quad-\frac{5}{2} \h(1,0,0;z)-2\h(1,0,1;z)-2 \h(1,1,0;z)-\frac{7}{4} \h(1,1,1;z)\bigr)\Bigr]\nl
&+\mathcal{O}\left(\epsilon\right)\,,
\end{align}
\begin{align}
\lefteqn{\mathcal{\tilde{E}}^{1}_{q^{\prime},q^{\prime}q}=}\nl
+&\frac{1}{\eps^3}\Bigl[\frac{1}{2}-\frac{1}{2 z}-\frac{z}{4}\Bigr]\nl
+&\frac{1}{\eps^2}\Bigl[\frac{7}{4}+\h(0;z)+\frac{1}{2} \h(1;z)-\frac{1}{z}\bigl(\frac{9}{4}+\h(0;z)+\frac{1}{2} \h(1;z)\bigr)-z \bigl(\frac{3}{8}+\frac{1}{2}\h(0;z)
+\frac{1}{4} \h(1;z)\bigr)\Bigr]\nl
+&\frac{1}{\eps}\Bigl[\frac{13}{2}-\frac{\pi^2}{4}+\frac{7}{2} \h(0;z)+\frac{7}{4} \h(1;z)+2 \h(0,0;z)+\h(0,1;z)+\h(1,0;z)+\frac{1}{2} \h(1,1;z)\nl
&\quad+\frac{1}{z}\bigl(-\frac{35}{4}+\frac{\pi^2}{4}-\frac{9}{2}\h(0;z)-\frac{9}{4} \h(1;z)-2 \h(0,0;z)-\h(0,1;z)-\h(1,0;z)\nl
&\quad-\frac{1}{2} \h(1,1;z)\bigr)+z\bigl(-1+\frac{\pi^2}{8}-\frac{3}{4} \h(0;z)-\frac{3}{8} \h(1;z)-\h(0,0;z)-\frac{1}{2} \h(0,1;z)\nl
&\quad-\frac{1}{2}\h(1,0;z)-\frac{1}{4} \h(1,1;z)\bigr)\Bigr]\nl
+&\Bigl[\frac{43}{2}-\frac{7 \pi^2}{8}-\frac{13 \zeta_{3}}{3}+13 \h(0;z)-\frac{1}{2} \pi^2 \h(0;z)+\frac{13}{2}\h(1;z)-\frac{1}{4} \pi^2 \h(1;z)+7 \h(0,0;z)\nl
&\quad+\frac{7}{2} \h(0,1;z)+\frac{7}{2} \h(1,0;z)+\frac{7}{4}\h(1,1;z)+4 \h(0,0,0;z)+2 \h(0,0,1;z)+2 \h(0,1,0;z)\nl
&\quad+\h(0,1,1;z)+2 \h(1,0,0;z)+\h(1,0,1;z)+\h(1,1,0;z)+\frac{1}{2} \h(1,1,1;z)+\frac{1}{z}\bigl(-\frac{121}{4}\nl
&\quad+\frac{9\pi^2}{8}+\frac{13 \zeta_{3}}{3}-\frac{35}{2} \h(0;z)+\frac{1}{2} \pi^2 \h(0;z)-\frac{35}{4} \h(1;z)+\frac{1}{4}\pi^2 \h(1;z)-9 \h(0,0;z)\nl
&\quad-\frac{9}{2} \h(0,1;z)-\frac{9}{2} \h(1,0;z)-\frac{9}{4} \h(1,1;z)-4\h(0,0,0;z)-2 \h(0,0,1;z)-2 \h(0,1,0;z)\nl
&\quad-\h(0,1,1;z)-2 \h(1,0,0;z)-\h(1,0,1;z)-\h(1,1,0;z)-\frac{1}{2}\h(1,1,1;z)\bigr)+z \bigl(-2\nl
&\quad+\frac{3 \pi^2}{16}+\frac{13 \zeta_{3}}{6}-2 \h(0;z)+\frac{1}{4} \pi^2 \h(0;z)-\h(1;z)+\frac{1}{8}\pi^2 \h(1;z)-\frac{3}{2} \h(0,0;z)\nl
&\quad-\frac{3}{4} \h(0,1;z)-\frac{3}{4} \h(1,0;z)-\frac{3}{8}\h(1,1;z)-2 \h(0,0,0;z)-\h(0,0,1;z)-\h(0,1,0;z)\nl
&\quad-\frac{1}{2} \h(0,1,1;z)-\h(1,0,0;z)-\frac{1}{2}\h(1,0,1;z)-\frac{1}{2} \h(1,1,0;z)-\frac{1}{4} \h(1,1,1;z)\bigr)\Bigr]\nl
&+\mathcal{O}\left(\epsilon\right)
\end{align}
and
\begin{align}
\lefteqn{\mathcal{\hat{E}}^{1}_{q^{\prime},q^{\prime}q}=}\nl
+&\frac{1}{\eps}\Bigl[\frac{13}{18}+\frac{1}{3}\h(0;z)-\frac{1}{z}\bigl(\frac{8}{9}+\frac{1}{3} \h(0;z)\bigr)-z \bigl(\frac{5}{18}+\frac{1}{6} \h(0;z)\bigr)\Bigr]\nl
+&\Bigl[\frac{85}{27}+\frac{\pi^2}{12}+\frac{35}{18} \h(0;z)+\frac{13}{18} \h(1;z)+\h(0,0;z)+\frac{1}{3}\h(0,1;z)+\frac{1}{3} \h(1,0;z)+\frac{1}{z}\bigl(-\frac{118}{27}\nl
&\quad-\frac{\pi^2}{12}-\frac{22}{9} \h(0;z)-\frac{8}{9} \h(1;z)-\h(0,0;z)-\frac{1}{3}\h(0,1;z)-\frac{1}{3} \h(1,0;z)\bigr)\nl
&\quad+z \bigl(-\frac{14}{27}-\frac{\pi^2}{24}-\frac{5}{9} \h(0;z)-\frac{5}{18}\h(1;z)-\frac{1}{2} \h(0,0;z)-\frac{1}{6} \h(0,1;z)-\frac{1}{6} \h(1,0;z)\bigr)\Bigr]\nl
&+\mathcal{O}\left(\epsilon\right)\,.
\end{align}
\subsection{Gluon-gluon antennae}
The gluon-gluon antenna functions can contribute to quark-initiated
processes, since they also contain configurations with
the splitting of a gluon into a
quark-antiquark pair.

At NLO, one has only the $(q,qg)$ gluon-gluon antenna function, whose integral
is
\begin{align}
\lefteqn{\mathcal{G}^{0}_{q,qg}=}\nl
&\frac{1}{\eps}\Bigl[-1+\frac{1}{z}+\frac{z}{2}\Bigr]\nl
+&\Bigl[-1-\h(0;z)-\h(1;z)+\frac{1}{z}\bigl(\frac{7}{4}+\h(0;z)+\h(1;z)\bigr)+z \bigl(\frac{1}{2} \h(0;z)+\frac{1}{2} \h(1;z)\bigr)\Bigr]\nl
+&\eps\Bigl[-2+\frac{\pi ^2}{4}-\h(0;z)-\h(1;z)-\h(0,0;z)-\h(0,1;z)-\h(1,0;z)-\h(1,1;z)\nl
&\quad+\frac{1}{z}\bigl(\frac{7}{2}-\frac{\pi ^2}{4}+\frac{7}{4} \h(0;z)+\frac{7}{4} \h(1;z)+\h(0,0;z)+\h(0,1;z)+\h(1,0;z)+\h(1,1;z)\bigr)\nl
&\quad+z \bigl(-\frac{\pi ^2}{8}+\frac{1}{2} \h(0,0;z)+\frac{1}{2} \h(0,1;z)+\frac{1}{2} \h(1,0;z)+\frac{1}{2} \h(1,1;z)\bigr)\Bigr]\nl
+&\eps^{2}\Bigl[-4+\frac{\pi ^2}{4}+\frac{7 \zeta_{3}}{3}-2 \h(0;z)+\frac{1}{4} \pi ^2 \h(0;z)-2 \h(1;z)+\frac{1}{4} \pi ^2 \h(1;z)-\h(0,0;z)\nl
&\quad-\h(0,1;z)-\h(1,0;z)-\h(1,1;z)-\h(0,0,0;z)-\h(0,0,1;z)-\h(0,1,0;z)\nl
&\quad-\h(0,1,1;z)-\h(1,0,0;z)-\h(1,0,1;z)-\h(1,1,0;z)-\h(1,1,1;z)+\frac{1}{z}\bigl(7-\frac{7 \pi ^2}{16}\nl
&\quad-\frac{7 \zeta_{3}}{3}+\frac{7}{2} \h(0;z)-\frac{1}{4} \pi ^2 \h(0;z)+\frac{7}{2} \h(1;z)-\frac{1}{4} \pi ^2 \h(1;z)+\frac{7}{4} \h(0,0;z)+\frac{7}{4} \h(0,1;z)\nl
&\quad+\frac{7}{4} \h(1,0;z)+\frac{7}{4} \h(1,1;z)+\h(0,0,0;z)+\h(0,0,1;z)+\h(0,1,0;z)+\h(0,1,1;z)\nl
&\quad+\h(1,0,0;z)+\h(1,0,1;z)+\h(1,1,0;z)+\h(1,1,1;z)\bigr)+z \bigl(-\frac{7 \zeta_{3}}{6}-\frac{1}{8} \pi ^2 \h(0;z)\nl
&\quad-\frac{1}{8} \pi ^2 \h(1;z)+\frac{1}{2} \h(0,0,0;z)+\frac{1}{2} \h(0,0,1;z)+\frac{1}{2} \h(0,1,0;z)+\frac{1}{2} \h(0,1,1;z)\nl
&\quad+\frac{1}{2} \h(1,0,0;z)+\frac{1}{2} \h(1,0,1;z)+\frac{1}{2} \h(1,1,0;z)+\frac{1}{2} \h(1,1,1;z)\bigr)\Bigr]\nl
&+\mathcal{O}\left(\epsilon^{3}\right)\,.
\end{align}

At NNLO, the double real radiation
$G_4^0$-type and $H_4^0$=type antenna functions allow for
crossings with initial state quarks.

The $(q,qgg)$ at leading colour integrates to
\begin{align}
\lefteqn{\mathcal{G}^{0}_{q,qgg}=}\nl
+&\frac{1}{\eps^3}\Bigl[-8+\frac{8}{z}+4 z\Bigr]\nl
+&\frac{1}{\eps^2}\Bigl[-\frac{82}{3}-9 \h(0;z)-16 \h(1;z)+\frac{1}{z}\bigl(\frac{104}{3}+16\h(0;z)+16 \h(1;z)\bigr)+z \bigl(\frac{53}{12}+\frac{21}{2} \h(0;z)\nl
&\quad+8 \h(1;z)\bigr)-\frac{4}{3} z^2\Bigr]\nl
+&\frac{1}{\eps}\Bigl[-\frac{2291}{36}+\frac{13 \pi^2}{3}-\frac{185}{3} \h(0;z)-\frac{164}{3} \h(1;z)+4 \h(-1,0;z)-15 \h(0,0;z)\nl
&\quad-18 \h(0,1;z)-32 \h(1,0;z)-32\h(1,1;z)+\frac{1}{z}\bigl(\frac{499}{6}-6 \pi^2+\frac{164}{3}\h(0;z)+\frac{208}{3} \h(1;z)\nl
&\quad+4 \h(-1,0;z)+32 \h(0,0;z)+32 \h(0,1;z)+32 \h(1,0;z)+32\h(1,1;z)\bigr)+z \bigl(\frac{301}{36}\nl
&\quad-\frac{23 \pi^2}{6}+\frac{61}{12} \h(0;z)+\frac{53}{6} \h(1;z)+2 \h(-1,0;z)+\frac{47}{2}\h(0,0;z)+21 \h(0,1;z)\nl
&\quad+16 \h(1,0;z)+16 \h(1,1;z)\bigr)+z^2 \bigl(\frac{32}{9}-\frac{16}{3} \h(0;z)-\frac{8}{3}\h(1;z)\bigr)\Bigr]\nl
+&\Bigl[-\frac{6625}{27}+\frac{256 \pi^2}{9}+\frac{286 \zeta_{3}}{3}-2 \pi^2 \h(-1;z)-\frac{3181}{36} \h(0;z)+\frac{55}{6} \pi^2\h(0;z)-\frac{2291}{18} \h(1;z)\nl
&\quad+\frac{38}{3} \pi^2 \h(1;z)-8 \h(-1,0;z)-\frac{373}{3} \h(0,0;z)-\frac{370}{3}\h(0,1;z)-\frac{298}{3} \h(1,0;z)\nl
&\quad-\frac{328}{3}\h(1,1;z)-8 \h(-1,-1,0;z)+16 \h(-1,0,0;z)+8 \h(-1,0,1;z)+16 \h(0,-1,0;z)\nl
&\quad-27 \h(0,0,0;z)-30\h(0,0,1;z)-28 \h(0,1,0;z)-36 \h(0,1,1;z)-60 \h(1,0,0;z)\nl
&\quad-64 \h(1,0,1;z)-64 \h(1,1,0;z)-64\h(1,1,1;z)+\frac{1}{z}\bigl(\frac{33167}{108}-\frac{250 \pi^2}{9}-\frac{268 \zeta_{3}}{3}\nl
&\quad-2 \pi^2 \h(-1;z)+\frac{1633}{9} \h(0;z)-16\pi^2 \h(0;z)+\frac{499}{3} \h(1;z)-\frac{38}{3} \pi^2 \h(1;z)\nl
&\quad-\frac{32}{3} \h(-1,0;z)+\frac{328}{3}\h(0,0;z)+\frac{328}{3} \h(0,1;z)+\frac{398}{3} \h(1,0;z)+\frac{416}{3} \h(1,1;z)\nl
&\quad-8 \h(-1,-1,0;z)+16\h(-1,0,0;z)+8 \h(-1,0,1;z)+64 \h(0,0,0;z)+64 \h(0,0,1;z)\nl
&\quad+56 \h(0,1,0;z)+64 \h(0,1,1;z)+60\h(1,0,0;z)+64 \h(1,0,1;z)+64 \h(1,1,0;z)\nl
&\quad+64 \h(1,1,1;z)\bigr)+z \bigl(\frac{7411}{216}-\frac{199\pi^2}{72}-\frac{149\zeta_{3}}{3}-\pi^2 \h(-1;z)+\frac{511}{18} \h(0;z)\nl
&\quad-\frac{131}{12} \pi^2 \h(0;z)+\frac{301}{18} \h(1;z)-\frac{19}{3}\pi^2 \h(1;z)+\frac{65}{12} \h(0,0;z)+\frac{61}{6} \h(0,1;z)\nl
&\quad+\frac{50}{3} \h(1,0;z)+\frac{53}{3}\h(1,1;z)-4 \h(-1,-1,0;z)+8 \h(-1,0,0;z)+4 \h(-1,0,1;z)\nl
&\quad+\frac{99}{2} \h(0,0,0;z)+47 \h(0,0,1;z)+38\h(0,1,0;z)+42 \h(0,1,1;z)+30 \h(1,0,0;z)\nl
&\quad+32 \h(1,0,1;z)+32 \h(1,1,0;z)+32 \h(1,1,1;z)\bigr)+z^2 \bigl(-\frac{208}{27}+2 \pi^2+\frac{128}{9} \h(0;z)\nl
&\quad+\frac{64}{9} \h(1;z)-\frac{8}{3}\h(-1,0;z)-\frac{40}{3} \h(0,0;z)-\frac{32}{3} \h(0,1;z)-\frac{16}{3} \h(1,0;z)-\frac{16}{3} \h(1,1;z)\bigr)\Bigr]\nl
&+\mathcal{O}\left(\epsilon\right)\,,
\end{align}
while the subleading colour is
\begin{align}
\lefteqn{\mathcal{\tilde{G}}^{0}_{q,qgg}=}\nl
+&\frac{1}{\eps^3}\Bigl[-4+\frac{4}{z}+2 z\Bigr]\nl
+&\frac{1}{\eps^2}\Bigl[-12-10 \h(0;z)-8 \h(1;z)+\frac{1}{z}\bigl(16+8 \h(0;z)+8\h(1;z)\bigr)+z \bigl(\frac{7}{2}+5 \h(0;z)+4 \h(1;z)\bigr)\Bigr]\nl
+&\frac{1}{\eps}\Bigl[-\frac{91}{2}+4 \pi^2-22 \h(0;z)-24 \h(1;z)-22 \h(0,0;z)-20 \h(0,1;z)-16 \h(1,0;z)\nl
&\quad-16\h(1,1;z)+\frac{1}{z}\bigl(67-\frac{10 \pi^2}{3}+32 \h(0;z)+32 \h(1;z)+16 \h(0,0;z)+16\h(0,1;z)\nl
&\quad+16 \h(1,0;z)+16 \h(1,1;z)\bigr)+z \bigl(\frac{11}{2}-2 \pi^2+\frac{19}{2} \h(0;z)+7 \h(1;z)+11 \h(0,0;z)\nl
&\quad+10 \h(0,1;z)+8\h(1,0;z)+8 \h(1,1;z)\bigr)\Bigr]\nl
+&\Bigl[-184+\frac{32 \pi^2}{3}+\frac{140 \zeta_{3}}{3}-\frac{4}{3} \pi^2 \h(-1;z)-\frac{185}{2} \h(0;z)+9 \pi^2 \h(0;z)-91\h(1;z)\nl
&\quad+8 \pi^2 \h(1;z)+8 \h(-1,0;z)-38 \h(0,0;z)-44 \h(0,1;z)-44 \h(1,0;z)\nl
&\quad-48\h(1,1;z)-16 \h(-1,-1,0;z)+8 \h(-1,0,0;z)+16 \h(0,-1,0;z)-46 \h(0,0,0;z)\nl
&\quad-44 \h(0,0,1;z)-40\h(0,1,0;z)-40 \h(0,1,1;z)-16 \h(1,0,0;z)-32 \h(1,0,1;z)\nl
&\quad-24 \h(1,1,0;z)-32 \h(1,1,1;z)+\frac{1}{z}\bigl(\frac{485}{2}-\frac{40\pi^2}{3}-\frac{80 \zeta_{3}}{3}-\frac{4}{3} \pi^2 \h(-1;z)+134 \h(0;z)\nl
&\quad-\frac{20}{3} \pi^2 \h(0;z)+134 \h(1;z)-8\pi^2 \h(1;z)+64 \h(0,0;z)+64 \h(0,1;z)+64 \h(1,0;z)\nl
&\quad+64 \h(1,1;z)-16 \h(-1,-1,0;z)+8\h(-1,0,0;z)+32 \h(0,0,0;z)+32 \h(0,0,1;z)\nl
&\quad+32 \h(0,1,0;z)+32 \h(0,1,1;z)+16 \h(1,0,0;z)+32\h(1,0,1;z)+24 \h(1,1,0;z)\nl
&\quad+32 \h(1,1,1;z)\bigr)+z \bigl(\frac{97}{4}-\frac{7 \pi^2}{4}-\frac{46 \zeta_{3}}{3}-\frac{2}{3}\pi^2 \h(-1;z)+11 \h(0;z)-\frac{9}{2} \pi^2 \h(0;z)\nl
&\quad+11 \h(1;z)-4 \pi^2 \h(1;z)+8\h(-1,0;z)+\frac{31}{2} \h(0,0;z)+19 \h(0,1;z)+16 \h(1,0;z)\nl
&\quad+14 \h(1,1;z)-8 \h(-1,-1,0;z)+4\h(-1,0,0;z)+23 \h(0,0,0;z)+22 \h(0,0,1;z)\nl
&\quad+20 \h(0,1,0;z)+20 \h(0,1,1;z)+8 \h(1,0,0;z)+16\h(1,0,1;z)+12 \h(1,1,0;z)\nl
&\quad+16 \h(1,1,1;z)\bigr)\Bigr]\nl
&+\mathcal{O}\left(\epsilon\right)\,.
\end{align}

The four-quark antenna function contains two different quark flavours. Since
it is symmetric under the interchange of the quark-antiquark pairs,
we have to consider only one crossing, resulting in
$(q,qq^{\prime}\bar{q}^{\prime})$, which integrates to
\begin{align}
\lefteqn{\mathcal{H}^{0}_{q,qq^{\prime}\bar{q}^{\prime}}=}\nl
+&\frac{1}{\eps^2}\Bigl[\frac{4}{3}-\frac{4}{3 z}-\frac{2 z}{3}\Bigr]\nl
+&\frac{1}{\eps}\Bigl[\frac{50}{9}+\frac{8}{3} \h(0;z)+\frac{8}{3} \h(1;z)+\frac{1}{z}\bigl(-\frac{50}{9}-\frac{8}{3} \h(0;z)-\frac{8}{3} \h(1;z)\bigr)+z
\bigl(-\frac{19}{9}-\frac{4}{3} \h(0;z)\nl
&\quad-\frac{4}{3} \h(1;z)\bigr)\Bigr]\nl
+&\Bigl[\frac{505}{27}-\frac{10 \pi^2}{9}+\frac{100}{9} \h(0;z)+\frac{100}{9} \h(1;z)+\frac{16}{3} \h(0,0;z)+\frac{16}{3}\h(0,1;z)+\frac{16}{3} \h(1,0;z)\nl
&\quad+\frac{16}{3} \h(1,1;z)+\frac{1}{z}\bigl(-\frac{505}{27}+\frac{10 \pi^2}{9}-\frac{100}{9} \h(0;z)-\frac{100}{9}\h(1;z)-\frac{16}{3} \h(0,0;z)\nl
&\quad-\frac{16}{3} \h(0,1;z)-\frac{16}{3} \h(1,0;z)-\frac{16}{3} \h(1,1;z)\bigr)+z\bigl(-\frac{355}{54}+\frac{5 \pi^2}{9}\nl
&\quad-\frac{38}{9} \h(0;z)-\frac{38}{9} \h(1;z)-\frac{8}{3} \h(0,0;z)-\frac{8}{3}\h(0,1;z)-\frac{8}{3} \h(1,0;z)-\frac{8}{3} \h(1,1;z)\bigr)\Bigr]\nl
&+\mathcal{O}\left(\epsilon\right)\,.
\end{align}

The quark-initiated one-loop virtual gluon-gluon antenna of the form
$(q,qg)$ have a leading and subleading colour contribution and a quark loop
piece. Their integrals are:
\begin{align}
\lefteqn{\mathcal{G}^{1}_{q,qg}=}\nl
+&\frac{1}{\eps^2}\Bigl[\frac{3}{4}+\h(0;z)+\frac{1}{2} \h(1;z)+\frac{1}{z}\bigl(-\frac{3}{4}-\h(0;z)-\frac{1}{2} \h(1;z)\bigr)-z \bigl(\frac{3}{8}+\frac{1}{2}\h(0;z)
+\frac{1}{4} \h(1;z)\bigr)\Bigr]\nl
+&\frac{1}{\eps}\Bigl[-\frac{53}{36}-\frac{\pi^2}{3}+\frac{2}{3} \h(0;z)+\frac{5}{4} \h(1;z)+4 \h(0,0;z)+2 \h(0,1;z)+2\h(1,0;z)+\frac{3}{2} \h(1,1;z)\nl
&\quad+\frac{1}{z}\bigl(\frac{79}{72}+\frac{\pi^2}{3}-\frac{13}{6} \h(0;z)-2 \h(1;z)-4 \h(0,0;z)-2 \h(0,1;z)-2\h(1,0;z)\nl
&\quad-\frac{3}{2} \h(1,1;z)\bigr)+z \bigl(\frac{10}{9}+\frac{\pi^2}{6}+\frac{1}{6} \h(0;z)-\frac{3}{8}\h(1;z)-2 \h(0,0;z)-\h(0,1;z)\nl
&\quad-\h(1,0;z)-\frac{3}{4} \h(1,1;z)\bigr)\Bigr]\nl
+&\Bigl[-\frac{571}{108}-\frac{35 \pi^2}{12}-5 \zeta_{3}-\frac{73}{36} \h(0;z)-\frac{11}{6} \pi^2 \h(0;z)+\frac{1}{36}\h(1;z)-\frac{3}{4} \pi^2 \h(1;z)-\h(0,0;z)\nl
&\quad+\frac{5}{3} \h(0,1;z)-\frac{5}{6} \h(1,0;z)+\frac{9}{4}\h(1,1;z)+10 \h(0,0,0;z)+6 \h(0,0,1;z)+4 \h(0,1,0;z)\nl
&\quad+4 \h(0,1,1;z)+5 \h(1,0,0;z)+4\h(1,0,1;z)+4 \h(1,1,0;z)+\frac{7}{2} \h(1,1,1;z)\nl
&\quad+\frac{1}{z}\bigl(\frac{3733}{432}+\frac{37 \pi^2}{12}+5 \zeta_{3}-\frac{43}{72} \h(0;z)+\frac{11}{6}\pi^2 \h(0;z)-\frac{25}{9} \h(1;z)+\frac{3}{4} \pi^2 \h(1;z)\nl
&\quad-6 \h(0,0;z)-\frac{14}{3} \h(0,1;z)-\frac{19}{6}\h(1,0;z)-\frac{9}{2} \h(1,1;z)-10 \h(0,0,0;z)\nl
&\quad-6 \h(0,0,1;z)-4 \h(0,1,0;z)-4 \h(0,1,1;z)-5\h(1,0,0;z)-4 \h(1,0,1;z)\nl
&\quad-4 \h(1,1,0;z)-\frac{7}{2} \h(1,1,1;z)\bigr)+z \bigl(\frac{121}{54}+\frac{23\pi^2}{24}+\frac{5 \zeta_{3}}{2}+\frac{20}{9} \h(0;z)+\frac{11}{12} \pi^2 \h(0;z)\nl
&\quad+\frac{10}{9} \h(1;z)+\frac{3}{8}\pi^2 \h(1;z)+\frac{3}{2} \h(0,0;z)+\frac{1}{6} \h(0,1;z)+\frac{5}{12} \h(1,0;z)-\frac{3}{8}\h(1,1;z)\nl
&\quad-5 \h(0,0,0;z)-3 \h(0,0,1;z)-2 \h(0,1,0;z)-2 \h(0,1,1;z)-\frac{5}{2} \h(1,0,0;z)\nl
&\quad-2\h(1,0,1;z)-2 \h(1,1,0;z)-\frac{7}{4} \h(1,1,1;z)\bigr)\Bigr]\nl
&+\mathcal{O}\left(\epsilon\right)\,,
\end{align}
\begin{align}
\lefteqn{\mathcal{\tilde{G}}^{1}_{q,qg}=}\nl
+&\frac{1}{\eps^3}\Bigl[\frac{1}{2}-\frac{1}{2 z}-\frac{z}{4}\Bigr]\nl
+&\frac{1}{\eps^2}\Bigl[\frac{5}{4}+\h(0;z)+\frac{1}{2} \h(1;z)-\frac{1}{z}\bigl(2+\h(0;z)+\frac{1}{2} \h(1;z)\bigr)-z \bigl(\frac{3}{8}+\frac{1}{2} \h(0;z)
+\frac{1}{4}\h(1;z)\bigr)\Bigr]\nl
+&\frac{1}{\eps}\Bigl[\frac{17}{4}-\frac{\pi^2}{4}+\frac{5}{2} \h(0;z)+\frac{5}{4} \h(1;z)+2 \h(0,0;z)+\h(0,1;z)+\h(1,0;z)+\frac{1}{2} \h(1,1;z)\nl
&\quad+\frac{1}{z}\bigl(-\frac{31}{4}+\frac{\pi^2}{4}-4\h(0;z)-2 \h(1;z)-2 \h(0,0;z)-\h(0,1;z)-\h(1,0;z)\nl
&\quad-\frac{1}{2} \h(1,1;z)\bigr)+z\bigl(-1+\frac{\pi^2}{8}-\frac{3}{4} \h(0;z)-\frac{3}{8} \h(1;z)-\h(0,0;z)-\frac{1}{2} \h(0,1;z)\nl
&\quad-\frac{1}{2}\h(1,0;z)-\frac{1}{4} \h(1,1;z)\bigr)\Bigr]\nl
+&\Bigl[\frac{51}{4}-\frac{19 \pi^2}{24}-\frac{10 \zeta_{3}}{3}+\frac{35}{4} \h(0;z)-\frac{1}{2} \pi^2 \h(0;z)+\frac{17}{4}\h(1;z)-\frac{7}{12} \pi^2 \h(1;z)
+\frac{9}{2} \h(0,0;z)\nl
&\quad+\frac{5}{2} \h(0,1;z)+2 \h(1,0;z)+\frac{5}{4}\h(1,1;z)+4 \h(0,0,0;z)+2 \h(0,0,1;z)+2 \h(0,1,0;z)\nl
&\quad+\h(0,1,1;z)+\h(1,0,0;z)+\h(1,0,1;z)+\frac{1}{2} \h(1,1,1;z)+\frac{1}{z}\bigl(-\frac{415}{16}+\pi^2+\frac{10\zeta_{3}}{3}\nl
&\quad-\frac{31}{2} \h(0;z)+\frac{1}{2} \pi^2 \h(0;z)-\frac{67}{8} \h(1;z)+\frac{7}{12} \pi^2\h(1;z)-8 \h(0,0;z)-4 \h(0,1;z)\nl
&\quad-4 \h(1,0;z)-2 \h(1,1;z)-4 \h(0,0,0;z)-2 \h(0,0,1;z)-2\h(0,1,0;z)-\h(0,1,1;z)\nl
&\quad-\h(1,0,0;z)-\h(1,0,1;z)-\frac{1}{2} \h(1,1,1;z)\bigr)+z \bigl(-2+\frac{5\pi^2}{48}+\frac{5 \zeta_{3}}{3}-2 \h(0;z)\nl
&\quad+\frac{1}{4} \pi^2 \h(0;z)-\h(1;z)+\frac{7}{24} \pi^2\h(1;z)-\frac{7}{4} \h(0,0;z)-\frac{3}{4} \h(0,1;z)-\h(1,0;z)\nl
&\quad-\frac{3}{8} \h(1,1;z)-2\h(0,0,0;z)-\h(0,0,1;z)-\h(0,1,0;z)-\frac{1}{2} \h(0,1,1;z)\nl
&\quad-\frac{1}{2} \h(1,0,0;z)-\frac{1}{2}\h(1,0,1;z)-\frac{1}{4} \h(1,1,1;z)\bigr)\Bigr]\nl
&+\mathcal{O}\left(\epsilon\right)\,,
\end{align}
\begin{align}
\lefteqn{\mathcal{\hat{G}}^{1}_{q,qg}=}\nl
+&\frac{1}{\eps}\Bigl[\frac{5}{9}+\frac{1}{3}\h(0;z)-\frac{1}{z}\bigl(\frac{29}{36}+\frac{1}{3} \h(0;z)\bigr)-z \bigl(\frac{5}{18}+\frac{1}{6} \h(0;z)\bigr)\Bigr]\nl
+&\Bigl[\frac{52}{27}+\frac{\pi^2}{4}+\frac{13}{9} \h(0;z)+\frac{5}{9} \h(1;z)+\h(0,0;z)+\frac{1}{3}\h(0,1;z)+\frac{1}{3} \h(1,0;z)+\frac{1}{z}\bigl(-\frac{415}{108}\nl
&\quad-\frac{\pi^2}{4}-\frac{79}{36} \h(0;z)-\frac{29}{36} \h(1;z)-\h(0,0;z)-\frac{1}{3}\h(0,1;z)-\frac{1}{3} \h(1,0;z)\bigr)+z \bigl(-\frac{14}{27}\nl
&\quad-\frac{\pi^2}{8}-\frac{5}{9} \h(0;z)-\frac{5}{18}\h(1;z)-\frac{1}{2} \h(0,0;z)-\frac{1}{6} \h(0,1;z)-\frac{1}{6} \h(1,0;z)\bigr)\Bigr]\nl
&+\mathcal{O}\left(\epsilon\right)\,.
\end{align}
\section{Gluon initiated antennae}
\label{sec:gi}

The gluon-initiated initial-final antenna functions are obtained from final-final antenna functions
listed in~\cite{ourant}
by crossing a gluon into the initial state. As in the quark case, their unintegrated forms
are pure kinematical crossings of the final-final expressions, with no symmetry factors
or polarization sums multiplied on them. Also, we do not provide a decomposition into
sub-antennae here. The precise decomposition depends on the requirements and symmetries of the process under consideration, and is normally performed in the context of an actual calculation.

As in the quark-initiated case, some attention has to be paid in the crossing of
the virtual one-loop antenna functions from the final-final kinematics~\cite{ourant} to
the initial-final kinematics. This crossing requires the analytical continuation of the
polylogarithmic functions to the relevant phase space, again consisting of
 three different Riemann
sheets~\cite{allMC,distensor}.

All gluon-initiated initial-final antenna functions are listed in Table~\ref{tab:two}, their
integrated forms are collected in the following.

\subsection{Quark-quark antennae}
The NLO quark-antiquark antenna function to order $\e^2$ is:
\begin{align}
\lefteqn{\mathcal{A}^{0}_{g,q\bar{q}}=}\nl
+&\frac{1}{\eps}\Bigl[1-2 z+2 z^2\Bigr]\nl
+&\Bigl[\bigl(1-2z+2z^{2}\bigr)\bigl(\h(0;z)+\h(1;z)\bigr)\Bigr]\nl
+&\eps\Bigl[2+\bigl(1-2z+2z^{2}\bigr)\bigl(-\frac{\pi ^2}{4}+\h(0,0;z)+\h(0,1;z)+\h(1,0;z)+\h(1,1;z)\bigr)\Bigr]\nl
+&\eps^{2}\Bigl[4+2 \h(0;z)+2 \h(1;z)+\bigl(1-2z+2z^{2}\bigr)\bigl(-\frac{7\zeta_{3}}{3}-\frac{1}{4} \pi ^2 \h(0;z)-\frac{1}{4} \pi ^2 \h(1;z)\nl
&\quad+\h(0,0,0;z)+\h(0,0,1;z)+\h(0,1,0;z)+\h(0,1,1;z)+\h(1,0,0;z)+\h(1,0,1;z)\nl
&\quad+\h(1,1,0;z)+\h(1,1,1;z)\bigr)\Bigr]\nl
&+\mathcal{O}\left(\epsilon^{3}\right)\,.
\end{align}
For the NNLO double real radiation antenna functions, we have the leading and
subleading colour contributions to $(g,gq\bar q)$, whose integrals read
\begin{align}
\lefteqn{\mathcal{A}^{0}_{g,gq\bar{q}}=}\nl
+&\frac{1}{\eps^3}\Bigl[3-6 z+6 z^2\Bigr]\nl
+&\frac{1}{\eps^2}\Bigl[\frac{13}{4}-\frac{4}{3 z}+\frac{7}{2} \h(0;z)+6 \h(1;z)+z\bigl(-15-19 \h(0;z)-12 \h(1;z)\bigr)\nl
&\quad+z^2 \bigl(\frac{49}{3}+12 \h(0;z)+12 \h(1;z)\bigr)\Bigr]\nl
+&\frac{1}{\eps}\Bigl[\frac{173}{12}-2 \pi^2+\frac{27}{4} \h(0;z)+\frac{13}{2} \h(1;z)-2 \h(-1,0;z)+\frac{11}{2} \h(0,0;z)+7 \h(0,1;z)\nl
&\quad+11 \h(1,0;z)+12 \h(1,1;z)+\frac{1}{z}\bigl(\frac{8}{9}-\frac{8}{3} \h(1;z)\bigr)+z \bigl(-\frac{89}{12}+\frac{22 \pi^2}{3}-16 \h(0;z)\nl
&\quad-30 \h(1;z)-4 \h(-1,0;z)-43 \h(0,0;z)-38 \h(0,1;z)-22 \h(1,0;z)-24 \h(1,1;z)\bigr)\nl
&\quad+z^2 \bigl(\frac{91}{9}-\frac{17 \pi^2}{3}+\frac{142}{3} \h(0;z)+\frac{98}{3}\h(1;z)-4 \h(-1,0;z)+26 \h(0,0;z)+24 \h(0,1;z)\nl
&\quad+22 \h(1,0;z)+24 \h(1,1;z)\bigr)\Bigr]\nl
+&\Bigl[\frac{1895}{36}-\frac{67 \pi^2}{24}-41 \zeta_{3}+\pi^2 \h(-1;z)+\frac{9}{4} \h(0;z)-\frac{15}{4} \pi^2 \h(0;z)+\frac{173}{6}\h(1;z)-5 \pi^2 \h(1;z)\nl
&\quad+2 \h(-1,0;z)+\frac{63}{4} \h(0,0;z)+\frac{27}{2}\h(0,1;z)+13 \h(1,0;z)+13 \h(1,1;z)\nl
&\quad+4 \h(-1,-1,0;z)-8 \h(-1,0,0;z)-4 \h(-1,0,1;z)-8 \h(0,-1,0;z)+\frac{19}{2} \h(0,0,0;z)\nl
&\quad+11 \h(0,0,1;z)+10\h(0,1,0;z)+14 \h(0,1,1;z)+17 \h(1,0,0;z)+22 \h(1,0,1;z)\nl
&\quad+22 \h(1,1,0;z)+24 \h(1,1,1;z)+\frac{2}{3 z^2} \h(-1,0;z)+\frac{1}{z}\bigl(-\frac{502}{27}+\frac{2 \pi^2}{3}-\frac{2}{3} \h(0;z)\nl
&\quad+\frac{16}{9} \h(1;z)+\frac{8}{3}\h(-1,0;z)-\frac{16}{3} \h(1,0;z)-\frac{16}{3} \h(1,1;z)\bigr)+z \bigl(-\frac{1135}{18}+\frac{185 \pi^2}{18}+86 \zeta_{3}\nl
&\quad+2 \pi^2 \h(-1;z)-\frac{217}{4} \h(0;z)+\frac{121}{6} \pi^2 \h(0;z)-\frac{89}{6} \h(1;z)+10 \pi^2 \h(1;z)\nl
&\quad+\frac{56}{3} \h(-1,0;z)-\frac{116}{3} \h(0,0;z)-32 \h(0,1;z)-62 \h(1,0;z)-60 \h(1,1;z)\nl
&\quad+8 \h(-1,-1,0;z)-16 \h(-1,0,0;z)-8 \h(-1,0,1;z)-91 \h(0,0,0;z)-86 \h(0,0,1;z)\nl
&\quad-68 \h(0,1,0;z)-76 \h(0,1,1;z)-34 \h(1,0,0;z)-44 \h(1,0,1;z)-44 \h(1,1,0;z)\nl
&\quad-48 \h(1,1,1;z)\bigr)+z^2 \bigl(\frac{2329}{27}-\frac{37 \pi^2}{2}-92 \zeta_{3}+2 \pi^2 \h(-1;z)-\frac{86}{9} \h(0;z)-\frac{40}{3} \pi^2 \h(0;z)\nl
&\quad+\frac{182}{9} \h(1;z)-10 \pi^2 \h(1;z)+\frac{44}{3} \h(-1,0;z)+\frac{328}{3} \h(0,0;z)+\frac{284}{3} \h(0,1;z)\nl
&\quad+\frac{196}{3}\h(1,0;z)+\frac{196}{3} \h(1,1;z)+8 \h(-1,-1,0;z)-16 \h(-1,0,0;z)-8 \h(-1,0,1;z)\nl
&\quad-16 \h(0,-1,0;z)+54 \h(0,0,0;z)+52 \h(0,0,1;z)+40 \h(0,1,0;z)+48 \h(0,1,1;z)\nl
&\quad+34 \h(1,0,0;z)+44 \h(1,0,1;z)+44\h(1,1,0;z)+48 \h(1,1,1;z)\bigr)\Bigr]\nl
&+\mathcal{O}\left(\epsilon\right)
\end{align}
and
\begin{align}
\lefteqn{\mathcal{\tilde{A}}^{0}_{g,gq\bar{q}}=}\nl
+&\frac{1}{\eps^3}\Bigl[4-8 z+8 z^2\Bigr]\nl
+&\frac{1}{\eps^2}\Bigl[\frac{13}{2}+7 \h(0;z)+8 \h(1;z)-z \bigl(14+14 \h(0;z)+16 \h(1;z)\bigr)\nl
&\quad+z^2 (12+16 \h(0;z)+16 \h(1;z))\Bigr]\nl
+&\frac{1}{\eps}\Bigl[\frac{39}{2}-\frac{7 \pi^2}{3}+\frac{15}{2} \h(0;z)+13 \h(1;z)+15 \h(0,0;z)+14 \h(0,1;z)+18 \h(1,0;z)\nl
&\quad+16 \h(1,1;z)+z \bigl(-\frac{75}{2}+\frac{14 \pi^2}{3}-32 \h(0;z)-28 \h(1;z)-30 \h(0,0;z)-28 \h(0,1;z)\nl
&\quad-36 \h(1,0;z)-32 \h(1,1;z)\bigr)+z^2 \bigl(38-\frac{16 \pi^2}{3}+24 \h(0;z)+24 \h(1;z)+36 \h(0,0;z)\nl
&\quad+32 \h(0,1;z)+36 \h(1,0;z)+32 \h(1,1;z)\bigr)\Bigr]\nl
+&\Bigl[\frac{301}{6}-\frac{9 \pi^2}{4}-\frac{86 \zeta_{3}}{3}+\frac{4}{3} \pi^2 \h(-1;z)+\frac{199}{6} \h(0;z)-\frac{37}{6} \pi^2 \h(0;z)+39 \h(1;z)-\frac{22}{3} \pi^2 \h(1;z)\nl
&\quad-24 \h(-1,0;z)+\frac{27}{2} \h(0,0;z)+15 \h(0,1;z)+30 \h(1,0;z)+26 \h(1,1;z)\nl
&\quad+16 \h(-1,-1,0;z)-8 \h(-1,0,0;z)-16 \h(0,-1,0;z)+31 \h(0,0,0;z)+30 \h(0,0,1;z)\nl
&\quad+28 \h(0,1,0;z)+28 \h(0,1,1;z)+26 \h(1,0,0;z)+36 \h(1,0,1;z)+28 \h(1,1,0;z)\nl
&\quad+32 \h(1,1,1;z)+\frac{4}{3 z^2}\h(-1,0;z)+\frac{1}{z}\bigl(\frac{4}{3}-\frac{4}{3} \h(0;z)\bigr)+z \bigl(-99+\frac{77 \pi^2}{9}+\frac{76 \zeta_{3}}{3}\nl
&\quad+\frac{8}{3} \pi^2 \h(-1;z)-\frac{153}{2} \h(0;z)+\frac{37}{3} \pi^2 \h(0;z)-75 \h(1;z)+\frac{44}{3} \pi^2 \h(1;z)-\frac{56}{3} \h(-1,0;z)\nl
&\quad-\frac{160}{3} \h(0,0;z)-64 \h(0,1;z)-60 \h(1,0;z)-56 \h(1,1;z)+32 \h(-1,-1,0;z)\nl
&\quad-16 \h(-1,0,0;z)-62 \h(0,0,0;z)-60 \h(0,0,1;z)-56 \h(0,1,0;z)-56 \h(0,1,1;z)\nl
&\quad-52 \h(1,0,0;z)-72 \h(1,0,1;z)-56 \h(1,1,0;z)-64 \h(1,1,1;z)\bigr)+z^2 \bigl(98-10 \pi^2\nl
&\quad-\frac{208 \zeta_{3}}{3}+\frac{4}{3} \pi^2 \h(-1;z)+76 \h(0;z)-\frac{44}{3} \pi^2 \h(0;z)+76 \h(1;z)-\frac{40}{3} \pi^2 \h(1;z)\nl
&\quad+48 \h(0,0;z)+48 \h(0,1;z)+48 \h(1,0;z)+48 \h(1,1;z)+16 \h(-1,-1,0;z)\nl
&\quad-8 \h(-1,0,0;z)-16 \h(0,-1,0;z)+76 \h(0,0,0;z)+72 \h(0,0,1;z)+64 \h(0,1,0;z)\nl
&\quad+64 \h(0,1,1;z)+68 \h(1,0,0;z)+72 \h(1,0,1;z)+64 \h(1,1,0;z)+64 \h(1,1,1;z)\bigr)\Bigr]\nl
&+\mathcal{O}\left(\epsilon\right)\,.
\end{align}
The one-loop corrections to the $(q,q\bar q)$ antenna function have a leading and
subleading colour term as well as a closed quark loop contribution. These integrate to
\begin{align}
\lefteqn{\mathcal{A}^{1}_{g,q\bar{q}}=}\nl
+&\frac{1}{\eps^3}\Bigl[-\frac{1}{2}+z-z^2\Bigr]\nl
+&\frac{1}{\eps^2}\Bigl[-\frac{7}{3}-\frac{3}{2} \h(0;z)-\frac{1}{2} \h(1;z)+z\bigl(\frac{11}{3}+3 \h(0;z)+\h(1;z)\bigr)\nl
&\quad+z^2 \bigl(-\frac{11}{3}-3 \h(0;z)-\h(1;z)\bigr)\Bigr]\nl
+&\frac{1}{\eps}\Bigl[-4+\frac{\pi^2}{2}-\frac{10}{3} \h(0;z)-\frac{7}{3} \h(1;z)-4 \h(0,0;z)-\frac{3}{2} \h(0,1;z)-\h(1,0;z)-\frac{1}{2}\h(1,1;z)\nl
&\quad+z \bigl(\frac{1}{4}-\pi^2+\frac{11}{3} \h(0;z)+\frac{11}{3} \h(1;z)+8 \h(0,0;z)+3\h(0,1;z)+2 \h(1,0;z)+\h(1,1;z)\bigr)\nl
&\quad+z^2\bigl(\pi^2-\frac{11}{3} \h(0;z)-\frac{11}{3} \h(1;z)-8 \h(0,0;z)-3 \h(0,1;z)-2 \h(1,0;z)-\h(1,1;z)\bigr)\Bigr]\nl
+&\Bigl[-\frac{227}{12}+\frac{19 \pi^2}{24}+\frac{53 \zeta_{3}}{6}-11 \h(0;z)+\frac{19}{12} \pi^2 \h(0;z)-4\h(1;z)+\frac{7}{12} \pi^2 \h(1;z)-\frac{19}{3} \h(0,0;z)\nl
&\quad-\frac{10}{3} \h(0,1;z)-\frac{10}{3}\h(1,0;z)-\frac{7}{3} \h(1,1;z)-9 \h(0,0,0;z)-4 \h(0,0,1;z)-2 \h(0,1,0;z)\nl
&\quad-\frac{3}{2}\h(0,1,1;z)-\h(1,0,0;z)-\h(1,0,1;z)-\frac{1}{2} \h(1,1,1;z)+z \bigl(\frac{1}{2}-\frac{7\pi^2}{12}-\frac{53 \zeta_{3}}{3}\nl
&\quad+\frac{1}{2} \h(0;z)-\frac{19}{6} \pi^2 \h(0;z)+\frac{1}{4} \h(1;z)-\frac{7}{6}\pi^2 \h(1;z)+\frac{14}{3} \h(0,0;z)+\frac{11}{3} \h(0,1;z)\nl
&\quad+\frac{14}{3} \h(1,0;z)+\frac{11}{3}\h(1,1;z)+18 \h(0,0,0;z)+8 \h(0,0,1;z)+4 \h(0,1,0;z)\nl
&\quad+3 \h(0,1,1;z)+2 \h(1,0,0;z)+2\h(1,0,1;z)+\h(1,1,1;z)\bigr)+z^2 \bigl(\frac{11 \pi^2}{12}+\frac{53 \zeta_{3}}{3}\nl
&\quad+\frac{19}{6}\pi^2 \h(0;z)+\frac{7}{6} \pi^2 \h(1;z)-\frac{11}{3} \h(0,0;z)-\frac{11}{3} \h(0,1;z)-\frac{11}{3}\h(1,0;z)\nl
&\quad-\frac{11}{3} \h(1,1;z)-18 \h(0,0,0;z)-8 \h(0,0,1;z)-4 \h(0,1,0;z)-3 \h(0,1,1;z)\nl
&\quad-2\h(1,0,0;z)-2 \h(1,0,1;z)-\h(1,1,1;z)\bigr)\Bigr]\nl
&+\mathcal{O}\left(\epsilon\right)
\end{align}
and
\begin{align}
\lefteqn{\mathcal{\tilde{A}}^{1}_{g,q\bar{q}}=}\nl
+&\frac{1}{\eps^2}\Bigl[-\frac{1}{2} \h(0;z)-\frac{1}{2} \h(1;z)+z \bigl(\h(0;z)+\h(1;z)\bigr)-z^2 \bigl(\h(0;z)+\h(1;z)\bigr)\Bigr]\nl
+&\frac{1}{\eps}\Bigl[\frac{\pi^2}{12}-\frac{1}{2} \h(0;z)-\frac{1}{2} \h(1;z)-2 \h(0,0;z)-\frac{3}{2} \h(0,1;z)-2\h(1,0;z)-\frac{3}{2}\h(1,1;z)\nl
&\quad+z \bigl(-\frac{1}{4}-\frac{\pi^2}{6}+4 \h(0,0;z)+3 \h(0,1;z)+4 \h(1,0;z)+3 \h(1,1;z)\bigr)\nl
&\quad+z^2 \bigl(\frac{\pi^2}{6}-4 \h(0,0;z)-3 \h(0,1;z)-4 \h(1,0;z)-3 \h(1,1;z)\bigr)\Bigr]\nl
+&\Bigl[\frac{5}{4}-\frac{\pi^2}{12}-\frac{3 \zeta_{3}}{2}-3 \h(0;z)+\frac{3}{4} \pi^2 \h(0;z)-4 \h(1;z)+\frac{13}{12}\pi^2 \h(1;z)-\frac{5}{2} \h(0,0;z)\nl
&\quad-\frac{3}{2} \h(0,1;z)-\frac{5}{2} \h(1,0;z)-\frac{3}{2}\h(1,1;z)-5 \h(0,0,0;z)-4 \h(0,0,1;z)\nl
&\quad-4 \h(0,1,0;z)-\frac{7}{2} \h(0,1,1;z)-4 \h(1,0,0;z)-4\h(1,0,1;z)-3 \h(1,1,0;z)\nl
&\quad-\frac{7}{2} \h(1,1,1;z)+z \bigl(-\frac{3}{2}+\frac{\pi^2}{3}+3
\zeta_{3}-\frac{1}{2} \h(0;z)-\frac{3}{2} \pi^2 \h(0;z)-\frac{1}{4} \h(1;z)-\frac{13}{6} \pi^2 \h(1;z)\nl
&\quad+\h(0,0;z)+\h(1,0;z)+10\h(0,0,0;z)+8 \h(0,0,1;z)+8 \h(0,1,0;z)+7 \h(0,1,1;z)\nl
&\quad+8 \h(1,0,0;z)+8 \h(1,0,1;z)+6\h(1,1,0;z)+7 \h(1,1,1;z)\bigr)+z^2 \bigl(-\zeta_{3}+\h(0;z)\nl
&\quad+\frac{3}{2} \pi^2 \h(0;z)+\h(1;z)+\frac{3}{2}\pi^2 \h(1;z)-10 \h(0,0,0;z)-8 \h(0,0,1;z)-8 \h(0,1,0;z)\nl
&\quad-7 \h(0,1,1;z)-10 \h(1,0,0;z)-8\h(1,0,1;z)-8 \h(1,1,0;z)-7 \h(1,1,1;z)\bigr)\Bigr]\nl
&+\mathcal{O}\left(\epsilon\right)
\end{align}
and
\begin{align}
\lefteqn{\mathcal{\hat{A}}^{1}_{g,q\bar{q}}=}\nl
+&\frac{1}{\eps^2}\Bigl[\frac{1}{3}-\frac{2 z}{3}+\frac{2 z^2}{3}\Bigr]\nl
+&\frac{1}{\eps}\Bigl[\frac{1}{3} \h(0;z)+\frac{1}{3} \h(1;z)+z \bigl(-\frac{2}{3} \h(0;z)-\frac{2}{3} \h(1;z)\bigr)+z^2 \bigl(\frac{2}{3}\h(0;z)+\frac{2}{3} \h(1;z)\bigr)\Bigr]\nl
+&\Bigl[\frac{2}{3}-\frac{\pi^2}{12}+\frac{1}{3} \h(0,0;z)+\frac{1}{3} \h(0,1;z)+\frac{1}{3} \h(1,0;z)+\frac{1}{3}\h(1,1;z)+z\bigl(\frac{\pi^2}{6}-\frac{2}{3} \h(0,0;z)\nl
&\quad-\frac{2}{3} \h(0,1;z)-\frac{2}{3} \h(1,0;z)-\frac{2}{3} \h(1,1;z)\bigr)+z^2\bigl(-\frac{\pi^2}{6}+\frac{2}{3} \h(0,0;z)+\frac{2}{3} \h(0,1;z)\nl
&\quad+\frac{2}{3} \h(1,0;z)+\frac{2}{3} \h(1,1;z)\bigr)\Bigr]\nl
&+\mathcal{O}\left(\epsilon\right)\,.
\end{align}

\subsection{Quark-gluon antennae}
The integrated gluon-initiated $(g,qq)$ antenna function at NLO is through to order $\e^2$:
\begin{align}
\lefteqn{\mathcal{D}^{0}_{g,gq}=}\nl
+&\frac{1}{\eps^2}\Bigl[\delta(1-z)\Bigr]\nl
+&\frac{1}{\eps}\Bigl[\frac{5}{2}+\frac{3}{4}\delta(1-z)-\dd_{0}(z)-\frac{1}{z}-2 z+2 z^2\Bigr]\nl
+&\Bigl[-\frac{1}{2}+\delta(1-z)\bigl(\frac{7}{4}-\frac{\pi^2}{4}\bigr)-\frac{3}{4} \dd_{0}(z)+\dd_{1}(z)+\frac{5}{2} \h(0;z)+\frac{5}{2} \h(1;z)-\frac{1}{1-z}\h(0;z)\nl
&\quad-\frac{1}{z}\bigl(\frac{3}{4}+\h(0;z)+\h(1;z)\bigr)+\bigl(-z+z^2\bigr)\bigl(2 \h(0;z)+2 \h(1;z)\bigr)\Bigr]\nl
+&\eps\Bigl[1-\frac{5 \pi ^2}{8}+\delta(1-z)\bigl(\frac{7}{2}-\frac{3 \pi ^2}{16}-\frac{7 \zeta_{3}}{3}\bigr)+\bigl(-\frac{7}{4}+\frac{\pi ^2}{4}\bigr) \dd_{0}(z)+\frac{3}{4} \dd_{1}(z)
-\frac{1}{2} \dd_{2}(z)-\frac{1}{2}\h(0;z)\nl
&\quad-\frac{1}{2} \h(1;z)+\frac{5}{2} \h(0,0;z)+\frac{5}{2} \h(0,1;z)+\frac{5}{2}\h(1,0;z)+\frac{5}{2}\h(1,1;z)-\frac{1}{1-z}\bigl(\frac{3}{4} \h(0;z)\nl
&\quad+\h(0,0;z)+\h(0,1;z)+\h(1,0;z)\bigr)+\frac{1}{z}\bigl(-\frac{7}{4}+\frac{\pi^2}{4}-\frac{3}{4} \h(0;z)-\frac{3}{4} \h(1;z)\nl
&\quad-\h(0,0;z)-\h(0,1;z)-\h(1,0;z)-\h(1,1;z)\bigr)+\bigl(z-z^2\bigr)\bigl(\frac{\pi ^2}{2}-2 \h(0,0;z)\nl
&\quad-2 \h(0,1;z)-2 \h(1,0;z)-2 \h(1,1;z)\bigr)\Bigr]\nl
+&\eps^{2}\Bigl[3+\frac{\pi ^2}{8}-\frac{35 \zeta_{3}}{6}+\delta(1-z)\bigl(7-\frac{7 \pi ^2}{16}-\frac{\pi ^4}{96}-\frac{7\zeta_{3}}{4}\bigr)+\bigl(-\frac{7}{2}+\frac{3 \pi ^2}{16}+\frac{7 \zeta_{3}}{3}\bigr)\dd_{0}(z)+\bigl(\frac{7}{4}-\frac{\pi ^2}{4}\bigr) \dd_{1}(z)\nl
&\quad-\frac{3}{8} \dd_{2}(z)+\frac{1}{6} \dd_{3}(z)+\h(0;z)-\frac{5}{8}\pi ^2 \h(0;z)+\h(1;z)-\frac{5}{8} \pi ^2 \h(1;z)-\frac{1}{2} \h(0,0;z)\nl
&\quad-\frac{1}{2} \h(0,1;z)-\frac{1}{2}\h(1,0;z)-\frac{1}{2} \h(1,1;z)+\frac{5}{2} \h(0,0,0;z)+\frac{5}{2} \h(0,0,1;z)+\frac{5}{2} \h(0,1,0;z)\nl
&\quad+\frac{5}{2}\h(0,1,1;z)+\frac{5}{2} \h(1,0,0;z)+\frac{5}{2} \h(1,0,1;z)+\frac{5}{2}\h(1,1,0;z)+\frac{5}{2} \h(1,1,1;z)\nl
&\quad+\frac{1}{1-z}\bigl(-\frac{7}{4} \h(0;z)+\frac{1}{4}\pi ^2 \h(0;z)-\frac{3}{4} \h(0,0;z)-\frac{3}{4} \h(0,1;z)-\frac{3}{4} \h(1,0;z)\nl
&\quad-\h(0,0,0;z)-\h(0,0,1;z)-\h(0,1,0;z)-\h(0,1,1;z)-\h(1,0,0;z)-\h(1,0,1;z)\nl
&\quad-\h(1,1,0;z)\bigr)+\frac{1}{z}\bigl(-\frac{7}{2}+\frac{3\pi ^2}{16}+\frac{7\zeta_{3}}{3}-\frac{7}{4} \h(0;z)+\frac{1}{4} \pi ^2 \h(0;z)-\frac{7}{4} \h(1;z)+\frac{1}{4} \pi ^2 \h(1;z)\nl
&\quad-\frac{3}{4}\h(0,0;z)-\frac{3}{4} \h(0,1;z)-\frac{3}{4} \h(1,0;z)-\frac{3}{4} \h(1,1;z)-\h(0,0,0;z)-\h(0,0,1;z)\nl
&\quad-\h(0,1,0;z)-\h(0,1,1;z)-\h(1,0,0;z)-\h(1,0,1;z)-\h(1,1,0;z)-\h(1,1,1;z)\bigr)\nl
&\quad+\bigl(z-z^2\bigr)\bigl(\frac{14\zeta_{3}}{3}+\frac{1}{2} \pi ^2 \h(0;z)+\frac{1}{2} \pi ^2 \h(1;z)-2 \h(0,0,0;z)-2 \h(0,0,1;z)-2\h(0,1,0;z)\nl
&\quad-2 \h(0,1,1;z)-2 \h(1,0,0;z)-2 \h(1,0,1;z)-2 \h(1,1,0;z)-2 \h(1,1,1;z)\bigr)\Bigr]\nl
&+\mathcal{O}\left(\epsilon^{3}\right)\,.
\end{align}

In crossing one of the gluons  in the $D_4^0$ final-final quark-gluon antenna function
into the initial state, one has to distinguish two cases, depending on whether the gluon is
colour-connected to the quark or not. If the crossed gluon is colour-connected to the quark,
we obtain the following integrated antenna function:
\begin{align}
\lefteqn{\mathcal{D}^{0}_{g,ggq}=}\nl
+&\frac{1}{\eps^4}\Bigl[3 \delta(1-z)\Bigr]\nl
+&\frac{1}{\eps^3}\Bigl[16+\frac{19}{3} \delta(1-z)-6 \dd_{0}(z)-\frac{6}{z}-14 z+14 z^2\Bigr]\nl
+&\frac{1}{\eps^2}\Bigl[\frac{85}{4}+\delta(1-z) \bigl(\frac{1429}{72}-\frac{7 \pi^2}{3}\bigr)-\frac{38}{3} \dd_{0}(z)+12 \dd_{1}(z)+26 \h(0;z)+32 \h(1;z)-\frac{12 \h(0;z)}{1-z}\nl
&\quad+z \bigl(-\frac{61}{2}-34 \h(0;z)-28 \h(1;z)\bigr)+\frac{1}{z}\bigl(-\frac{62}{3}-12 \h(0;z)-12 \h(1;z)\bigr)\nl
&\quad+z^2 (31+28 \h(0;z)+28 \h(1;z))\Bigr]\nl
+&\frac{1}{\eps}\Bigl[\frac{223}{4}-\frac{21 \pi^2}{2}+\delta(1-z) \bigl(\frac{23959}{432}-\frac{181 \pi^2}{36}-25 \zeta_{3}\bigr)+\bigl(-\frac{1429}{36}+\frac{14 \pi^2}{3}\bigr) \dd_{0}(z)+\frac{76}{3} \dd_{1}(z)-12 \dd_{2}(z)\nl
&\quad+\frac{595}{12} \h(0;z)+\frac{85}{2} \h(1;z)-2 \h(-1,0;z)+49 \h(0,0;z)+52 \h(0,1;z)+65 \h(1,0;z)\nl
&\quad+64 \h(1,1;z)-\frac{1}{1-z}\bigl(\frac{76}{3} \h(0;z)+24 \h(0,0;z)+24 \h(0,1;z)+24 \h(1,0;z)\bigr)\nl
&\quad+\frac{1}{z}\bigl(-\frac{1823}{36}+\frac{14 \pi^2}{3}-\frac{98}{3} \h(0;z)-\frac{124}{3} \h(1;z)-24 \h(0,0;z)-24 \h(0,1;z)-24 \h(1,0;z)\nl
&\quad-24 \h(1,1;z)\bigr)+z \bigl(-\frac{265}{4}+\frac{37 \pi^2}{3}-\frac{307}{6} \h(0;z)-61 \h(1;z)-4 \h(-1,0;z)-76 \h(0,0;z)\nl
&\quad-68 \h(0,1;z)-58 \h(1,0;z)-56 \h(1,1;z)\bigr)+z^2 \bigl(\frac{565}{9}-11 \pi^2+\frac{230}{3} \h(0;z)+62 \h(1;z)\nl
&\quad-4 \h(-1,0;z)+62 \h(0,0;z)+56 \h(0,1;z)+58 \h(1,0;z)+56 \h(1,1;z)\bigr)\Bigr]\nl
+&\Bigl[\frac{7793}{36}-\frac{1385 \pi^2}{72}-\frac{479 \zeta_{3}}{3}+\delta(1-z)\bigl(\frac{389623}{2592}-\frac{6857 \pi^2}{432}+\frac{163 \pi^4}{360}-\frac{1027 \zeta_{3}}{18}\bigr)\nl
&\quad+\bigl(-\frac{23959}{216}+\frac{181 \pi^2}{18}+50 \zeta_{3}\bigr)\dd_{0}(z)+\bigl(\frac{1429}{18}-\frac{28 \pi^2}{3}\bigr) \dd_{1}(z)-\frac{76}{3} \dd_{2}(z)+8 \dd_{3}(z)+3 \pi^2 \h(-1;z)\nl
&\quad+\frac{2801}{36} \h(0;z)-24 \pi^2 \h(0;z)+\frac{223}{2} \h(1;z)-27 \pi^2 \h(1;z)+10 \h(-1,0;z)+\frac{317}{4} \h(0,0;z)\nl
&\quad+\frac{595}{6} \h(0,1;z)+78 \h(1,0;z)+85 \h(1,1;z)+28 \h(-1,-1,0;z)-20 \h(-1,0,0;z)\nl
&\quad-4 \h(-1,0,1;z)-44 \h(0,-1,0;z)+95 \h(0,0,0;z)+98 \h(0,0,1;z)+94 \h(0,1,0;z)\nl
&\quad+104 \h(0,1,1;z)+101 \h(1,0,0;z)+130 \h(1,0,1;z)+118 \h(1,1,0;z)+128 \h(1,1,1;z)\nl
&\quad+\frac{1}{1-z}\bigl(\frac{5 \pi^2}{6}+16 \zeta_{3}-\frac{1501}{18} \h(0;z)+\frac{34}{3} \pi^2 \h(0;z)-\frac{122}{3} \h(0,0;z)-\frac{152}{3}\h(0,1;z)\nl
&\quad-\frac{137}{3} \h(1,0;z)+8 \h(0,-1,0;z)-48 \h(0,0,0;z)-48 \h(0,0,1;z)-44 \h(0,1,0;z)\nl
&\quad-48 \h(0,1,1;z)-44 \h(1,0,0;z)-48 \h(1,0,1;z)-48 \h(1,1,0;z)\bigr)+z \bigl(-\frac{9395}{36}\nl
&\quad+\frac{701 \pi^2}{36}+\frac{376 \zeta_{3}}{3}+4 \pi^2 \h(-1;z)-\frac{6349}{36}\h(0;z)+\frac{101}{3} \pi^2 \h(0;z)-\frac{265}{2} \h(1;z)\nl
&\quad+24 \pi^2 \h(1;z)-\frac{201}{2} \h(0,0;z)-\frac{307}{3} \h(0,1;z)-130 \h(1,0;z)-122 \h(1,1;z)\nl
&\quad+32 \h(-1,-1,0;z)-28 \h(-1,0,0;z)-8\h(-1,0,1;z)-160 \h(0,0,0;z)\nl
&\quad-152 \h(0,0,1;z)-128 \h(0,1,0;z)-136 \h(0,1,1;z)-94 \h(1,0,0;z)-116 \h(1,0,1;z)\nl
&\quad-104 \h(1,1,0;z)-112 \h(1,1,1;z)\bigr)+\frac{1}{z}\bigl(-\frac{46025}{216}+\frac{151 \pi^2}{9}+54 \zeta_{3}+\frac{2}{3}\pi^2 \h(-1;z)\nl
&\quad-\frac{2179}{18} \h(0;z)+\frac{34}{3} \pi^2 \h(0;z)-\frac{1823}{18} \h(1;z)+10\pi^2 \h(1;z)+\frac{68}{3} \h(-1,0;z)-\frac{196}{3} \h(0,0;z)\nl
&\quad-\frac{196}{3} \h(0,1;z)-\frac{239}{3} \h(1,0;z)-\frac{248}{3} \h(1,1;z)+8 \h(-1,-1,0;z)-4 \h(-1,0,0;z)\nl
&\quad+8 \h(0,-1,0;z)-48 \h(0,0,0;z)-48 \h(0,0,1;z)-44 \h(0,1,0;z)-48 \h(0,1,1;z)\nl
&\quad-36 \h(1,0,0;z)-48 \h(1,0,1;z)-44 \h(1,1,0;z)-48 \h(1,1,1;z)\bigr)+z^2 \bigl(\frac{6668}{27}-\frac{553 \pi^2}{18}\nl
&\quad-\frac{484 \zeta_{3}}{3}+\frac{10}{3} \pi^2 \h(-1;z)+\frac{862}{9} \h(0;z)-28 \pi^2 \h(0;z)+\frac{1130}{9} \h(1;z)-\frac{70}{3} \pi^2 \h(1;z)\nl
&\quad+\frac{44}{3} \h(-1,0;z)+168 \h(0,0;z)+\frac{460}{3} \h(0,1;z)+124 \h(1,0;z)+124 \h(1,1;z)\nl
&\quad+24 \h(-1,-1,0;z)-24 \h(-1,0,0;z)-8 \h(-1,0,1;z)-32 \h(0,-1,0;z)\nl
&\quad+130 \h(0,0,0;z)+124 \h(0,0,1;z)+104 \h(0,1,0;z)+112 \h(0,1,1;z)\nl
&\quad+102\h(1,0,0;z)+116 \h(1,0,1;z)+108 \h(1,1,0;z)+112 \h(1,1,1;z)\bigr)\Bigr]\nl
&+\mathcal{O}\left(\epsilon\right)\,,
\end{align}
while the crossing the middle gluon (which is colour-connected to the other two gluons)
results in:
\begin{align}
\lefteqn{\mathcal{D}^{0}_{g^{\prime},ggq}=}\nl
+&\frac{1}{\eps^4}\Bigl[4 \delta(1-z)\Bigr]\nl
+&\frac{1}{\eps^3}\Bigl[18+6 \delta(1-z)-8 \dd_{0}(z)-\frac{8}{z}-12 z+12 z^2\Bigr]\nl
+&\frac{1}{\eps^2}\Bigl[\frac{45}{2}+\delta(1-z) \bigl(\frac{75}{4}-\frac{10 \pi^2}{3}\bigr)-12 \dd_{0}(z)+16 \dd_{1}(z)+26 \h(0;z)-\frac{16 \h(0;z)}{1-z}+\frac{1}{z}\bigl(-28\nl
&\quad-16 \h(0;z)-16 \h(1;z)\bigr)+36 \h(1;z)+z \bigl(-29-40 \h(0;z)-24 \h(1;z)\bigr)\nl
&\quad+z^2 (34+24 \h(0;z)+24 \h(1;z))\Bigr]\nl
+&\frac{1}{\eps}\Bigl[\frac{85}{2}-\frac{40\pi^2}{3}+\delta(1-z) \bigl(\frac{417}{8}-5\pi^2-\frac{116\zeta_{3}}{3}\bigr)+\bigl(-\frac{75}{2}+\frac{20\pi^2}{3}\bigr) \dd_{0}(z)+24 \dd_{1}(z)-16 \dd_{2}(z)\nl
&\quad+\frac{133}{2} \h(0;z)+45 \h(1;z)-16 \h(-1,0;z)+56\h(0,0;z)+52 \h(0,1;z)+70\h(1,0;z)\nl
&\quad+72 \h(1,1;z)+\frac{1}{1-z}\bigl(-24 \h(0;z)-36 \h(0,0;z)-32 \h(0,1;z)-32 \h(1,0;z)\bigr)\nl
&\quad+\frac{1}{1+z}\bigl(\frac{2 \pi^2}{3}+8 \h(-1,0;z)-4 \h(0,0;z)\bigr)+\frac{1}{z}\bigl(-\frac{785}{18}+6 \pi^2-\frac{116}{3} \h(0;z)-56 \h(1;z)\nl
&\quad-8 \h(-1,0;z)-32 \h(0,0;z)-32 \h(0,1;z)-32 \h(1,0;z)-32 \h(1,1;z)\bigr)+z \bigl(-\frac{59}{2}\nl
&\quad+\frac{46 \pi^2}{3}-29 \h(0;z)-58 \h(1;z)-8 \h(-1,0;z)-92 \h(0,0;z)-80\h(0,1;z)\nl
&\quad-44 \h(1,0;z)-48 \h(1,1;z)\bigr)+z^2 \bigl(\frac{280}{9}-\frac{34 \pi^2}{3}+\frac{292}{3} \h(0;z)+68 \h(1;z)\nl
&\quad-8 \h(-1,0;z)+52 \h(0,0;z)+48 \h(0,1;z)+44 \h(1,0;z)+48 \h(1,1;z)\bigr)\Bigr]\nl
+&\Bigl[\frac{3493}{18}-\frac{1121 \pi^2}{36}-274 \zeta_{3}+\delta(1-z) \bigl(\frac{2275}{16}-\frac{125 \pi^2}{8}+\frac{61\pi^4}{90}-64 \zeta_{3}\bigr)+\dd_{0}(z) \bigl(-\frac{417}{4}+10 \pi^2\nl
&\quad+\frac{232 \zeta_{3}}{3}\bigr)+\bigl(75-\frac{40 \pi^2}{3}\bigr) \dd_{1}(z)-24 \dd_{2}(z)+\frac{32}{3} \dd_{3}(z)+\frac{26}{3}\pi^2 \h(-1;z)+\frac{137}{6} \h(0;z)\nl
&\quad-\frac{89}{3} \pi^2 \h(0;z)+85 \h(1;z)-\frac{86}{3}\pi^2 \h(1;z)+\frac{1055}{6} \h(0,0;z)+133 \h(0,1;z)\nl
&\quad+\frac{262}{3} \h(1,0;z)+90 \h(1,1;z)+40\h(-1,-1,0;z)-68 \h(-1,0,0;z)-32 \h(-1,0,1;z)\nl
&\quad-36 \h(0,-1,0;z)+116 \h(0,0,0;z)+112 \h(0,0,1;z)+84 \h(0,1,0;z)+104 \h(0,1,1;z)\nl
&\quad+146 \h(1,0,0;z)+140 \h(1,0,1;z)+148 \h(1,1,0;z)+144 \h(1,1,1;z)\nl
&\quad+\frac{1}{1-z}\bigl(+24 \zeta_{3}-79 \h(0;z)+\frac{52}{3} \pi^2 \h(0;z)-\frac{166}{3} \h(0,0;z)-48 \h(0,1;z)-48 \h(1,0;z)\nl
&\quad-76 \h(0,0,0;z)-72 \h(0,0,1;z)-56 \h(0,1,0;z)-64 \h(0,1,1;z)-64 \h(1,0,0;z)\nl
&\quad-64 \h(1,0,1;z)-64 \h(1,1,0;z)\bigr)+\frac{1}{1+z}\bigl(+28 \zeta_{3}-\frac{16}{3}\pi^2 \h(-1;z)+\frac{4}{3} \pi^2 \h(0;z)\nl
&\quad-32 \h(-1,-1,0;z)+40 \h(-1,0,0;z)+16 \h(-1,0,1;z)+24 \h(0,-1,0;z)\nl
&\quad-12 \h(0,0,0;z)-8 \h(0,0,1;z)\bigr) +\frac{1}{z}\bigl(-\frac{8407}{36}+\frac{206 \pi^2}{9}+\frac{14}{3} \pi^2 \h(-1;z)-\frac{1141}{9} \h(0;z)\nl
&\quad+16 \pi^2 \h(0;z)-\frac{785}{9} \h(1;z)+\frac{38}{3} \pi^2 \h(1;z)-\frac{20}{3} \h(-1,0;z)-\frac{232}{3} \h(0,0;z)\nl
&\quad-\frac{232}{3} \h(0,1;z)-\frac{314}{3} \h(1,0;z)-112 \h(1,1;z)+24 \h(-1,-1,0;z)-36 \h(-1,0,0;z)\nl
&\quad-16 \h(-1,0,1;z)-24 \h(0,-1,0;z)-64 \h(0,0,0;z)-64 \h(0,0,1;z)-56 \h(0,1,0;z)\nl
&\quad-64 \h(0,1,1;z)-72 \h(1,0,0;z)-64 \h(1,0,1;z)-68 \h(1,1,0;z)-64 \h(1,1,1;z)\nl
&\quad+\frac{256 \zeta_{3}}{3}\bigr)+z \bigl(-\frac{3571}{18}+\frac{377 \pi^2}{18}+188 \zeta_{3}+\frac{10}{3} \pi^2 \h(-1;z)-\frac{857}{6} \h(0;z)+\frac{128}{3}\pi^2 \h(0;z)\nl
&\quad-59 \h(1;z)+\frac{58}{3} \pi^2 \h(1;z)+36 \h(-1,0;z)-\frac{191}{3} \h(0,0;z)-58\h(0,1;z)\nl
&\quad-\frac{344}{3} \h(1,0;z)-116 \h(1,1;z)+8 \h(-1,-1,0;z)-28 \h(-1,0,0;z)-16 \h(-1,0,1;z)\nl
&\quad-196 \h(0,0,0;z)-184 \h(0,0,1;z)-144 \h(0,1,0;z)-160 \h(0,1,1;z)-76 \h(1,0,0;z)\nl
&\quad-88 \h(1,0,1;z)-92 \h(1,1,0;z)-96 \h(1,1,1;z)\bigr) +z^2 \bigl(\frac{2026}{9}-\frac{343 \pi^2}{9}-184 \zeta_{3}\nl
&\quad+4 \pi^2 \h(-1;z)+\frac{8}{3} \h(0;z)-\frac{80}{3} \pi^2 \h(0;z)+\frac{560}{9} \h(1;z)-20 \pi^2 \h(1;z)+\frac{88}{3} \h(-1,0;z)\nl
&\quad+224 \h(0,0;z)+\frac{584}{3} \h(0,1;z)+136 \h(1,0;z)+136 \h(1,1;z)+16 \h(-1,-1,0;z)\nl
&\quad-32 \h(-1,0,0;z)-16 \h(-1,0,1;z)-32 \h(0,-1,0;z)+108 \h(0,0,0;z)\nl
&\quad+104 \h(0,0,1;z)+80 \h(0,1,0;z)+96 \h(0,1,1;z)+68 \h(1,0,0;z)+88 \h(1,0,1;z)\nl
&\quad+88 \h(1,1,0;z)+96 \h(1,1,1;z)\bigr)\Bigr]\nl
&+\mathcal{O}\left(\epsilon\right)\,.
\end{align}

Crossing the gluon in the $E_4^0$-type antenna functions into the initial state results in
$(g,qq^{\prime}\bar{q}^{\prime})$-antenna at leading and subleading colour, which integrate to:
\begin{align}
\lefteqn{\mathcal{E}^{0}_{g,qq^{\prime}\bar{q}^{\prime}}=}\nl
+&\frac{1}{\eps^3}\Bigl[-\frac{1}{3}\delta(1-z)\Bigr]\nl
+&\frac{1}{\eps^2}\Bigl[-\frac{3}{2}-\frac{19}{18} \delta(1-z)+\frac{2}{3} \dd_{0}(z)+\h(0;z)+\frac{4}{3 z}+z \bigl(\frac{3}{2}+\h(0;z)\bigr)-\frac{8}{3} z^2\Bigr]\nl
+&\frac{1}{\eps}\Bigl[-\frac{7}{3}-\frac{\pi^2}{3}+\delta(1-z)\bigl(-\frac{373}{108}+\frac{5\pi^2}{18}\bigr)+\frac{19}{9}\dd_{0}(z)-\frac{4}{3}\dd_{1}(z)-\frac{5}{6}\h(0;z)+\frac{4}{3(1-z)} \h(0;z)\nl
&\quad-3\h(1;z)+3 \h(0,0;z)+2 \h(0,1;z)+\frac{1}{z}\bigl(\frac{52}{9}+\frac{8}{3} \h(0;z)+\frac{8}{3} \h(1;z)\bigr)+z\bigl(\frac{13}{3}-\frac{\pi^2}{3}\nl
&\quad+\frac{25}{6} \h(0;z)+3 \h(1;z)+3 \h(0,0;z)+2 \h(0,1;z)\bigr)-z^2 \bigl(8+\frac{16}{3} \h(0;z)+\frac{16}{3} \h(1;z)\bigr)\Bigr]\nl
+&\Bigl[-\frac{17}{6}+\frac{19\pi^2}{36}-2\zeta_{3}+\delta(1-z)\bigl(-\frac{7081}{648}+\frac{95\pi^2}{108}+\frac{32\zeta_{3}}{9}\bigr)+\bigl(\frac{373}{54}-\frac{5\pi^2}{9}\bigr)\dd_{0}(z)
-\frac{38}{9}\dd_{1}(z)\nl
&\quad+\frac{4}{3}\dd_{2}(z)+\frac{43}{9} \h(0;z)-\frac{7}{6} \pi^2 \h(0;z)-\frac{14}{3} \h(1;z)+\frac{1}{2} \h(0,0;z)-\frac{5}{3}\h(0,1;z)-6 \h(1,0;z)\nl
&\quad-6 \h(1,1;z)+7 \h(0,0,0;z)+6 \h(0,0,1;z)+4 \h(0,1,0;z)+4 \h(0,1,1;z)\nl
&\quad+\frac{1}{1-z}\bigl(\frac{38}{9} \h(0;z)+\frac{8}{3} \h(0,0;z)+\frac{8}{3} \h(0,1;z)+\frac{8}{3}\h(1,0;z)\bigr)+\frac{1}{z}\bigl(23-\frac{10 \pi^2}{9}\nl
&\quad+\frac{104}{9}\h(0;z)+\frac{104}{9} \h(1;z)+\frac{16}{3} \h(0,0;z)+\frac{16}{3} \h(0,1;z)+\frac{16}{3} \h(1,0;z)+\frac{16}{3}\h(1,1;z)\bigr)\nl
&\quad+z \bigl(\frac{16}{3}-\frac{59\pi^2}{36}-2 \zeta_{3}+\frac{115}{9} \h(0;z)-\frac{7}{6} \pi^2 \h(0;z)+\frac{26}{3} \h(1;z)+\frac{19}{2} \h(0,0;z)\nl
&\quad+\frac{25}{3}\h(0,1;z)+6 \h(1,0;z)+6 \h(1,1;z)+7 \h(0,0,0;z)+6 \h(0,0,1;z)+4 \h(0,1,0;z)\nl
&\quad+4\h(0,1,1;z)\bigr)+z^2 \bigl(-\frac{580}{27}+\frac{20 \pi^2}{9}-16 \h(0;z)-16 \h(1;z)-\frac{32}{3} \h(0,0;z)\nl
&\quad-\frac{32}{3}\h(0,1;z)-\frac{32}{3} \h(1,0;z)-\frac{32}{3} \h(1,1;z)\bigr)\Bigr]\nl
&+\mathcal{O}\left(\epsilon\right)
\end{align}
and
\begin{align}
\lefteqn{\mathcal{\tilde{E}}^{0}_{g,qq^{\prime}\bar{q}^{\prime}}=}\nl
+&\frac{1}{\eps^2}\Bigl[1+2 \h(0;z)+\frac{4}{3 z}+z \bigl(-1+2 \h(0;z)\bigr)-\frac{4}{3}z^2\Bigr]\nl
+&\frac{1}{\eps}\Bigl[\frac{10}{3}-\frac{2 \pi^2}{3}+5 \h(0;z)+2\h(1;z)+\frac{1}{z}\bigl(\frac{44}{9}+\frac{8}{3} \h(0;z)+\frac{8}{3} \h(1;z)\bigr)+6 \h(0,0;z)\nl
&\quad+4 \h(0,1;z)+z\bigl(-\frac{16}{3}-\frac{2 \pi^2}{3}-\h(0;z)-2 \h(1;z)+6 \h(0,0;z)+4 \h(0,1;z)\bigr)\nl
&\quad+z^2 \bigl(-\frac{26}{9}-\frac{8}{3} \h(0;z)-\frac{8}{3} \h(1;z)\bigr)\Bigr]\nl
+&\Bigl[\frac{200}{9}-\frac{11 \pi^2}{6}-4 \zeta_{3}+\frac{68}{3} \h(0;z)-\frac{7}{3} \pi^2 \h(0;z)+\frac{20}{3} \h(1;z)+13 \h(0,0;z)+10 \h(0,1;z)\nl
&\quad+4 \h(1,0;z)+14 \h(0,0,0;z)+12\h(0,0,1;z)+8 \h(0,1,0;z)+8 \h(0,1,1;z)\nl
&\quad+4\h(1,1;z)+\frac{1}{z}\bigl(\frac{457}{27}-\frac{10 \pi^2}{9}+\frac{88}{9} \h(0;z)+\frac{88}{9} \h(1;z)+\frac{16}{3}\h(0,0;z)+\frac{16}{3} \h(0,1;z)\nl
&\quad+\frac{16}{3} \h(1,0;z)+\frac{16}{3} \h(1,1;z)\bigr)+z \bigl(-\frac{299}{9}+\frac{\pi^2}{2}-\frac{14}{3} \h(0;z)-\frac{7}{3}\pi^2 \h(0;z)-\frac{32}{3} \h(1;z)\nl
&\quad-\h(0,0;z)-2 \h(0,1;z)-4 \h(1,0;z)-4 \h(1,1;z)+14\h(0,0,0;z)+12 \h(0,0,1;z)\nl
&\quad+8 \h(0,1,0;z)+8 \h(0,1,1;z)-4 \zeta_{3}\bigr)+z^2 \bigl(-\frac{160}{27}+\frac{10 \pi^2}{9}-\frac{52}{9} \h(0;z)-\frac{52}{9}\h(1;z)\nl
&\quad-\frac{16}{3} \h(0,0;z)-\frac{16}{3} \h(0,1;z)-\frac{16}{3} \h(1,0;z)-\frac{16}{3} \h(1,1;z)\bigr)\Bigr]\nl
&+\mathcal{O}\left(\epsilon\right)\,.
\end{align}

The integrals of the
one-loop virtual corrections to the $(g,gq)$ antenna functions at leading colour and
for a closed quark loop yield:
\begin{align}
\lefteqn{\mathcal{D}^{1}_{g,gq}=}\nl
+&\frac{1}{\eps^4}\Bigl[-\frac{1}{4}\delta(1-z)\Bigr]\nl
+&\frac{1}{\eps^3}\Bigl[-\frac{5}{4}-\frac{53}{24} \delta(1-z)+\frac{1}{2} \dd_{0}(z)+\frac{1}{2 z}+z-z^2\Bigr]\nl
+&\frac{1}{\eps^2}\Bigl[-\frac{13}{3}+\delta(1-z) \bigl(-3+\frac{7 \pi^2}{24}\bigr)+\frac{31}{12} \dd_{0}(z)-\dd_{1}(z)-5 \h(0;z)-\frac{5}{2} \h(1;z)+\frac{2\h(0;z)}{1-z}
+\frac{1}{z}\bigl(\frac{31}{12}\nl
&\quad+2\h(0;z)+\h(1;z)\bigr)+z \bigl(\frac{11}{3}+4 \h(0;z)+2 \h(1;z)\bigr)+z^2 \bigl(-\frac{11}{3}-4 \h(0;z)-2 \h(1;z)\bigr)\Bigr]\nl
+&\frac{1}{\eps}\Bigl[-\frac{1}{12}+\frac{35 \pi^2}{24}+\delta(1-z) \bigl(-\frac{197}{24}+\frac{39 \pi^2}{16}+\frac{11 \zeta_{3}}{3}\bigr)+\bigl(\frac{37}{8}-\frac{7\pi^2}{12}\bigr) \dd_{0}(z)
-\frac{10}{3} \dd_{1}(z)+\dd_{2}(z)\nl
&\quad-\frac{43}{12} \h(0;z)-\frac{49}{12} \h(1;z)-15\h(0,0;z)-\frac{15}{2} \h(0,1;z)-\frac{15}{2} \h(1,0;z)-5 \h(1,1;z)\nl
&\quad+\frac{1}{1-z}\bigl(\frac{29}{6} \h(0;z)+6 \h(0,0;z)+3\h(0,1;z)+3 \h(1,0;z)\bigr)+\frac{1}{z}\bigl(\frac{37}{8}-\frac{7\pi^2}{12}+\frac{29}{6} \h(0;z)\nl
&\quad+\frac{10}{3} \h(1;z)+6 \h(0,0;z)+3 \h(0,1;z)+3 \h(1,0;z)+2\h(1,1;z)\bigr)+z \bigl(\frac{1}{12}-\frac{7 \pi^2}{6}\nl
&\quad+\frac{11}{3} \h(0;z)+\frac{11}{3} \h(1;z)+12 \h(0,0;z)+6\h(0,1;z)+6 \h(1,0;z)+4 \h(1,1;z)\bigr)\nl
&\quad+z^2 \bigl(\frac{7 \pi^2}{6}-\frac{11}{3} \h(0;z)-\frac{11}{3} \h(1;z)-12\h(0,0;z)-6 \h(0,1;z)-6 \h(1,0;z)-4 \h(1,1;z)\bigr)\Bigr]\nl
+&\Bigl[-7+\frac{241 \pi^2}{48}+\frac{125 \zeta_{3}}{6}+\delta(1-z) \bigl(-\frac{239}{12}+\frac{283 \pi^2}{96}-\frac{11 \pi^4}{144}+\frac{793\zeta_{3}}{36}\bigr)+\bigl(\frac{317}{24}
-\frac{11 \pi^2}{4}\nl
&\quad-\frac{22 \zeta_{3}}{3}\bigr) \dd_{0}(z)+\bigl(-\frac{63}{8}+\frac{7\pi^2}{6}\bigr) \dd_{1}(z)+\frac{29}{12} \dd_{2}(z)-\frac{2}{3} \dd_{3}(z)-\frac{37}{12} \h(0;z)+\frac{35}{6} \pi^2\h(0;z)
-\frac{13}{12} \h(1;z)\nl
&\quad+\frac{10}{3} \pi^2 \h(1;z)-\frac{19}{12} \h(0,0;z)-\frac{37}{12}\h(0,1;z)-\frac{37}{12} \h(1,0;z)-\frac{43}{12} \h(1,1;z)\nl
&\quad-35 \h(0,0,0;z)-20 \h(0,0,1;z)-15\h(0,1,0;z)-\frac{25}{2} \h(0,1,1;z)-15 \h(1,0,0;z)\nl
&\quad-\frac{25}{2} \h(1,0,1;z)-10 \h(1,1,0;z)-10 \h(1,1,1;z)+\frac{1}{1-z}\bigl(\frac{\pi^2}{4}-\zeta_{3}+\frac{91}{8}\h(0;z)\nl
&\quad-\frac{7}{3} \pi^2 \h(0;z)-\frac{1}{6} \pi^2 \h(1;z)+\frac{37}{3} \h(0,0;z)+\frac{19}{3}\h(0,1;z)+\frac{47}{6} \h(1,0;z)+14 \h(0,0,0;z)\nl
&\quad+8 \h(0,0,1;z)+6 \h(0,1,0;z)+5 \h(0,1,1;z)+6\h(1,0,0;z)+5 \h(1,0,1;z)\nl
&\quad+4 \h(1,1,0;z)\bigr)+\frac{1}{z}\bigl(\frac{107}{8}-\frac{7\pi^2}{2}-\frac{25 \zeta_{3}}{3}+\frac{91}{8} \h(0;z)-\frac{7}{3} \pi^2 \h(0;z)+\frac{63}{8} \h(1;z)\nl
&\quad-\frac{4}{3}\pi^2 \h(1;z)+\frac{28}{3} \h(0,0;z)+\frac{19}{3} \h(0,1;z)+\frac{29}{6} \h(1,0;z)+\frac{29}{6}\h(1,1;z)\nl
&\quad+14 \h(0,0,0;z)+8 \h(0,0,1;z)+6 \h(0,1,0;z)+5 \h(0,1,1;z)+6 \h(1,0,0;z)\nl
&\quad+5\h(1,0,1;z)+4 \h(1,1,0;z)+4 \h(1,1,1;z)\bigr)+z \bigl(\frac{11}{36}-\frac{17 \pi^2}{4}-\frac{50 \zeta_{3}}{3}+\frac{1}{6}\h(0;z)\nl
&\quad-\frac{14}{3} \pi^2 \h(0;z)+\frac{1}{12} \h(1;z)-\frac{8}{3} \pi^2 \h(1;z)+\frac{11}{3}\h(0,0;z)+\frac{11}{3} \h(0,1;z)+\frac{11}{3} \h(1,0;z)\nl
&\quad+\frac{11}{3} \h(1,1;z)+28 \h(0,0,0;z)+16\h(0,0,1;z)+12 \h(0,1,0;z)+10 \h(0,1,1;z)\nl
&\quad+12 \h(1,0,0;z)+10 \h(1,0,1;z)+8 \h(1,1,0;z)+8\h(1,1,1;z)\bigr)+z^2 \bigl(\frac{17 \pi^2}{4}+\frac{50 \zeta_{3}}{3}\nl
&\quad+\frac{14}{3}\pi^2 \h(0;z)+\frac{8}{3} \pi^2 \h(1;z)-\frac{11}{3} \h(0,0;z)-\frac{11}{3} \h(0,1;z)-\frac{11}{3}\h(1,0;z)-\frac{11}{3} \h(1,1;z)\nl
&\quad-28 \h(0,0,0;z)-16 \h(0,0,1;z)-12 \h(0,1,0;z)-10 \h(0,1,1;z)-12\h(1,0,0;z)\nl
&\quad-10 \h(1,0,1;z)-8 \h(1,1,0;z)-8 \h(1,1,1;z)\bigr)\Bigr]\nl
&+\mathcal{O}\left(\epsilon\right)
\end{align}
and
\begin{align}
\lefteqn{\mathcal{\hat{D}}^{1}_{g,gq}=}\nl
+&\frac{1}{\eps^3}\Bigl[\frac{1}{3}\delta(1-z)\Bigr]\nl
+&\frac{1}{\eps^2}\Bigl[\frac{5}{6}+\frac{1}{4}\delta(1-z)-\frac{1}{3} \dd_{0}(z)-\frac{1}{3 z}-\frac{2}{3}z+\frac{2}{3} z^2\Bigr]\nl
+&\frac{1}{\eps}\Bigl[-\frac{1}{6}+\delta(1-z) \bigl(\frac{7}{12}-\frac{\pi^2}{4}\bigr)-\frac{1}{4} \dd_{0}(z)+\frac{1}{3} \dd_{1}(z)+\frac{5}{6}
\h(0;z)-\frac{\h(0;z)}{3 (1-z)}+\frac{1}{z}\bigl(-\frac{1}{4}-\frac{1}{3}\h(0;z)\nl
&\quad-\frac{1}{3} \h(1;z)\bigr)+z \bigl(-\frac{1}{12}-\frac{2}{3} \h(0;z)-\frac{2}{3} \h(1;z)\bigr)+z^2 \bigl(\frac{2}{3} \h(0;z)+\frac{2}{3} \h(1;z)\bigr)+\frac{5}{6}\h(1;z)\Bigr]\nl
+&\Bigl[\frac{1}{2}-\frac{5 \pi^2}{8}+\delta(1-z) \bigl(\frac{7}{6}-\frac{3 \pi^2}{16}-\frac{7 \zeta_{3}}{9}\bigr)+\bigl(-\frac{7}{12}+\frac{\pi^2}{4}\bigr)\dd_{0}(z)+\frac{1}{4} \dd_{1}(z)
-\frac{1}{6} \dd_{2}(z)-\frac{1}{6} \h(0;z)\nl
&\quad-\frac{1}{6} \h(1;z)+\frac{5}{6}\h(0,0;z)+\frac{5}{6} \h(0,1;z)+\frac{5}{6} \h(1,0;z)+\frac{5}{6} \h(1,1;z)+\frac{1}{1-z}\bigl(-\frac{1}{4} \h(0;z)\nl
&\quad-\frac{1}{3} \h(0,0;z)-\frac{1}{3} \h(0,1;z)-\frac{1}{3}\h(1,0;z)\bigr)+\frac{1}{z}\bigl(-\frac{3}{4}+\frac{\pi^2}{4}-\frac{1}{4}\h(0;z)-\frac{1}{4} \h(1;z)\nl
&\quad-\frac{1}{3} \h(0,0;z)-\frac{1}{3} \h(0,1;z)-\frac{1}{3} \h(1,0;z)-\frac{1}{3}\h(1,1;z)\bigr)+z \bigl(-\frac{7}{18}+\frac{\pi^2}{2}-\frac{1}{6} \h(0;z)\nl
&\quad-\frac{1}{12}\h(1;z)-\frac{2}{3} \h(0,0;z)-\frac{2}{3} \h(0,1;z)-\frac{2}{3} \h(1,0;z)-\frac{2}{3} \h(1,1;z)\bigr)+z^2 \bigl(-\frac{\pi^2}{2}\nl
&\quad+\frac{2}{3} \h(0,0;z)+\frac{2}{3} \h(0,1;z)+\frac{2}{3} \h(1,0;z)+\frac{2}{3}\h(1,1;z)\bigr)\Bigr]\nl
&+\mathcal{O}\left(\epsilon\right)\,.
\end{align}
\subsection{Gluon-gluon antennae}
The NLO gluon-gluon antenna functions crossed to initial-final kinematics integrate to:
\begin{align}
\lefteqn{\mathcal{F}^{0}_{g,gg}=}\nl
+&\frac{1}{\eps^2}\Bigl[2\delta(1-z)\Bigr]\nl
+&\frac{1}{\eps}\Bigl[4+\frac{11}{6}\delta(1-z)-\frac{2}{z}-2 z+2 z^2-2 \dd_{0}(z)\Bigr]\nl
+&\Bigl[\delta(1-z) \bigl(\frac{67}{18}-\frac{\pi ^2}{2}\bigr)-\frac{11}{6} \dd_{0}(z)+2 \dd_{1}(z)+4 \h(0;z)+4 \h(1;z)-\frac{2}{1-z}\h(0;z)\nl
&\quad-\frac{1}{z}\bigl(\frac{11}{6}+2\h(0;z)+2 \h(1;z)\bigr)+\bigl(-z+z^2\bigr)(2 \h(0;z)+2 \h(1;z))\Bigr]\nl
+&\eps\Bigl[2-\pi ^2+\delta(1-z)\bigl(\frac{202}{27}-\frac{11 \pi ^2}{24}-\frac{14 \zeta_{3}}{3}\bigr)+\bigl(-\frac{67}{18}+\frac{\pi ^2}{2}\bigr) \dd_{0}(z)+\frac{11}{6} \dd_{1}(z)-\dd_{2}(z)+4 \h(0,0;z)\nl
&\quad+4\h(0,1;z)+4 \h(1,0;z)+4 \h(1,1;z)+\frac{1}{1-z}\bigl(-\frac{11}{6} \h(0;z)-2 \h(0,0;z)-2 \h(0,1;z)\nl
&\quad-2 \h(1,0;z)\bigr)+\frac{1}{z}\bigl(-\frac{67}{18}+\frac{\pi^2}{2}-\frac{11}{6} \h(0;z)-\frac{11}{6} \h(1;z)-2 \h(0,0;z)-2 \h(0,1;z)\nl
&\quad-2 \h(1,0;z)-2 \h(1,1;z)\bigr)+\bigl(z-z^2\bigr)\bigl( \frac{\pi ^2}{2}-2 \h(0,0;z)-2 \h(0,1;z)-2 \h(1,0;z)\nl
&\quad-2 \h(1,1;z)\bigr)\Bigr]\nl
+&\eps^{2}\Bigl[6-\frac{28\zeta_{3}}{3}+\delta(1-z) \bigl(\frac{1214}{81}-\frac{67 \pi ^2}{72}-\frac{\pi ^4}{48}-\frac{77 \zeta_{3}}{18}\bigr)+\dd_{0}(z) \bigl(-\frac{202}{27}+\frac{11 \pi ^2}{24}+\frac{14 \zeta_{3}}{3}\bigr)\nl
&\quad+\bigl(\frac{67}{18}-\frac{\pi ^2}{2}\bigr) \dd_{1}(z)-\frac{11}{12} \dd_{2}(z)+\frac{1}{3} \dd_{3}(z)+2 \h(0;z)-\pi^2 \h(0;z)+2 \h(1;z)-\pi ^2 \h(1;z)\nl
&\quad+4 \h(0,0,0;z)+4 \h(0,0,1;z)+4 \h(0,1,0;z)+4\h(0,1,1;z)+4 \h(1,0,0;z)\nl
&\quad+4 \h(1,0,1;z)+4\h(1,1,0;z)+4 \h(1,1,1;z)+\frac{1}{1-z}\bigl(-\frac{67}{18} \h(0;z)+\frac{1}{2} \pi^2 \h(0;z)\nl
&\quad-\frac{11}{6} \h(0,0;z)-\frac{11}{6} \h(0,1;z)-\frac{11}{6} \h(1,0;z)-2 \h(0,0,0;z)-2\h(0,0,1;z)\nl
&\quad-2 \h(0,1,0;z)-2 \h(0,1,1;z)-2 \h(1,0,0;z)-2 \h(1,0,1;z)-2 \h(1,1,0;z)\bigr)\nl
&\quad+\frac{1}{z}\bigl(-\frac{202}{27}+\frac{11 \pi ^2}{24}+\frac{14 \zeta_{3}}{3}-\frac{67}{18} \h(0;z)+\frac{1}{2} \pi ^2 \h(0;z)-\frac{67}{18} \h(1;z)+\frac{1}{2} \pi ^2 \h(1;z)\nl
&\quad-\frac{11}{6} \h(0,0;z)-\frac{11}{6} \h(0,1;z)-\frac{11}{6} \h(1,0;z)-\frac{11}{6} \h(1,1;z)-2 \h(0,0,0;z)\nl
&\quad-2 \h(0,0,1;z)-2 \h(0,1,0;z)-2 \h(0,1,1;z)-2 \h(1,0,0;z)-2 \h(1,0,1;z)\nl
&\quad-2 \h(1,1,0;z)-2\h(1,1,1;z)\bigr)+\bigl(z-z^2\bigr) \bigl(+\frac{14 \zeta_{3}}{3}+\frac{1}{2} \pi ^2 \h(0;z)+\frac{1}{2} \pi ^2 \h(1;z)\nl
&\quad-2 \h(0,0,0;z)-2 \h(0,0,1;z)-2 \h(0,1,0;z)-2 \h(0,1,1;z)-2 \h(1,0,0;z)\nl
&\quad-2 \h(1,0,1;z)-2 \h(1,1,0;z)-2 \h(1,1,1;z)\bigr)\Bigr]\nl
&+\mathcal{O}\left(\epsilon^{3}\right)
\end{align}
for the $(g,gg)$ case and to
\begin{align}
\lefteqn{\mathcal{G}^{0}_{g,q\bar{q}}=}\nl
+&\frac{1}{\eps}\Bigl[-\frac{\delta(1-z)}{3}\Bigr]\nl
+&\Bigl[-\frac{5}{9} \delta(1-z)+\frac{1}{3} \dd_{0}(z)+\frac{1}{3 z}\Bigr]\nl
+&\eps\Bigl[\delta(1-z) \bigl(-\frac{28}{27}+\frac{\pi ^2}{12}\bigr)+\frac{5}{9} \dd_{0}(z)-\frac{1}{3} \dd_{1}(z)+\frac{1}{3(1-z)}\h(0;z)+\frac{1}{z}\bigl(\frac{5}{9}+\frac{1}{3} \h(0;z)+\frac{1}{3} \h(1;z)\bigr)\Bigr]\nl
+&\eps^{2}\Bigl[\delta(1-z) \bigl(-\frac{164}{81}+\frac{5 \pi ^2}{36}+\frac{7 \zeta_{3}}{9}\bigr)+\bigl(\frac{28}{27}-\frac{\pi ^2}{12}\bigr) \dd_{0}(z)-\frac{5}{9} \dd_{1}(z)+\frac{1}{6} \dd_{2}(z)
+\frac{1}{1-z}\bigl(\frac{5}{9} \h(0;z)\nl
&\quad+\frac{1}{3} \h(0,0;z)+\frac{1}{3} \h(0,1;z)+\frac{1}{3} \h(1,0;z)\bigr)+\frac{1}{z}\bigl(\frac{28}{27}-\frac{\pi ^2}{12}+\frac{5}{9} \h(0;z)+\frac{5}{9} \h(1;z)\nl
&\quad+\frac{1}{3} \h(0,0;z)+\frac{1}{3} \h(0,1;z)+\frac{1}{3} \h(1,0;z)+\frac{1}{3} \h(1,1;z)\bigr)\Bigr]\nl
&+\mathcal{O}\left(\epsilon^{3}\right)
\end{align}
for the $(g,q\bar q)$ case.

The integrated form of the NNLO double real radiation antenna function for $(g,ggg)$ is:
\begin{align}
\lefteqn{\mathcal{F}^{0}_{g,ggg}=}\nl
+&\frac{1}{\eps^4}\Bigl[10 \delta(1-z)\Bigr]\nl
+&\frac{1}{\eps^3}\Bigl[40+22 \delta(1-z)-20 \dd_{0}(z)-\frac{20}{z}-20 z+20 z^2\Bigr]\nl
+&\frac{1}{\eps^2}\Bigl[68+\delta(1-z) \bigl(\frac{590}{9}-8 \pi^2\bigr)-44 \dd_{0}(z)+40 \dd_{1}(z)+64 \h(0;z)+80 \h(1;z)-\frac{40}{1-z} \h(0;z)\nl
&\quad-\frac{1}{z}\bigl(\frac{220}{3}+40 \h(0;z)+40 \h(1;z)\bigr)+z\bigl(-46-56 \h(0;z)-40 \h(1;z)\bigr)\nl
&\quad+z^2\bigl(\frac{154}{3}+40 \h(0;z)+40 \h(1;z)\bigr)\Bigr]\nl
+&\frac{1}{\eps}\Bigl[\frac{461}{3}-28 \pi^2+\delta(1-z) \bigl(\frac{4868}{27}-\frac{154\pi^2}{9}-\frac{272 \zeta_{3}}{3}\bigr)+\bigl(-\frac{1180}{9}+16 \pi^2\bigr) \dd_{0}(z)+88 \dd_{1}(z)-40 \dd_{2}(z)\nl
&\quad+\frac{482}{3}\h(0;z)+136 \h(1;z)-16 \h(-1,0;z)+128\h(0,0;z)+128 \h(0,1;z)\nl
&\quad+160\h(1,0;z)+160 \h(1,1;z)+\frac{1}{1-z}\bigl(-88 \h(0;z)-84 \h(0,0;z)-80 \h(0,1;z)\nl
&\quad-80 \h(1,0;z)\bigr)+\frac{1}{1+z}\bigl(\frac{2 \pi^2}{3}+8 \h(-1,0;z)-4 \h(0,0;z)\bigr)+\frac{1}{z}\bigl(-\frac{1556}{9}+\frac{46 \pi^2}{3}\nl
&\quad-\frac{352}{3}\h(0;z)-\frac{440}{3} \h(1;z)-8 \h(-1,0;z)-80 \h(0,0;z)-80 \h(0,1;z)-80 \h(1,0;z)\nl
&\quad-80\h(1,1;z)\bigr)+z \bigl(-\frac{257}{3}+\frac{62 \pi^2}{3}-\frac{202}{3} \h(0;z)-92 \h(1;z)-8 \h(-1,0;z)\nl
&\quad-128\h(0,0;z)-112 \h(0,1;z)-80 \h(1,0;z)-80 \h(1,1;z)\bigr)+z^2 \bigl(\frac{778}{9}-\frac{50 \pi^2}{3}\nl
&\quad+132 \h(0;z)+\frac{308}{3} \h(1;z)-8 \h(-1,0;z)+88 \h(0,0;z)+80 \h(0,1;z)+80 \h(1,0;z)\nl
&\quad+80 \h(1,1;z)\bigr)\Bigr]\nl
+&\Bigl[\frac{5249}{9}-\frac{592\pi^2}{9}-\frac{1424\zeta_{3}}{3}+\delta(1-z)\bigl(\frac{25811}{54}-\frac{1408\pi^2}{27}+\frac{319\pi^4}{180}-\frac{934\zeta_{3}}{3}\bigr)
+\dd_{0}(z)\bigl(-\frac{9736}{27}\nl
&\quad+\frac{308 \pi^2}{9}+\frac{544 \zeta_{3}}{3}\bigr)+\bigl(\frac{2360}{9}-32 \pi^2\bigr) \dd_{1}(z)-88 \dd_{2}(z)+\frac{80}{3} \dd_{3}(z)+\frac{32}{3}\pi^2 \h(-1;z)+\frac{1759}{9} \h(0;z)\nl
&\quad-64 \pi^2 \h(0;z)+\frac{922}{3} \h(1;z)-\frac{200}{3}\pi^2 \h(1;z)+16 \h(-1,0;z)+\frac{1030}{3} \h(0,0;z)\nl
&\quad+\frac{964}{3} \h(0,1;z)+\frac{808}{3} \h(1,0;z)+272 \h(1,1;z)+64 \h(-1,-1,0;z)-80 \h(-1,0,0;z)\nl
&\quad-32 \h(-1,0,1;z)-96 \h(0,-1,0;z)+256 \h(0,0,0;z)+256\h(0,0,1;z)\nl
&\quad+224 \h(0,1,0;z)+256 \h(0,1,1;z)+272 \h(1,0,0;z)+320 \h(1,0,1;z)+304 \h(1,1,0;z)\nl
&\quad+320\h(1,1,1;z)+\frac{1}{1-z}\bigl(-\frac{22 \pi^2}{9}+44 \zeta_{3}-\frac{2360}{9} \h(0;z)+40 \pi^2 \h(0;z)+\frac{4}{3}\pi^2 \h(1;z)\nl
&\quad-\frac{616}{3} \h(0,0;z)-176 \h(0,1;z)-\frac{572}{3} \h(1,0;z)+24 \h(0,-1,0;z)-172\h(0,0,0;z)\nl
&\quad-168 \h(0,0,1;z)-144 \h(0,1,0;z)-160 \h(0,1,1;z)-136 \h(1,0,0;z)-160 \h(1,0,1;z)\nl
&\quad-152\h(1,1,0;z)\bigr)+\frac{1}{1+z}\bigl(+28\zeta_{3}-\frac{16}{3} \pi^2 \h(-1;z)+\frac{4}{3} \pi^2 \h(0;z)-32\h(-1,-1,0;z)\nl
&\quad+40 \h(-1,0,0;z)+16 \h(-1,0,1;z)+24 \h(0,-1,0;z)-12 \h(0,0,0;z)-8 \h(0,0,1;z)\bigr)\nl
&\quad+\frac{1}{z}\bigl(-\frac{2102}{3}+\frac{550\pi^2}{9}+\frac{592 \zeta_{3}}{3}+\frac{16}{3} \pi^2 \h(-1;z)-\frac{1216}{3} \h(0;z)+\frac{116}{3} \pi^2 \h(0;z)\nl
&\quad-\frac{3112}{9}\h(1;z)+\frac{100}{3} \pi^2 \h(1;z)+\frac{88}{3} \h(-1,0;z)-\frac{704}{3} \h(0,0;z)-\frac{704}{3}\h(0,1;z)\nl
&\quad-\frac{836}{3} \h(1,0;z)-\frac{880}{3} \h(1,1;z)+32 \h(-1,-1,0;z)-40 \h(-1,0,0;z)-16\h(-1,0,1;z)\nl
&\quad-160 \h(0,0,0;z)-160 \h(0,0,1;z)-144 \h(0,1,0;z)-160 \h(0,1,1;z)-136 \h(1,0,0;z)\nl
&\quad-160\h(1,0,1;z)-152 \h(1,1,0;z)-160 \h(1,1,1;z)\bigr)+z \bigl(-\frac{3305}{9}+\frac{311\pi^2}{9}+\frac{688 \zeta_{3}}{3}\nl
&\quad+\frac{16}{3} \pi^2 \h(-1;z)-\frac{2363}{9} \h(0;z)+\frac{172}{3} \pi^2 \h(0;z)-\frac{514}{3}\h(1;z)+\frac{100}{3} \pi^2 \h(1;z)\nl
&\quad+16 \h(-1,0;z)-\frac{362}{3} \h(0,0;z)-\frac{404}{3} \h(0,1;z)-\frac{536}{3}\h(1,0;z)-184 \h(1,1;z)\nl
&\quad+32 \h(-1,-1,0;z)-40 \h(-1,0,0;z)-16 \h(-1,0,1;z)-272 \h(0,0,0;z)\nl
&\quad-256\h(0,0,1;z)-208 \h(0,1,0;z)-224 \h(0,1,1;z)-136 \h(1,0,0;z)-160 \h(1,0,1;z)\nl
&\quad-152 \h(1,1,0;z)-160\h(1,1,1;z)\bigr)+z^2\bigl(\frac{10382}{27}-\frac{473 \pi^2}{9}-\frac{760\zeta_{3}}{3}+\frac{16}{3} \pi^2 \h(-1;z)\nl
&\quad+\frac{340}{3} \h(0;z)-\frac{124}{3} \pi^2\h(0;z)+\frac{1556}{9} \h(1;z)-\frac{100}{3} \pi^2 \h(1;z)+\frac{88}{3} \h(-1,0;z)\nl
&\quad+\frac{880}{3}\h(0,0;z)+264 \h(0,1;z)+\frac{616}{3} \h(1,0;z)+\frac{616}{3} \h(1,1;z)\nl
&\quad+32 \h(-1,-1,0;z)-40\h(-1,0,0;z)-16 \h(-1,0,1;z)-48 \h(0,-1,0;z)\nl
&\quad+184 \h(0,0,0;z)+176 \h(0,0,1;z)+144 \h(0,1,0;z)+160\h(0,1,1;z)+136 \h(1,0,0;z)\nl
&\quad+160 \h(1,0,1;z)+152 \h(1,1,0;z)+160 \h(1,1,1;z)\bigr)\Bigr]\nl
&+\mathcal{O}\left(\epsilon\right)\,.
\end{align}

The $(g,q\bar{q}g)$ antenna function has a leading and subleading colour contribution. They
integrate to:
\begin{align}
\lefteqn{\mathcal{G}^{0}_{g,q\bar{q}g}=}\nl
+&\frac{1}{\eps^3}\Bigl[-\frac{5}{3} \delta(1-z)\Bigr]\nl
+&\frac{1}{\eps^2}\Bigl[-\frac{7}{2}-\frac{47}{9} \delta(1-z)+\frac{10}{3} \dd_{0}(z)+\h(0;z)+\frac{4}{z}+z \bigl(\frac{3}{2}+\h(0;z)\bigr)-\frac{8}{3} z^2\Bigr]\nl
+&\frac{1}{\eps}\Bigl[-\frac{59}{6}-\frac{\pi^2}{3}+\delta(1-z) \bigl(-\frac{557}{36}+\frac{23 \pi^2}{18}\bigr)+\frac{94}{9} \dd_{0}(z)-\frac{20}{3}\dd_{1}(z)
-\frac{35}{6} \h(0;z)-7 \h(1;z)\nl
&\quad+\frac{20 \h(0;z)}{3 (1-z)}+\frac{1}{z}\bigl(\frac{148}{9}+8 \h(0;z)+8 \h(1;z)\bigr)+3 \h(0,0;z)+2 \h(0,1;z)+z\bigl(\frac{41}{6}-\frac{\pi^2}{3}\nl
&\quad+\frac{25}{6} \h(0;z)+3 \h(1;z)+3 \h(0,0;z)+2 \h(0,1;z)\bigr)+z^2 \bigl(-\frac{28}{3}-\frac{16}{3} \h(0;z)-\frac{16}{3}\h(1;z)\bigr)\Bigr]\nl
+&\Bigl[-\frac{259}{12}+\frac{11 \pi^2}{4}-2 \zeta_{3}+\delta(1-z)\bigl(-\frac{28613}{648}+\frac{25 \pi^2}{6}+\frac{118 \zeta_{3}}{9}\bigr)+\bigl(\frac{557}{18}-\frac{23 \pi^2}{9}\bigr) \dd_{0}(z)
-\frac{188}{9} \dd_{1}(z)\nl
&\quad+\frac{20}{3}\dd_{2}(z)-\frac{113}{9} \h(0;z)-\frac{7}{6} \pi^2 \h(0;z)-\frac{59}{3} \h(1;z)-\frac{59}{6} \h(0,0;z)-\frac{35}{3}\h(0,1;z)\nl
&\quad-\frac{40}{3} \h(1,0;z)-14\h(1,1;z)+\frac{1}{1-z}\bigl(\frac{2 \pi^2}{9}+\frac{188}{9} \h(0;z)+\frac{44}{3} \h(0,0;z)+\frac{40}{3}\h(0,1;z)\nl
&\quad+\frac{44}{3} \h(1,0;z)\bigr)+\frac{1}{z}\bigl(\frac{3065}{54}-\frac{10 \pi^2}{3}+\frac{296}{9} \h(0;z)+\frac{296}{9} \h(1;z)+16 \h(0,0;z)\nl
&\quad+16\h(0,1;z)+16 \h(1,0;z)+16 \h(1,1;z)\bigr)+7 \h(0,0,0;z)+6 \h(0,0,1;z)\nl
&\quad+4 \h(0,1,0;z)+4\h(0,1,1;z)+z \bigl(\frac{71}{4}-\frac{25 \pi^2}{12}-2 \zeta_{3}+\frac{160}{9} \h(0;z)-\frac{7}{6} \pi^2 \h(0;z)\nl
&\quad+\frac{41}{3}\h(1;z)+\frac{49}{6} \h(0,0;z)+\frac{25}{3} \h(0,1;z)+\frac{14}{3} \h(1,0;z)+6 \h(1,1;z)+7\h(0,0,0;z)\nl
&\quad+6 \h(0,0,1;z)+4 \h(0,1,0;z)+4 \h(0,1,1;z)\bigr)+z^2 \bigl(-\frac{757}{27}+\frac{20 \pi^2}{9}-\frac{56}{3} \h(0;z)\nl
&\quad-\frac{56}{3}\h(1;z)-\frac{32}{3} \h(0,0;z)-\frac{32}{3} \h(0,1;z)-\frac{32}{3} \h(1,0;z)-\frac{32}{3} \h(1,1;z)\bigr)\Bigr]\nl
&+\mathcal{O}\left(\epsilon\right)
\end{align}
and
\begin{align}
\lefteqn{\mathcal{\tilde{G}}^{0}_{g,q\bar{q}g}=}\nl
+&\frac{1}{\eps^3}\Bigl[-\frac{2}{3} \delta(1-z)\Bigr]\nl
+&\frac{1}{\eps^2}\Bigl[1-\frac{19}{9} \delta(1-z)+\frac{4}{3} \dd_{0}(z)+2 \h(0;z)+\frac{8}{3 z}+z \bigl(-1+2 \h(0;z)\bigr)-\frac{4}{3} z^2\Bigr]\nl
+&\frac{1}{\eps}\Bigl[\frac{7}{3}-\frac{2 \pi^2}{3}+\delta(1-z) \bigl(-\frac{373}{54}+\frac{5 \pi^2}{9}\bigr)+\frac{38}{9} \dd_{0}(z)-\frac{8}{3}\dd_{1}(z)+3 \h(0;z)+2\h(1;z)+6 \h(0,0;z)\nl
&\quad+4 \h(0,1;z)+\frac{8 \h(0;z)}{3 (1-z)}+\frac{1}{z}\bigl(\frac{76}{9}+\frac{16}{3} \h(0;z)+\frac{16}{3} \h(1;z)\bigr)+z\bigl(-\frac{7}{3}-\frac{2 \pi^2}{3}-\h(0;z)\nl
&\quad-2 \h(1;z)+6 \h(0,0;z)+4 \h(0,1;z)\bigr)+z^2 \bigl(-\frac{38}{9}-\frac{8}{3} \h(0;z)-\frac{8}{3} \h(1;z)\bigr)\Bigr]\nl
+&\Bigl[\frac{391}{18}-\frac{13 \pi^2}{18}-4 \zeta_{3}+\delta(1-z) \bigl(-\frac{6973}{324}+\frac{95 \pi^2}{54}+\frac{64\zeta_{3}}{9}\bigr)+\bigl(\frac{373}{27}-\frac{10 \pi^2}{9}\bigr) \dd_{0}(z)
-\frac{76}{9} \dd_{1}(z)\nl
&\quad+\frac{8}{3}\dd_{2}(z)+\frac{52}{3} \h(0;z)-\frac{7}{3} \pi^2 \h(0;z)+\frac{14}{3} \h(1;z)+\frac{25}{3} \h(0,0;z)+6\h(0,1;z)\nl
&\quad+\frac{16}{3} \h(1,0;z)+4 \h(1,1;z)+14 \h(0,0,0;z)+12 \h(0,0,1;z)+8 \h(0,1,0;z)\nl
&\quad+8 \h(0,1,1;z)+\frac{1}{1-z}\bigl(\frac{76}{9} \h(0;z)+8 \h(0,0;z)+\frac{16}{3} \h(0,1;z)+\frac{16}{3}\h(1,0;z)\bigr)\nl
&\quad+\frac{1}{z}\bigl(\frac{809}{27}-\frac{28\pi^2}{9}+\frac{152}{9} \h(0;z)+\frac{152}{9} \h(1;z)+\frac{32}{3} \h(0,0;z)\nl
&\quad+\frac{32}{3} \h(0,1;z)+8\h(1,0;z)+\frac{32}{3} \h(1,1;z)\bigr)+z\bigl(-\frac{325}{18}-\frac{7 \pi^2}{18}-4 \zeta_{3}+\frac{4}{3} \h(0;z)\nl
&\quad-\frac{7}{3} \pi^2 \h(0;z)-\frac{14}{3} \h(1;z)-\frac{11}{3}\h(0,0;z)-2 \h(0,1;z)-\frac{20}{3} \h(1,0;z)-4 \h(1,1;z)\nl
&\quad+14 \h(0,0,0;z)+12 \h(0,0,1;z)+8\h(0,1,0;z)+8 \h(0,1,1;z)\bigr)+z^2 \bigl(-\frac{346}{27}+\frac{10 \pi^2}{9}\nl
&\quad-\frac{76}{9} \h(0;z)-\frac{76}{9} \h(1;z)-\frac{16}{3}\h(0,0;z)-\frac{16}{3} \h(0,1;z)-\frac{16}{3} \h(1,0;z)-\frac{16}{3} \h(1,1;z)\bigr)\Bigr]\nl
&+\mathcal{O}\left(\epsilon\right)\,.
\end{align}

Finally, the integrated one-loop antenna functions for $(g,gg)$ at leading colour and
for a closed quark loop read:
\begin{align}
\lefteqn{\mathcal{F}^{1}_{g,gg}=}\nl
+&\frac{1}{\eps^4}\Bigl[-\frac{\delta(1-z)}{2}\Bigr]\nl
+&\frac{1}{\eps^3}\Bigl[-2-\frac{55}{12} \delta(1-z)+\dd_{0}(z)+\frac{1}{z}+z-z^2\Bigr]\nl
+&\frac{1}{\eps^2}\Bigl[-\frac{22}{3}+\delta(1-z) \bigl(-\frac{85}{12}+\frac{7 \pi^2}{12}\bigr)+\frac{11}{2} \dd_{0}(z)-2 \dd_{1}(z)-8 \h(0;z)-4 \h(1;z)+\frac{4\h(0;z)}{1-z}\nl
&\quad+\frac{1}{z}\bigl(\frac{11}{2}+4 \h(0;z)+2 \h(1;z)\bigr)+z \bigl(\frac{11}{3}+4\h(0;z)+2 \h(1;z)\bigr)\nl
&\quad+z^2 \bigl(-\frac{11}{3}-4 \h(0;z)-2 \h(1;z)\bigr)\Bigr]\nl
+&\frac{1}{\eps}\Bigl[-2+\frac{7 \pi^2}{3}+\delta(1-z) \bigl(-\frac{1967}{108}+\frac{385 \pi^2}{72}+\frac{22 \zeta_{3}}{3}\bigr)+\bigl(\frac{389}{36}-\frac{7\pi^2}{6}\bigr) \dd_{0}(z)
-\frac{22}{3} \dd_{1}(z)+2 \dd_{2}(z)\nl
&\quad-\frac{22}{3} \h(0;z)-\frac{22}{3} \h(1;z)-24\h(0,0;z)-12 \h(0,1;z)-12 \h(1,0;z)-8 \h(1,1;z)\nl
&\quad+\frac{1}{1-z}\bigl(11 \h(0;z)+12 \h(0,0;z)+6 \h(0,1;z)+6\h(1,0;z)\bigr)+z \bigl(\frac{1}{6}-\frac{7 \pi^2}{6}+\frac{11}{3}\h(0;z)\nl
&\quad+\frac{11}{3} \h(1;z)+12 \h(0,0;z)+6 \h(0,1;z)+6 \h(1,0;z)+4 \h(1,1;z)\bigr)+z^2 \bigl(\frac{7 \pi^2}{6}\nl
&\quad-\frac{11}{3} \h(0;z)-\frac{11}{3} \h(1;z)-12 \h(0,0;z)-6\h(0,1;z)-6 \h(1,0;z)-4 \h(1,1;z)\bigr)\nl
&\quad+\frac{1}{z}\bigl(\frac{389}{36}-\frac{7\pi^2}{6}+11 \h(0;z)+\frac{22}{3} \h(1;z)+12 \h(0,0;z)+6 \h(0,1;z)+6 \h(1,0;z)\nl
&\quad+4\h(1,1;z)\bigr)\Bigr]\nl
+&\Bigl[-13+\frac{169 \pi^2}{18}+\frac{100 \zeta_{3}}{3}+\delta(1-z) \bigl(-\frac{14453}{324}+\frac{3155 \pi^2}{432}-\frac{11 \pi^4}{72}+\frac{269\zeta_{3}}{6}\bigr)+\bigl(\frac{3197}{108}
-\frac{55 \pi^2}{9}\nl
&\quad-\frac{44 \zeta_{3}}{3}\bigr) \dd_{0}(z)+\bigl(-\frac{73}{4}+\frac{7\pi^2}{3}\bigr) \dd_{1}(z)+\frac{11}{2} \dd_{2}(z)-\frac{4}{3} \dd_{3}(z)-\frac{20}{3} \h(0;z)+\frac{28}{3} \pi^2\h(0;z)-4 \h(1;z)\nl
&\quad+\frac{16}{3} \pi^2 \h(1;z)-\frac{20}{3} \h(0,0;z)-\frac{22}{3} \h(0,1;z)-\frac{20}{3}\h(1,0;z)-\frac{22}{3} \h(1,1;z)\nl
&\quad-56 \h(0,0,0;z)-32 \h(0,0,1;z)-24 \h(0,1,0;z)-20 \h(0,1,1;z)-24\h(1,0,0;z)\nl
&\quad-20 \h(1,0,1;z)-16 \h(1,1,0;z)-16 \h(1,1,1;z)+\frac{1}{1-z}\bigl(\frac{11 \pi^2}{18}-2 \zeta_{3}+\frac{925}{36}\h(0;z)\nl
&\quad-\frac{14}{3} \pi^2 \h(0;z)-\frac{1}{3} \pi^2 \h(1;z)+\frac{88}{3} \h(0,0;z)+\frac{44}{3}\h(0,1;z)+\frac{55}{3} \h(1,0;z)\nl
&\quad+28 \h(0,0,0;z)+16 \h(0,0,1;z)+12 \h(0,1,0;z)+10 \h(0,1,1;z)+12\h(1,0,0;z)\nl
&\quad+10 \h(1,0,1;z)+8 \h(1,1,0;z)\bigr)+\frac{1}{z}\bigl(\frac{3179}{108}-\frac{143 \pi^2}{18}-\frac{50\zeta_{3}}{3}+\frac{925}{36} \h(0;z)\nl
&\quad-\frac{14}{3} \pi^2 \h(0;z)+\frac{73}{4} \h(1;z)-\frac{8}{3} \pi^2\h(1;z)+22 \h(0,0;z)+\frac{44}{3} \h(0,1;z)+11 \h(1,0;z)\nl
&\quad+11 \h(1,1;z)+28 \h(0,0,0;z)+16\h(0,0,1;z)+12 \h(0,1,0;z)+10 \h(0,1,1;z)\nl
&\quad+12 \h(1,0,0;z)+10 \h(1,0,1;z)+8 \h(1,1,0;z)+8\h(1,1,1;z)\bigr)+z \bigl(\frac{11}{18}-\frac{181\pi^2}{36}\nl
&\quad-\frac{50 \zeta_{3}}{3}+\frac{1}{3} \h(0;z)-\frac{14}{3} \pi^2 \h(0;z)+\frac{1}{6} \h(1;z)-\frac{8}{3}\pi^2 \h(1;z)+\frac{7}{3} \h(0,0;z)+\frac{11}{3} \h(0,1;z)\nl
&\quad+\frac{7}{3} \h(1,0;z)+\frac{11}{3}\h(1,1;z)+28 \h(0,0,0;z)+16 \h(0,0,1;z)+12 \h(0,1,0;z)\nl
&\quad+10 \h(0,1,1;z)+12 \h(1,0,0;z)+10\h(1,0,1;z)+8 \h(1,1,0;z)+8 \h(1,1,1;z)\bigr)\nl
&\quad+z^2 \bigl(\frac{55 \pi^2}{12}+\frac{50 \zeta_{3}}{3}+\frac{14}{3}\pi^2 \h(0;z)+\frac{8}{3} \pi^2 \h(1;z)-\frac{11}{3} \h(0,0;z)-\frac{11}{3} \h(0,1;z)\nl
&\quad-\frac{11}{3}\h(1,0;z)-\frac{11}{3} \h(1,1;z)-28 \h(0,0,0;z)-16 \h(0,0,1;z)-12 \h(0,1,0;z)\nl
&\quad-10 \h(0,1,1;z)-12\h(1,0,0;z)-10 \h(1,0,1;z)-8 \h(1,1,0;z)-8 \h(1,1,1;z)\bigr)\Bigr]\nl
&+\mathcal{O}\left(\epsilon\right)
\end{align}
and
\begin{align}
\lefteqn{\mathcal{\hat{F}}^{1}_{g,gg}=}\nl
+&\frac{1}{\eps^3}\Bigl[\frac{2}{3} \delta(1-z)\Bigr]\nl
+&\frac{1}{\eps^2}\Bigl[\frac{4}{3}+\frac{11}{18} \delta(1-z)-\frac{2}{3} \dd_{0}(z)-\frac{2}{3 z}-\frac{2}{3}z+\frac{2}{3} z^2\Bigr]\nl
+&\frac{1}{\eps}\Bigl[\delta(1-z) \bigl(\frac{143}{108}-\frac{5 \pi^2}{6}\bigr)-\frac{11}{18} \dd_{0}(z)+\frac{2}{3} \dd_{1}(z)+\frac{4}{3} \h(0;z)+\frac{4}{3}\h(1;z)-\frac{2\h(0;z)}{3 (1-z)}
-\frac{1}{z}\bigl(\frac{11}{18}\nl
&\quad+\frac{2}{3} \h(0;z)+\frac{2}{3} \h(1;z)\bigr)-z \bigl(\frac{1}{6}+\frac{2}{3}\h(0;z)+\frac{2}{3} \h(1;z)\bigr)+z^2 \bigl(\frac{2}{3} \h(0;z)+\frac{2}{3} \h(1;z)\bigr)\Bigr]\nl
+&\Bigl[\frac{4}{3}-\frac{5 \pi^2}{3}+\delta(1-z) \bigl(\frac{979}{324}-\frac{55 \pi^2}{72}-\frac{14 \zeta_{3}}{9}\bigr)+\bigl(-\frac{38}{27}+\frac{5\pi^2}{6}\bigr) \dd_{0}(z)+\frac{11}{18} \dd_{1}(z)
-\frac{1}{3} \dd_{2}(z)\nl
&\quad+\frac{4}{3} \h(0,0;z)+\frac{4}{3} \h(0,1;z)+\frac{4}{3} \h(1,0;z)+\frac{4}{3}\h(1,1;z)+\frac{1}{1-z}\bigl(-\frac{11}{18}\h(0;z)\nl
&\quad-\frac{2}{3} \h(0,0;z)-\frac{2}{3} \h(0,1;z)-\frac{2}{3} \h(1,0;z)\bigr)+\frac{1}{z}\bigl(-\frac{85}{54}+\frac{5\pi^2}{6}-\frac{11}{18} \h(0;z)-\frac{11}{18} \h(1;z)\nl
&\quad-\frac{2}{3} \h(0,0;z)-\frac{2}{3} \h(0,1;z)-\frac{2}{3}\h(1,0;z)-\frac{2}{3} \h(1,1;z)\bigr)+z \bigl(-\frac{7}{9}+\frac{5 \pi^2}{6}-\frac{1}{3} \h(0;z)\nl
&\quad-\frac{1}{6}\h(1;z)-\frac{2}{3} \h(0,0;z)-\frac{2}{3} \h(0,1;z)-\frac{2}{3} \h(1,0;z)-\frac{2}{3} \h(1,1;z)\bigr)+z^2\bigl(-\frac{5 \pi^2}{6}\nl
&\quad+\frac{2}{3} \h(0,0;z)+\frac{2}{3} \h(0,1;z)+\frac{2}{3} \h(1,0;z)+\frac{2}{3} \h(1,1;z)\bigr)\Bigr]\nl
&+\mathcal{O}\left(\epsilon\right)\,,
\end{align}
while the leading and subleading colour and quark loop contributions to the one-loop
$(g,q\bar q)$ antenna function integrate to:
\begin{align}
\lefteqn{\mathcal{G}^{1}_{g,q\bar{q}}=}\nl
+&\frac{1}{\eps^2}\Bigl[\frac{11}{18} \delta(1-z)\Bigr]\nl
+&\frac{1}{\eps}\Bigl[\frac{119}{108} \delta(1-z)-\frac{11}{18} \dd_{0}(z)-\frac{2 \h(0;z)}{3(1-z)}+\frac{1}{z}\bigl(-\frac{11}{18}-\frac{2}{3} \h(0;z)\bigr)\Bigr]\nl
+&\Bigl[-\frac{2}{3}-\frac{\pi^2}{9}+\delta(1-z) \bigl(\frac{787}{324}-\frac{55 \pi^2}{72}\bigr)-\frac{32}{27} \dd_{0}(z)+\frac{11}{18}\dd_{1}(z)-\frac{2}{3} \h(0;z)-\frac{1}{3} \h(0,0;z)\nl
&\quad-\frac{1}{3} \h(1,0;z)+\frac{1}{1-z}\bigl(-\frac{\pi^2}{9}-\frac{31}{18} \h(0;z)-\frac{8}{3}\h(0,0;z)-\frac{2}{3} \h(0,1;z)-\frac{4}{3} \h(1,0;z)\bigr)\nl
&\quad+\frac{1}{z}\bigl(-\frac{83}{108}+\frac{\pi^2}{9}-\frac{31}{18}\h(0;z)-\frac{11}{18} \h(1;z)-2 \h(0,0;z)-\frac{2}{3} \h(0,1;z)-\frac{2}{3} \h(1,0;z)\bigr)\nl
&\quad+z\bigl(\frac{2 \pi^2}{9}+\frac{2}{3} \h(0,0;z)+\frac{2}{3} \h(1,0;z)\bigr)\Bigr]\nl
&+\mathcal{O}\left(\epsilon\right)
\end{align}
and
\begin{align}
\lefteqn{\mathcal{\tilde{G}}^{1}_{g,q\bar{q}}=}\nl
+&\frac{1}{\eps^3}\Bigl[\frac{1}{6}\delta(1-z)\Bigr]\nl
+&\frac{1}{\eps^2}\Bigl[\frac{19}{36} \delta(1-z)-\frac{1}{3} \dd_{0}(z)-\frac{1}{3 z}\Bigr]\nl
+&\frac{1}{\eps}\Bigl[\delta(1-z) \bigl(\frac{173}{108}-\frac{5 \pi^2}{36}\bigr)-\frac{19}{18} \dd_{0}(z)+\frac{2}{3} \dd_{1}(z)-\frac{2 \h(0;z)}{3(1-z)}
+\frac{1}{z}\bigl(-\frac{19}{18}-\frac{2}{3} \h(0;z)-\frac{2}{3} \h(1;z)\bigr)\Bigr]\nl
+&\Bigl[-\frac{2}{3}-\frac{\pi^2}{9}+\delta(1-z) \bigl(\frac{343}{81}-\frac{95 \pi^2}{216}-\frac{7 \zeta_{3}}{9}\bigr)+\bigl(-\frac{173}{54}+\frac{5\pi^2}{18}\bigr) \dd_{0}(z)
+\frac{19}{9} \dd_{1}(z)-\frac{2}{3} \dd_{2}(z)\nl
&\quad-\frac{2}{3} \h(0;z)-\frac{1}{3} \h(0,0;z)-\frac{1}{3} \h(1,0;z)+\frac{1}{1-z}\bigl(-\frac{19}{9}\h(0;z)-2 \h(0,0;z)-\frac{4}{3} \h(0,1;z)\nl
&\quad-\frac{4}{3} \h(1,0;z)\bigr)+\frac{1}{z}\bigl(-\frac{355}{108}+\frac{\pi^2}{2}-\frac{19}{9}\h(0;z)-\frac{19}{9} \h(1;z)-\frac{4}{3} \h(0,0;z)-\frac{4}{3} \h(0,1;z)\nl
&\quad-\frac{2}{3} \h(1,0;z)-\frac{4}{3}\h(1,1;z)\bigr)+z \bigl(\frac{2 \pi^2}{9}+\frac{2}{3}\h(0,0;z)+\frac{2}{3} \h(1,0;z)\bigr)\Bigr]\nl
&+\mathcal{O}\left(\epsilon\right)
\end{align}
and
\begin{align}
\lefteqn{\mathcal{\hat{G}}^{1}_{g,q\bar{q}}=}\nl
+&\frac{1}{\eps}\Bigl[\frac{5}{27} \delta(1-z)-\frac{1}{9} \dd_{0}(z)-\frac{1}{9 z}\Bigr]\nl
+&\Bigl[\delta(1-z) \bigl(\frac{53}{81}+\frac{5 \pi^2}{108}\bigr)-\frac{5}{9} \dd_{0}(z)+\frac{1}{3} \dd_{1}(z)-\frac{\h(0;z)}{3(1-z)}
+\frac{1}{z}\bigl(-\frac{5}{9}-\frac{1}{3} \h(0;z)-\frac{1}{3} \h(1;z)\bigr)\Bigr]\nl
&+\mathcal{O}\left(\epsilon\right)\,.
\end{align}

\section{Rederivation of NNLO coefficient functions}
\label{sec:discoeff}
Being derived from physical matrix elements, the integrated antenna
functions can be compared to  results from literature for inclusive cross
sections or coefficient functions, as was done previously
for the final-final antennae in~\cite{our2j,ritzmann}.
In the case of the initial-final antennae, we can compare the
integrated
quark-antiquark antennae and gluon-gluon antennae against
NNLO corrections to
deep inelastic coefficient functions known in the literature.
The former ones can be checked against DIS structure function
calculations~\cite{zv} whereas the latter can be compared
to the $\phi$-DIS structure functions computed in~\cite{moch,moch2}.
The quark-gluon antennae, derived from neutralino decay, cannot be
associated to any physical process and only the deepest pole could
be checked against a combination of Altarelli-Parisi splitting  functions.

\subsection{Deep inelastic scattering: $\gamma$ induced}
In deep inelastic scattering (DIS) the hadronic tensor $W^{\mu\nu}$ is contracted with the metric tensor $-g_{\mu\nu}$. This corresponds to the trace of the hadronic tensor, which in terms of the structure functions $\mathcal{F}_{L}$ and $\mathcal{F}_{2}$ is given by
\begin{equation}
-W^{\mu}_{\mu}=-\frac{d-1}{2}\mathcal{F}_{L}\l(z,Q^{2}\r)+\frac{d-2}{2}\mathcal{F}_{2}\l(z,Q^2\r)\,,
\end{equation}
where the structure functions can be expanded in powers of the strong coupling constant, following the notation of~\cite{zv} (but using $d=4-2\e$, as
throughout the rest of this work):
\begin{eqnarray}
\mathcal{F}_{L}&=&\mathcal{F}_{L}^{\l(0\r)}+\frac{\alpha_{s}}{4\pi}S_{\eps}\l(\frac{\mu^{2}}{Q^{2}}\r)^{\eps}\mathcal{F}_{L}^{\l(1\r)}+\l(\frac{\alpha_{s}}{4\pi}\r)^{2}S_{\eps}^{2}\l(\frac{\mu^{2}}{Q^{2}}\r)^{2\eps}\mathcal{F}_{L}^{\l(2\r)}+\mathcal{O}\l(\alpha_{s}^{3}\r),\nonumber\\
\mathcal{F}_{2}&=&\mathcal{F}_{2}^{\l(0\r)}+\frac{\alpha_{s}}{4\pi}S_{\eps}\l(\frac{\mu^{2}}{Q^{2}}\r)^{\eps}\mathcal{F}_{2}^{\l(1\r)}+\l(\frac{\alpha_{s}}{4\pi}\r)^{2}S_{\eps}^{2}\l(\frac{\mu^{2}}{Q^{2}}\r)^{2\eps}\mathcal{F}_{2}^{\l(2\r)}+\mathcal{O}\l(\alpha_{s}^{3}\r)\,,
\end{eqnarray}
and $S_{\eps}=\exp\left[\eps\l(\ln\l(4\pi\r)-\gamma_{E}\r)\right]$.
For clarity, we drop the dependence on $z$ and $Q^{2}$.
The factors $S_{\eps }$ and $\mu$ are the conventional
factors appearing in dimensional regularization.

To zeroth order in $\alpha_{s}$ the structure functions are given by the simple parton model result
\begin{equation}
\mathcal{F}_{L,q}^{\l(0\r)}\,=\,\mathcal{F}_{L,g}^{\l(0\r)}\,=\,0\,,\qquad\qquad\mathcal{F}_{2,q}^{\l(0\r)}\,=\,\delta\l(1-z\r)\,,\qquad\qquad\mathcal{F}_{2,g}^{\l(0\r)}\,=\,0\,.
\end{equation}
Since the overall normalization of the antenna functions is given by the leading order antenna $\mathcal{A}_{q,q}\,=\,\delta(1-z)$,
we find that the correct normalization of $W^{\mu}_{\mu}$ to be checked against the antennae is
\begin{equation}\label{eq:haronictensortocheck}
-\frac{2}{{d-2}} W^{\mu}_{\mu}=\mathcal{F}_{2}-\frac{d-1}{d-2}\mathcal{F}_{L}\,.
\end{equation}
This last equation at order $\alpha_{s}$ and $\alpha_{s}^{2}$ can be
 compared to a linear combinations of NLO and NNLO antennae respectively.
For completeness and clarity, before giving the explicit relations
between the structure functions and the antennae, we report the expressions
of the structure functions in terms of the Altarelli-Parisi splitting functions and the other coefficient functions $c_{i}$ and $a_{i}$.
We take them from~\cite{zv}, adjusting to $d=4-2\eps$ and $\beta_{0}=\frac{11}{6}\ca-\frac{1}{3}\nf$.
\begin{align}
\mathcal{F}_{L,g}^{\l(1\r)}=&\nf\l(c_{L,g}^{\l(1\r)}-\eps\,2a_{L,g}^{\l(1\r)}\r)\label{ali:fvanNerveen1}\,,\\
\mathcal{F}_{2,g}^{\l(1\r)}=&\nf\l(-\frac{1}{2\eps}P_{gq}^{\l(0\r)}+c_{2,g}^{\l(1\r)}-\eps\,2a_{2,g}^{\l(1\r)}\r)\,,\\
\mathcal{F}_{L,q}^{\l(1\r)}=&\l(c_{L,q}^{\l(1\r)}-\eps\,2a_{L,q}^{\l(1\r)}\r)\,,\\
\mathcal{F}_{2,q}^{\l(1\r)}=&\l(-\frac{1}{2\eps}P_{qq}^{\l(0\r)}+c_{2,q}^{\l(1\r)}-\eps\,2a_{2,q}^{\l(1\r)}\r)\,,\\
\mathcal{F}_{L,g}^{\l(2\r)}=&\nf\l(\frac{1}{\eps}\l(2\beta_{0}c_{L,g}^{\l(1\r)}-\frac{1}{2}P_{gg}^{\l(0\r)}\otimes c_{L,g}^{\l(1\r)}-\frac{1}{2}P_{qg}^{\l(0\r)}\otimes c_{L,q}^{\l(1\r)}\r)\right.\nonumber\\
                &\hspace{0.1cm}\left.\phantom{\frac{1}{1}}+c_{L,g}^{\l(2\r)}-4\beta_{0}a_{L,g}^{\l(1\r)}+P_{gg}^{\l(0\r)}\otimes a_{L,g}^{\l(1\r)}+P_{qg}^{\l(0\r)}\otimes a_{L,q}^{\l(1\r)}\r)\,,\\
\mathcal{F}_{2,g}^{\l(2\r)}=&\nf\l(\frac{1}{\eps^{2}}\l(\frac{1}{8}P_{qg}^{\l(0\r)}\otimes\l(P_{gg}^{\l(0\r)}+P_{qq}^{\l(0\r)}\r)-\frac{\beta_{0}}{2}P_{qq}^{\l(0\r)}\r)\right.\nonumber\\
                &\hspace{0.6cm}+\frac{1}{\eps}\l(-\frac{1}{4}P_{qg}^{\l(1\r)}+2\beta_{0}c_{2,g}^{\l(1\r)}-\frac{1}{2}P_{gg}^{\l(0\r)}\otimes c_{2,g}^{\l(1\r)}-\frac{1}{2}P_{qg}^{\l(0\r)}\otimes c_{2,q}^{\l(1\r)}\r)\nonumber\\
                &\hspace{0.2cm}\left.\phantom{\frac{1}{1}}+c_{2,g}^{\l(2\r)}-4\beta_{0}a_{2,g}^{\l(1\r)}+P_{gg}^{\l(0\r)}\otimes a_{2,g}^{\l(1\r)}+P_{qg}^{\l(0\r)}\otimes a_{2,q}^{\l(1\r)}\r)\,,\\
\mathcal{F}_{L,q}^{\l(2\r)\mathrm{NS}}=&\l(\frac{1}{\eps}\l(2\beta_{0}c_{L,q}^{\l(1\r)}-\frac{1}{2}P_{qq}^{\l(0\r)}\otimes c_{L,q}^{\l(1\r)}\r)+c_{L,q}^{\l(2\r)\mathrm{NS}}-4\beta_{0}a_{L,q}^{\l(1\r)}+P_{qq}^{\l(0\r)}\otimes a_{L,q}^{\l(1\r)}\r)\,,\nonumber\\ \\
\mathcal{F}_{2,q}^{\l(2\r)\mathrm{NS}}=&\l(\frac{1}{\eps^{2}}\l(\frac{1}{8}P_{qq}^{\l(0\r)}\otimes P_{qq}^{\l(0\r)}-\frac{\beta_{0}}{2}P_{qq}^{\l(0\r)}\r)\right.\nonumber\\
                &\hspace{0.2cm}+\frac{1}{\eps}\l(-\frac{1}{4}\l(P_{qq}^{\l(1\r)\mathrm{NS}}+P_{q\bar{q}}^{\l(1\r)\mathrm{NS}}\r)+2\beta_{0}c_{2,q}^{\l(1\r)}-\frac{1}{2}P_{qq}^{\l(0\r)}\otimes c_{2,q}^{\l(1\r)}\r)\nonumber\\
                &\hspace{-0.2cm}\left.\phantom{\frac{1}{1}}+c_{2,q}^{\l(2\r)\mathrm{NS}}-4\beta_{0}a_{2,q}^{\l(1\r)}+P_{qq}^{\l(0\r)}\otimes a_{2,q}^{\l(1\r)}\r)\,,\\
\mathcal{F}_{L,q}^{\l(2\r)\mathrm{PS}}=&\l(-\frac{1}{2\eps}P_{gq}^{\l(0\r)}\otimes c_{L,g}^{\l(1\r)}+c_{L,q}^{\l(2\r)\mathrm{PS}}+P_{gq}^{\l(0\r)}\otimes a_{L,g}^{\l(1\r)}\r),\\
\mathcal{F}_{2,q}^{\l(2\r)\mathrm{PS}}=&\l(\frac{1}{\eps^{2}}\l(\frac{1}{8}P_{qg}^{\l(0\r)}\otimes P_{gq}^{\l(0\r)}\r)-\frac{1}{\eps}\l(\frac{1}{4}P_{qq}^{\l(1\r)\mathrm{PS}}+\frac{1}{2}P_{gq}^{\l(0\r)}\otimes c_{2,g}^{\l(1\r)}\r)\right.\nonumber\\
&\hspace{0.2cm}\left.+c_{2,q}^{\l(2\r)\mathrm{PS}}+P_{gq}^{\l(0\r)}\otimes a_{2,g}^{\l(1\r)}\r)\label{ali:fvanNerveen2}\,.
\end{align}
The previous expressions are needed to check all the initial-final quark-antiquark antennae. They are still unrenormalized and are thus to be compared to the unrenormalized antenna functions. The following identities hold at NLO:
\begin{align}
\frac{1}{\cf}\l(\mathcal{F}_{2,q}^{\l(1\r)}-\frac{3-2\eps}{2-2\eps}\mathcal{F}_{L,q}^{\l(1\r)}\r)= &\,4\mathcal{A}_{q,gq}+8\delta\l(1-z\r)\left.F_{q}^{\l(1\r)}\right|_{N}\,,\\
\frac{\l(2-2\eps\r)}{\nf}\l(\mathcal{F}_{2,g}^{\l(1\r)}-\frac{3-2\eps}{2-2\eps}\mathcal{F}_{L,g}^{\l(1\r)}\r)= &\,-4\mathcal{A}_{g,q\bar{q}}\,,
\end{align}
where $F_{q}^{\l(1\r)}$ is the one-loop quark form-factor given in~\cite{formfactors}. The notation $\left.X\right|_{a}$ means that only the terms of $X$ proportional to $a$ are considered. The corresponding identities at NNLO are
\begin{align}
\left.\frac{1}{\cf}\l(\mathcal{F}_{2,q}^{\l(2,\textrm{NS}\r)}-\frac{3-2\eps}{2-2\eps}\mathcal{F}_{L,q}^{\l(2,\textrm{NS}\r)}\r)\right|_{\frac{1}{N}}=\, &-\mathcal{\tilde{A}}^{0}_{q,ggq}+8\mathcal{\tilde{A}}^{1,U}_{q,gq}-4\mathcal{C}^{0}_{q,\bar{q}q\bar{q}}-2\mathcal{C}^{0}_{\bar{q},\bar{q}q\bar{q}}-2\mathcal{C}^{0}_{\bar{q},q\bar{q}\bar{q}}\nonumber\\
&\,+\left.16\delta\l(1-z\r)\l(F_{q}^{\l(2,U\r)}+\cf F_{q}^{\l(1\r)\,2}\r)\right|_{\frac{1}{N}}\,,\\
\left.\frac{1}{\cf}\l(\mathcal{F}_{2,q}^{\l(2,\textrm{NS}\r)}-\frac{3-2\eps}{2-2\eps}\mathcal{F}_{L,q}^{\l(2,\textrm{NS}\r)}\r)\right|_{N}=\, &\,2\mathcal{A}^{0}_{q,ggq}+8\mathcal{A}^{1,U}_{q,gq}+\left.16\delta\l(1-z\r)\l(F_{q}^{\l(2,U\r)}+\cf F_{q}^{\l(1\r)\,2}\r)\right|_{N}\,,\\
\left.\frac{1}{\cf}\l(\mathcal{F}_{2,q}^{\l(2,\textrm{NS}\r)}-\frac{3-2\eps}{2-2\eps}\mathcal{F}_{L,q}^{\l(2,\textrm{NS}\r)}\r)\right|_{\nf}=\, &\,2\mathcal{B}^{0}_{q,q^{\prime}\bar{q}^{\prime}q}+\left.16\delta\l(1-z\r)\l(F_{q}^{\l(2,U\r)}+\cf F_{q}^{\l(1\r)\,2}\r)\right|_{\nf}\,,\\
\frac{1}{\cf\,\nf}\l(\mathcal{F}_{2,q}^{\l(2,\textrm{PS}\r)}-\frac{3-2\eps}{2-2\eps}\mathcal{F}_{L,q}^{\l(2,\textrm{PS}\r)}\r)=\,&2\mathcal{\bar{B}}^{0}_{q,q^{\prime}\bar{q}^{\prime}q}\,,\\
\left.\frac{\l(2-2\eps\r)}{\nf}\l(\mathcal{F}_{2,g}^{\l(2\r)}-\frac{3-2\eps}{2-2\eps}\mathcal{F}_{L,g}^{\l(2\r)}\r)\right|_{\frac{1}{N}}=&\,2\mathcal{\tilde{A}}^{0}_{g,gq\bar{q}}-8\mathcal{\tilde{A}}^{1,U}_{g,q\bar{q}}\,,\\
\left.\frac{\l(2-2\eps\r)}{\nf}\l(\mathcal{F}_{2,g}^{\l(2\r)}-\frac{3-2\eps}{2-2\eps}\mathcal{F}_{L,g}^{\l(2\r)}\r)\right|_{N}=&\,\,-4\mathcal{A}^{0}_{g,gq\bar{q}}-8\mathcal{A}^{1,U}_{g,q\bar{q}}\,.
\end{align}
The structure functions (\ref{ali:fvanNerveen1})-(\ref{ali:fvanNerveen2})
are given in~\cite{zv} in unrenormalized form. To compare with our
results, we thus considered the one-loop unrenormalized antenna functions
(obtained by undoing the renormalization)
the two-loop unrenormalized form factor, both characterized by the label $U$.
Full agreement is found with~\cite{zv}.

\subsection{Deep inelastic scattering: $\phi$-scalar induced}
The gluon-gluon $\mathcal{F}$-,  $\mathcal{G}$- and $\mathcal{H}$-type
antenna functions
 can be checked with the deep inelastic scalar induced structure functions
derived in~\cite{moch,moch2}. These structure functions are obtained in an
effective theory with a scalar
 $\phi$ coupled to the square of the gluon field strength tensor. This
effective theory is very instrumental in deriving gluonic splitting functions,
and is also the effective field theory describing the coupling of the Higgs
boson to gluons through a loop of asymptotically heavy coloured particles.
Since $\phi$ is a scalar particle the corresponding hadronic tensor
has only one structure function which can be either quark- or
 gluon-initiated.
Their perturbative expansion reads:

\begin{eqnarray}
\mathcal{T}_{\phi,g}&=&\mathcal{T}_{\phi,g}^{\l(0\r)}+\frac{\alpha_{s}}{4\pi
}S_{\eps}\l(\frac{\mu^{2}}{Q^{2}}\r)^{\eps}\mathcal{T}_{\phi,g}^{\l(1\r)}+\l
(\frac{\alpha_{s}}{4\pi}\r)^{2}S_{\eps}^{2}\l(\frac{\mu^{2}}{Q^{2}}\r)^{2\eps}\mathcal{T}_{\phi,g}^{\l(2\r)}+\mathcal{O}\l(\alpha_{s}^{3}\r),\nonumber\\
\mathcal{T}_{\phi,q}&=&\mathcal{T}_{\phi,q}^{\l(0\r)}+\frac{\alpha_{s}}{4\pi
}S_{\eps}\l(\frac{\mu^{2}}{Q^{2}}\r)^{\eps}\mathcal{T}_{\phi,q}^{\l(1\r)}+\l
(\frac{\alpha_{s}}{4\pi}\r)^{2}S_{\eps}^{2}\l(\frac{\mu^{2}}{Q^{2}}\r)^{2\eps}\mathcal{T}_{\phi,q}^{\l(2\r)}+\mathcal{O}\l(\alpha_{s}^{3}\r)\,.
\end{eqnarray}

At NLO and NNLO the structure functions have the
following form:
\begin{align}
\mathcal{T}_{\phi,g}^{\l(1\r)}=&\l(-\frac{1}{2\epsilon}P_{gg}^{\l(0\r)}+c_{\phi,g}^{\l(1\r)}-\epsilon\,2a_{\phi,g}^{\l(1\r)}\r)\,,\\
\mathcal{T}_{\phi,q}^{\l(1\r)}=&\l(-\frac{1}{2\epsilon}P_{gq}^{\l(0\r)}+c_{\phi,q}^{\l(1\r)}-\epsilon\,2a_{\phi,q}^{\l(1\r)}\r)\,,
\end{align}
\begin{align}
\mathcal{T}_{\phi,g}^{\l(2\r)}=&\l(\frac{1}{\epsilon^{2}}\l(\frac{1}{8}P_{gg}^{\l(0\r)}\otimes P_{gg}^{\l(0\r)}+\frac{\nf}{8}P_{gq}^{\l(0\r)}\otimes P_{qg}^{\l(0\r)}+\frac{\beta_{0}}{2}P_{gg}^{\l(0\r)}\r)\right.\nonumber\\
                  &\hspace{0.4cm}+\frac{1}{\epsilon}\l(-\frac{1}{4}P_{gg}^{\l(1\r)}-\frac{1}{2}P_{gg}^{\l(0\r)}\otimes c_{\phi,g}^{\l(1\r)}-\frac{\nf}{2}P_{qg}^{\l(0\r)}\otimes c_{\phi,q}^{\l(1\r)}\r)\nonumber\\
                  &\hspace{0.0cm}\left.\phantom{\frac{1}{1}}+c_{\phi,g}^{\l(2\r)}+P_{gg}^{\l(0\r)}\otimes a_{\phi,g}^{\l(1\r)}+\nf P_{qg}^{\l(0\r)}\otimes a_{\phi,q}^{\l(1\r)}\r)\,,\\
\mathcal{T}_{\phi,q}^{\l(2\r)}=&\l(\frac{1}{\epsilon^{2}}\l(\frac{1}{8}P_{qq}^{\l(0\r)}\otimes P_{gq}^{\l(0\r)}+\frac{1}{8}P_{gg}^{\l(0\r)}\otimes P_{gq}^{\l(0\r)}+\frac{\beta_{0}}{2}P_{gq}^{\l(0\r)}\r)\right.\nonumber\\
                  &\hspace{0.4cm}+\frac{1}{\epsilon}\l(-\frac{1}{4}P_{gq}^{\l(1\r)}-\frac{1}{2}P_{gq}^{\l(0\r)}\otimes c_{\phi,g}^{\l(1\r)}-\frac{1}{2}P_{qq}^{\l(0\r)}\otimes c_{\phi,q}^{\l(1\r)}\r)\nonumber\\
                  &\hspace{0.0cm}\left.\phantom{\frac{1}{1}}+c_{\phi,q}^{\l(2\r)}+P_{gq}^{\l(0\r)}\otimes a_{\phi,g}^{\l(1\r)}+P_{qq}^{\l(0\r)}\otimes a_{\phi,q}^{\l(1\r)}\r)\,.
\end{align}
The explicit expressions for the coefficients $c_{\phi,i}^{\l(1\r)}$, $a_{\phi,i}^{\l(1\r)}$ and $c_{\phi,i}^{\l(2\r)}$, with $i=q,g$ are given in Appendix~\ref{app:phidiscoeffunc}. They were derived independently in~\cite{moch2}, and
we find full agreement. In our calculation, they follow
from the
following identities between structure functions and antennae:
\begin{align}
\mathcal{T}_{\phi,g}^{\l(1\r)}=&\,2\l(\ca \mathcal{F}^{0}_{g,gg} + \nf \mathcal{G}^{0}_{g,q\bar{q}}\r)+4\delta\l(1-z\r)F_{g}^{\l(1\r)}\,,\\
\frac{1}{\cf\l(1-\eps\r)}\mathcal{T}_{\phi,q}^{\l(1\r)}=&-4 \mathcal{G}^{0}_{q,qg}\,,\\
\left.\mathcal{T}_{\phi,g}^{\l(2\r)}\right|_{N^{2}}=&\,\mathcal{F}^{0}_{g,ggg} + 4\mathcal{F}^{0}_{g,gg} + 4\delta\l(1-z\r)\left.\l(2 F_{g}^{\l(2\r)} + F_{g}^{\l(1\r)\,2}\r)\right|_{N^{2}}\,,\\
\left.\mathcal{T}_{\phi,g}^{\l(2\r)}\right|_{N\,\nf}=&\,2\mathcal{G}^{0}_{g,q\bar{q}g}+4\mathcal{G}^{1,R}_{g,q\bar{q}}+ 4\mathcal{\hat{F}}^{1,R}_{g,gg}+ 4\delta\l(1-z\r)\left.\l(2 F_{g}^{\l(2\r)} + F_{g}^{\l(1\r)\,2}\r)\right|_{N\,\nf}\,,\\
\left.\mathcal{T}_{\phi,g}^{\l(2\r)}\right|_{\frac{\nf}{N}}=&-\mathcal{\tilde{G}}^{0}_{g,q\bar{q}g}+4\mathcal{\tilde{G}}^{1,R}_{g,q\bar{q}}+ 4\delta\l(1-z\r)\left.\l(2 F_{g}^{\l(2\r)} + F_{g}^{\l(1\r)\,2}\r)\right|_{\frac{\nf}{N}}\,,\\
\left.\mathcal{T}_{\phi,g}^{\l(2\r)}\right|_{\nf^{2}}=&\,4\hat{\mathcal{G}}^{1,R}_{g,q\bar{q}}+ 4\delta\l(1-z\r)\left.\l(2 F_{g}^{\l(2\r)} + F_{g}^{\l(1\r)\,2}\r)\right|_{\nf^{2}}\,,\\
\left.\frac{1}{\cf\l(1-\eps\r)}\mathcal{T}_{\phi,q}^{\l(2\r)}\right|_{N}=&\,-2\mathcal{G}^{0}_{q,qgg}-8\mathcal{G}^{1,R}_{q,qg}\,,\\
\left.\frac{1}{\cf\l(1-\eps\r)}\mathcal{T}_{\phi,q}^{\l(2\r)}\right|_{\frac{1}{N}}=&\,\mathcal{\tilde{G}}^{0}_{q,qgg}+ 8\mathcal{\tilde{G}}^{1,R}_{q,qg}-4 \mathcal{\tilde{J}}^{0}_{q,qq\bar{q}}\,,\\
\left.\frac{1}{\cf\l(1-\eps\r)}\mathcal{T}_{\phi,q}^{\l(2\r)}\right|_{\nf}=&\,-2\mathcal{H}^{0}_{q,qq^{\prime}\bar{q}^{\prime}}-8\mathcal{\hat{G}}^{1,R}_{q,qg}\,.
\end{align}
The antenna $\mathcal{\tilde{J}}^{0}_{q,qq\bar{q}}$ comes from the interference of four quark final states with
identical quark flavour, which
is not contained in Table~\ref{tab:one} since it is finite in all limits. It is given by
\begin{equation}
\mathcal{\tilde{J}}^{0}_{q,qq\bar{q}}=\Bigl(-\frac{7}{4}+\frac{7}{4 z}+\frac{9 z}{8}+\bigl(2-\frac{2}{z}-z\bigr) \zeta_{3}\Bigr)
+\mathcal{O}\left(\epsilon\right)\,.
\end{equation}
The functions $F_{g}^{\l(i\r)}$ ($i=1,2$) are the gluon form factor at one- respectively two-loop given in~\cite{formfactors}.
\section{Conclusions and Outlook}
\label{sec:conc}
In this paper, we have extended the NNLO antenna subtraction
formalism~\cite{ourant} to include initial-final antenna configurations, where
one of the hard radiator partons is in the initial state. We described the
construction of  initial-final antenna subtraction terms from antenna
functions, including the required phase space factorization and mappings.

The analytic integration of the NNLO initial-final antenna functions
requires the derivation of inclusive $2\to 3$ phase space integrals
at tree-level and of $2\to 2$ phase space integrals at one loop. We
reduced all these integrals  to a small set of master integrals, which
we computed using differential equations or by direct integration.
With these, we provide integrals of all NNLO initial-final antenna functions,
as required for the implementation into a parton-level event generator
programme.
A highly non-trivial
check of our results is provided by the rederivation of deep inelastic
coefficient functions~\cite{zv,moch,moch2},
which can be expressed as particular linear
combinations of antenna functions.

The subtraction terms  presented here allow the
construction of a parton-level event generator program for the
calculation of NNLO corrections to jet production observables in deeply
inelastic electron-proton scattering. Of foremost importance in this
context is the
 dijet production cross section
in deep inelastic scattering~\cite{dijethera}, which has been used
 at HERA to measure
the gluon distribution in the proton and the strong coupling constant
$\alpha_s$. This cross section is at present only known to
NLO~\cite{allMC}, which is insufficient for various reasons.
The error on the extraction of $\alpha_s$ from HERA  (2+1)
jet  data is dominated~\cite{ashera}  not by the statistical uncertainty on the
data, but  by the uncertainty inherent to the NLO calculation, as estimated by
varying renormalization and factorization scales. Given that the  statistical
precision of the data will further improve once all data from  the HERA-II run
are analysed, the theoretical description requires the inclusion of the
next-to-next-to-leading order (NNLO), i.e.\ ${\cal O}(\alpha_s^3)$,
corrections. Moreover, present determinations
of parton distribution functions~\cite{nnlopdf} at NNLO
accuracy~\cite{nnlosplit} can only include
data sets for observables where NNLO corrections are
known~\cite{zv,nnlocoeff}. Consequently, the precise HERA deep inelastic
dijet data could be used in global NNLO determinations only once the
NNLO QCD corrections to this observable are computed.
Using the subtraction formalism derived here, and
the already known
matrix elements~\cite{tree5p,onel4p,3jme,distensor},
such a calculation can now be envisaged.

Moreover, the initial-final antenna functions derived here are an important
ingredient to the calculation of NNLO corrections to jet observables at
hadron colliders. In this case, all radiator configurations contribute:
final-final (derived in~\cite{ourant}), initial-final (derived here) and
initial-initial. The initial-initial antenna functions are at
present known only to NLO, and work on their extension to NNLO is ongoing.

\section*{Acknowledgements}
We would like to thank Andreas Vogt and Sven Moch for providing us with the
results of~\cite{moch2}
on the NNLO $\phi$-DIS coefficient functions prior to publication,
and for useful discussions.
This research was supported in part by
the Swiss National Science Foundation (SNF) under contracts PP0022-118864 and
200020-126691.

\appendix
\section{Master Integrals}
\label{app:mi}

The common prefactor $S_{\Gamma}$ is defined as:
\begin{equation}
S_{\Gamma}=\left(\frac{(4\pi )^{\epsilon}}{16\pi^{2}\Gamma(1-\epsilon)}\right)^{2}\,.
\end{equation}

\subsection{Master integrals for double real radiation}
\label{app:realmasters}
The double real master integrals expanded to the needed order in $\epsilon$ are given by:
\begin{equation}
\begin{split}
I_{0}=&\,S_{\Gamma}\left(Q^{2}\right)^{1-2\epsilon}(1-x)^{1-2\epsilon}x^{-1}\pi\\
&\Bigl[1+\epsilon\Bigl(\frac{13}{2}+2\,\h(0;x)\Bigr)
+\epsilon^{2}\Bigl(\frac{115}{4}-\frac{2}{3}\pi^{2}+13\,\h(0;x)+4\,\h(0,0;x)\Bigr)\\
&+\epsilon^{3}\Bigl(\frac{865}{8}-\frac{13}{3}\pi^{2}-10\zeta_{3}+\frac{115}{2}\,\h(0;x)-\frac{4}{3}\pi^{2}\,\h(0;x)+26\,\h(0,0;x)+8\,\h(0,0,0;x)\Bigr)\\
&+\epsilon^{4}\Bigl(\frac{5971}{16}-\frac{115}{6}\pi^{2}-65\zeta_{3}-\frac{\pi^4}{30}+\frac{865}{4}\,\h(0;x)-\frac{26}{3}\pi^{2}\,\h(0;x)-20\,\h(0;x)\zeta_{3}\\
&+115\,\h(0,0;x)-\frac{8}{3}\pi^{2}\,\h(0,0;x)+52\,\h(0,0,0;x)+16\,\h(0,0,0,0;x)\Bigr)
+\mathcal{O}\left(\epsilon^{5}\right)\Bigr]\,,
\end{split}
\end{equation}

\begin{equation}
\begin{split}
I_{2}=&S_{\Gamma}\left(Q^{2}\right)^{-2\epsilon}(1-x)^{-2\epsilon}\pi\\
&\Bigl[2\,\h(0;x)-\epsilon\Bigl(\frac{2}{3}\pi^{2}-10\,\h(0;x)-6\,\h(0,0;x)+4\,\h(1,0;x)\Bigr)-\epsilon^2\Bigl(\frac{10}{3}\pi^{2}+4\zeta_{3}\\
&-38\,\h(0;x)+2\pi^{2}\,\h(0;x)-\frac{4}{3}\pi^{2}\h(1;x)-30\,\h(0,0;x)+20\h(1,0;x)+4\,\h(0,1,0;x)\\
&-14\,\h(0,0,0;x)+12\,\h(1,0,0;x)-8\,\h(1,1,0;x)\Bigr)+\epsilon^{3}\Bigl(-\frac{38}{3}\pi^{2}-20\zeta_{3}+\frac{11}{45}\pi^{4}\\
&+130\,\h(0;x)-10\pi^{2}\,\h(0;x)-24\,\h(0;x)\,\zeta_{3}+\frac{20}{3}\pi^{2}\,\h(1;x)+8\,\h(1;x)\,\zeta_{3}\\
&+\frac{4}{3}\pi^{2}\,\h(0,1;x)+114\,\h(0,0;x)-\frac{14}{3}\pi^{2}\,\h(0,0;x)-76\,\h(1,0;x)+4\pi^{2}\,\h(1,0;x)\\
&-\frac{8}{3}\pi^{2}\,\h(1,1;x)-20\,\h(0,1,0;x)-4\,\h(0,0,1,0;x)+70\,\h(0,0,0;x)-60\,\h(1,0,0;x)\\
&+40\,\h(1,1,0;x)+8\,\h(1,0,1,0;x)-12\,\h(0,1,0,0;x)+8\,\h(0,1,1,0;x)\\
&+30\,\h(0,0,0,0;x)-28\,\h(1,0,0,0;x)+24\,\h(1,1,0,0;x)-16\,\h(1,1,1,0;x)\Bigr)\\
&+\mathcal{O}\left(\epsilon^{4}\right)\Bigr]\,,
\end{split}
\end{equation}

\begin{equation}
\begin{split}
I_{26}=&S_{\Gamma}\left(Q^{2}\right)^{-1-2\epsilon}(1-x)^{-2\epsilon}x\,\pi\\
&\Bigl[\frac{1}{\epsilon}\Bigl(\frac{2\pi^2}{3}+2\,\h(0,0;x)+2\,\h(1,0;x)\Bigr)+4\zeta_{3}-\frac{2}{3}\pi^2\,\h(0;x)-\frac{2}{3}\pi^2\,\h(1;x)\\
&-4\h(0,1,0;x)+6\,\h(0,0,0;x)+6\,\h(1,0,0;x)-4\,\h(1,1,0;x)+\epsilon\Bigl(\frac{5}{9}\pi^4\\
&-4\h(0;x)\zeta_{3}-4\h(1;x)\zeta_{3}+\frac{4}{3}\pi^2\,\h(0,1;x)-2\pi^2\,\h(0,0;x)-2\pi^2\,\h(1,0;x)\\
&+\frac{4}{3}\pi^2\,\h(1,1;x)-4\h(0,0,1,0;x)-4\h(1,0,1,0;x)-12\h(0,1,0,0;x)\\
&+8\h(0,1,1,0;x)+14\h(0,0,0,0;x)+14\h(1,0,0,0;x)-12\h(1,1,0,0;x)\\
&+8\h(1,1,1,0;x)\Bigr)
+\mathcal{O}\left(\epsilon^{2}\right)\Bigr]\,,
\end{split}
\end{equation}

\begin{equation}
\begin{split}
I_{249}=&S_{\Gamma}\left(Q^{2}\right)^{-2\epsilon-2}(1-x)^{-2\epsilon}x^2\pi \\
&\Bigl[-\frac{2}{\epsilon^3}-\frac{8}{\epsilon^{2}}\,\h(0;x)+\frac{1}{\epsilon}\Bigl(\frac{10}{3}\pi^2+10\,\h(1,0;x)-22\,\h(0,0;x)\Bigr)+56\zeta_{3}+10\pi^2\,\h(0;x)\\
&-\frac{10}{3}\pi^2\h(1;x)+20\,\h(0,1,0;x)-50\,\h(0,0,0;x)+38\,\h(1,0,0;x)-20\,\h(1,1,0;x)\\
&+\epsilon\Bigl(-\frac{56}{45}\pi^4+156\zeta_{3}\h(0;x)-68\,\h(1;x)\zeta_{3}-\frac{20}{3}\pi^2\,\h(0,1;x)+22\pi^2\,\h(0,0;x)\\
&-18\pi^2\,\h(1,0;x)+\frac{20}{3}\pi^2\,\h(1,1;x)+28\,\h(0,0,1,0;x)-52\,\h(1,0,1,0;x)\\
&+76\,\h(0,1,0,0;x)-40\,\h(0,1,1,0;x)-106\,\h(0,0,0,0;x)+94\,\h(1,0,0,0;x)\\
&-76\,\h(1,1,0,0;x)+40\,\h(1,1,1,0;x)\Bigr)
+\mathcal{O}\left(\epsilon^{2}\right)\Bigr]\,,
\end{split}
\end{equation}

\begin{equation}
\begin{split}
I_{125}=&\,S_{\Gamma}\left(Q^{2}\right)^{-2-2\epsilon}(1-x)^{-2\epsilon}(1+x)^{-1}x^{2}\pi\\
&\Bigl[\frac{2}{\epsilon^2}\h(0;x)+\frac{1}{\epsilon}\Bigl(-\pi^2-4\h(-1,0;x)+8\h(0,0;x)-4\h(1,0;x)\Bigr)-14\zeta_{3}\\
&-\frac{8}{3}\pi^2\,\h(0;x)+2\pi^2\,\h(1;x)-8\h(-2,0;x)+2\pi^2\,\h(-1;x)-8\h(0,1,0;x)\\
&+8\h(-1,-1,0;x)-16\h(-1,0,0;x)+8\h(-1,1,0;x)+20\h(0,0,0;x)\\
&+8\h(1,-1,0;x)-16\h(1,0,0;x)+8\h(1,1,0;x)+\epsilon \Bigl(-\frac{\pi^4}{30}+4\pi^2 \h(0,-1;x)\\
&+4\pi^2\h(0,1;x)-4\pi^2 \h(-1,-1;x)+\frac{16}{3}\pi^2 \h(-1,0;x)-4\pi^2 \h(-1,1;x)\\
&-\frac{20}{3}\pi^2 \h(0,0;x)-4\pi^2 \h(1,-1;x)+\frac{16}{3}\pi^2\h(1,0;x)-4\pi^2 \h(1,1;x)\\
&+28 \zeta_{3}\h(-1;x)-28 \zeta_{3}\h(0;x)+28 \zeta_{3}\h(1;x)-32\h(0,0,-1,0;x)\\
&-8 \h(0,0,1,0;x)+16 \h(0,-1,-1,0;x)-32 \h(0,-1,0,0;x)+16 \h(0,-1,1,0;x)\\
&+16\h(-1,0,-1,0;x)+16 \h(-1,0,1,0;x)+16 \h(1,0,-1,0;x)+16 \h(1,0,1,0;x)\\
&+16\h(0,1,-1,0;x)-32 \h(0,1,0,0;x)+16 \h(0,1,1,0;x)-16 \h(-1,-1,-1,0;x)\\
&+32\h(-1,-1,0,0;x)-16 \h(-1,-1,1,0;x)-40 \h(-1,0,0,0;x)\\
&-16 \h(-1,1,-1,0;x)+32\h(-1,1,0,0;x)-16 \h(-1,1,1,0;x)+44 \h(0,0,0,0;x)\\
&-16 \h(1,-1,-1,0;x)+32 \h(1,-1,0,0;x)-16 \h(1,-1,1,0;x)-40 \h(1,0,0,0;x)\\
&-16\h(1,1,-1,0;x)+32\h(1,1,0,0;x)-16\h(1,1,1,0;x)\Bigr)
+\mathcal{O}\left(\epsilon^{2}\right)\Bigr]\,,
\end{split}
\end{equation}

\begin{equation}
\begin{split}
I_{235}=&S_{\Gamma}\left(Q^{2}\right)^{-2-2\epsilon}(1-x)^{-1-2\epsilon}x^2\pi\\
&\Bigl[-\frac{\h(0;x)}{\epsilon^2}-\frac{7\h(0,0;x)}{\epsilon}+8\zeta_{3}+\frac{7}{3}\pi^2\,\h(0;x)-8\h(0,-1,0;x)+10\h(0,1,0;x)\\
&-19\h(0,0,0;x)+\epsilon\Bigl(-\frac{2 \pi^4}{5}+20\zeta_{3}\h(0;x)-32 \h(0,0,-1,0;x)+4\pi^2 \h(0,-1;x)\\
&-2\pi^2\h(0,1;x)+\frac{19}{3}\pi^2\h(0,0;x)+10 \h(0,0,1,0;x)+16 \h(0,-1,-1,0;x)\\
&-32 \h(0,-1,0,0;x)+16\h(0,-1,1,0;x)+16 \h(0,1,-1,0;x)+22 \h(0,1,0,0;x)\\
&-20 \h(0,1,1,0;x)-43\h(0,0,0,0;x)\Bigr)
+\mathcal{O}\left(\epsilon^{2}\right)\Bigr]\,,
\end{split}
\end{equation}

\begin{equation}
\begin{split}
I_{1245}=&S_{\Gamma}\left(Q^{2}\right)^{-3-2\epsilon}(1-x)^{-2\epsilon}(1+x)^{-1}x^3\pi\\
&\Bigl[-\frac{8\h(0;x)}{\epsilon^2}+\frac{1}{\epsilon}\Bigl(\frac{10\pi^2}{3}+8\h(-1,0;x)-28\h(0,0;x)+16\h(1,0;x)\Bigr)+28\zeta_{3}\\
&-\frac{8}{3}\pi^2\,\h(-1;x)+\frac{28}{3}\pi^2\,\h(0;x)-\frac{20}{3}\pi^2\,\h(1;x)+8\h(0,-1,0;x)+24\h(0,1,0;x)\\
&+24\h(-1,0,0;x)-16\h(-1,1,0;x)-68\h(0,0,0;x)-16\h(1,-1,0;x)\\
&+56\h(1,0,0;x)-32\h(1,1,0;x)+\epsilon\Bigl(-\frac{49 \pi^4}{45}+88 \h(0;x)\zeta_{3}-56\h(1;x)\zeta_{3}\\
&-\frac{4}{3} \pi^2 \h(0,-1;x)-\frac{28}{3} \pi^2 \h(0,1;x)+16 \h(0,0,-1,0;x)-\frac{8}{3} \pi^2\h(-1,-1;x)\\
&-8 \pi^2 \h(-1,0;x)+\frac{16}{3} \pi^2 \h(-1,1;x)+\frac{68}{3} \pi^2 \h(0,0;x)+\frac{16}{3}\pi^2 \h(1,-1;x)\\
&-\frac{56}{3} \pi^2 \h(1,0;x)+\frac{40}{3} \pi^2 \h(1,1;x)+24\h(0,0,1,0;x)+16 \h(0,-1,-1,0;x)\\
&+16 \h(0,-1,0,0;x)-16 \h(0,-1,1,0;x)+16\h(-1,0,-1,0;x)-16 \h(-1,0,1,0;x)\\
&-16 \h(1,0,-1,0;x)-48 \h(1,0,1,0;x)-16\h(0,1,-1,0;x)+80 \h(0,1,0,0;x)\\
&-48 \h(0,1,1,0;x)-32 \h(-1,-1,-1,0;x)+16\h(-1,-1,0,0;x)\\
&+56 \h(-1,0,0,0;x)-48 \h(-1,1,0,0;x)+32 \h(-1,1,1,0;x)-148\h(0,0,0,0;x)\\
&-48 \h(1,-1,0,0;x)+32 \h(1,-1,1,0;x)+136 \h(1,0,0,0;x)+32\h(1,1,-1,0;x)\\
&-112 \h(1,1,0,0;x)+64 \h(1,1,1,0;x)\Bigr)
+\mathcal{O}\left(\epsilon^{2}\right)\Bigr]\,,
\end{split}
\end{equation}

\begin{equation}
\begin{split}
I_{2356}=&S_{\Gamma}\left(Q^{2}\right)^{-3-2\epsilon}(1-x)^{-1-2\epsilon}x^3\pi\\
&\Bigl[-\frac{2}{\epsilon^3}-\frac{8\h(0;x)}{\epsilon^2}+\frac{1}{\epsilon}\Bigl(2\pi^2-28\h(0,0;x)+4\h(1,0;x)\Bigr)+32\zeta_{3}+\frac{28}{3}\pi^2\,\h(0;x)\\
&-8\h(0,-1,0;x)+24\h(0,1,0;x)-68\h(0,0,0;x)+28\h(1,0,0;x)+\epsilon \Bigl(256\zeta_{3}\\
&-\frac{43\pi^4}{18}+4\pi^2\h(0,-1;x)-4\pi^2 \h(0,1;x)+\frac{68}{3}\pi^2 \h(0,0;x)-\frac{28}{3}\pi^2 \h(1,0;x)\\
&+88\zeta_{3}\h(0;x)-32\zeta_{3}\h(1;x)-16 \h(0,0,-1,0;x)+24 \h(0,0,1,0;x)\\
&+16 \h(0,-1,-1,0;x)-32 \h(0,-1,0,0;x)+16\h(0,-1,1,0;x)+32 \h(1,0,-1,0;x)\\
&-40 \h(1,0,1,0;x)+16 \h(0,1,-1,0;x)+96\h(0,1,0,0;x)-32 \h(0,1,1,0;x)\\
&-148 \h(0,0,0,0;x)+76 \h(1,0,0,0;x)\Bigr)
+\mathcal{O}\left(\epsilon^{2}\right)\Bigr]\,,
\end{split}
\end{equation}

\begin{equation}
\begin{split}
I_{1346}=&S_{\Gamma}\left(Q^{2}\right)^{-3-2\epsilon}(1-x)^{-1-2\epsilon}x^3\pi\\
&\Bigl[-\frac{10}{\epsilon^3}-\frac{20\h(0;x)}{\epsilon^2}+\frac{1}{\epsilon}\Bigl(\frac{20\pi^2}{3}-40\h(0,0;x)\Bigr)+92\zeta_{3}+\frac{40}{3}\pi^2\,\h(0;x)-80\h(0,0,0;x)\\
&+\epsilon\Bigl(-\frac{1}{3}\pi^4+184\zeta_{3}\h(0;x)+\frac{80}{3}\pi^2 \h(0,0;x)-160 \h(0,0,0,0;x)\Bigr)
+\mathcal{O}\left(\epsilon^{2}\right)\Bigr]\,.
\end{split}
\end{equation}

\subsection{Master integrals for one-loop single real radiation}
\label{app:virtmasters}
\noindent The virtual master integrals are given by:

\begin{equation}
\begin{split}
V_{14}=&2\pi i \,S_{\Gamma}\left(Q^{2}\right)^{-2\epsilon}(1-x)^{-\epsilon}\\
&\Bigl[\frac{1}{\epsilon}+4+\h(0;x)+\epsilon\Bigl(12-\frac{\pi^2}{6}+4\,\h(0;x)+\h(0,0;x)\Bigr)+\epsilon^2\Bigl(32-\frac{2\pi^2}{3}-4\zeta_{3}\\
&+12\,\h(0;x)-\frac{1}{6}\pi^2\,\h(0;x)+4\,\h(0,0;x)+\h(0,0,0;x)\Bigr)+\epsilon^3\Bigl(80-2\pi^2-16\zeta_{3}\\
&-\frac{7\pi^4}{120}+32\,\h(0;x)-\frac{2}{3}\pi^2\,\h(0;x)-4\,\h(0;x)\,\zeta_{3}+12\,\h(0,0;x)-\frac{1}{6}\pi^2\,\h(0,0;x)\\
&+4\,\h(0,0,0;x)+\h(0,0,0,0;x)\Bigr)
+\mathcal{O}\left(\epsilon^{4}\right)\Bigr]\,,
\end{split}
\end{equation}

\begin{equation}
\begin{split}
V_{24}=&V_{14}(1-x)^{-\epsilon}\\ &\Bigl[1+\epsilon\, \h(0;x)+\epsilon^{2}\left(\h(0,0;x)-\frac{\pi^2}{2}\right)+\epsilon^{3}\Bigl(\h(0,0,0;x)-\frac{\pi^2}{2}\,\h(0;x)\Bigr)
+\mathcal{O}\left(\epsilon^{4}\right)\Bigr]\,,
\end{split}
\end{equation}

\begin{equation}
\begin{split}
V_{13}=&2\pi i \,S_{\Gamma}\,\left(Q^{2}\right)^{-2\epsilon}(1-x)^{-\epsilon}\\
&\Bigl[\frac{1}{\epsilon}+5+2\,\h(0;x)+\epsilon\Bigl(19-\frac{\pi^2}{3}+10\,\h(0;x)+4\,\h(0,0;x)\Bigr)+\epsilon^2\Bigl(65-\frac{5\pi^2}{3}-8\zeta_{3}\\
&+38\,\h(0;x)-\frac{2}{3}\pi^2\,\h(0;x)+20\,\h(0,0;x)+8\,\h(0,0,0;x)\Bigr)+\epsilon^3\,\Bigl(211-\frac{19}{3}\pi^2-40\zeta_{3}\\
&-\frac{7}{45}\pi^4+130\,\h(0;x)-\frac{10}{3}\pi^2\,\h(0;x)-16\,\h(0;x)\,\zeta_{3}+76\,\h(0,0;x)-\frac{4}{3}\pi^2\,\h(0,0;x)\\
&+40\,\h(0,0,0;x)+16\,\h(0,0,0,0;x)\Bigr)
+\mathcal{O}\left(\epsilon^{4}\right)\Bigr]\,,
\end{split}
\end{equation}

\begin{equation}
\begin{split}
V_{134}=&2\pi i \,S_{\Gamma}\,\left(Q^{2}\right)^{-1-2\epsilon}(1-x)^{-\epsilon}x\\
&\Bigl[\frac{1}{\epsilon}\Bigl(\frac{\pi^2}{3}+\h(0,0;x)+\h(1,0;x)\Bigr)-\zeta_{3}-\frac{1}{6}\pi^2\,\h(0;x)+\frac{1}{6}\pi^2\,\h(1;x)+3\,\h(0,0,0;x)\\
&+4\,\h(1,0,0;x)+\h(1,1,0;x)+\epsilon\Bigl(\frac{43}{360}\pi^4-4\,\h(0;x)\,\zeta_{3}-5\,\h(1;x)\,\zeta_{3}\\
&-\frac{1}{2}\pi^2\,\h(0,0;x)-\frac{2}{3}\pi^2\,\h(1,0;x)+\frac{1}{6}\pi^2\,\h(1,1;x)+7\,\h(0,0,0,0;x)\\
&+10\,\h(1,0,0,0;x)+4\,\h(1,1,0,0;x)+\,\h(1,1,1,0;x)\Bigr)
+\mathcal{O}\left(\epsilon^{2}\right)\Bigr]\,,
\end{split}
\end{equation}

\begin{equation}
\begin{split}
C_{1234}=&2\pi i \,S_{\Gamma}\,\left(Q^{2}\right)^{-2-2\epsilon}(1-x)^{-\epsilon}x^2\\ &\Bigl[-\frac{2}{\epsilon^3}-\frac{6}{\epsilon^2}\,\h(0;x)+\frac{1}{\epsilon}\Bigl(\frac{5\pi^2}{3}-16\,\h(0,0;x)+2\,\h(1,0;x)\Bigr)+34\zeta_{3}+\frac{16}{3}\pi^2\,\h(0;x)\\
&+\frac{1}{3}\pi^2\,\h(1;x)+8\,\h(0,1,0;x)-36\,\h(0,0,0;x)+12\,\h(1,0,0;x)+2\,\h(1,1,0;x)\\
&+\,\epsilon\,\Bigl(-\frac{4}{9}\pi^4+80\,\h(0;x)\,\zeta_{3}-22\,\h(1;x)\zeta_{3}+\frac{4}{3}\pi^2\,\h(0,1;x)+\frac{34}{3}\pi^2\,\h(0,0;x)\\
&-\frac{14}{3}\pi^2\,\h(1,0;x)+\frac{1}{3}\pi^2\,\h(1,1;x)+16\,\h(0,0,1,0;x)-8\,\h(1,0,1,0;x)\\
&+40\,\h(0,1,0,0;x)+8\,\h(0,1,1,0;x)-76\,\h(0,0,0,0;x)+32\,\h(1,0,0,0;x)\\
&+4\,\h(1,1,0,0;x)+2\,\h(1,1,1,0;x)\Bigr)
+\mathcal{O}\left(\epsilon^{2}\right)\Bigr]\,,
\end{split}
\end{equation}

\begin{equation}
\begin{split}
V_{12345}=&2\pi i \,S_{\Gamma}\,\left(Q^{2}\right)^{-3-2\epsilon}(1-x)^{-1-2\epsilon}x^3\\
&\Bigl[-\frac{3}{\epsilon^3}-\frac{9}{\epsilon^2}\,\h(0;x)+\frac{1}{\epsilon}\Bigl(\frac{13\pi^2}{6}+3\,\h(0,1;x)-22\,\h(0,0;x)+2\,\h(1,0;x)\Bigr)+23\zeta_{3}\\
&+\frac{17}{3}\pi^2\,\h(0;x)+\frac{1}{3}\pi^2\,\h(1;x)+10\,\h(0,0,1;x)-2\,\h(1,0,1;x)+10\,\h(0,1,0;x)\\
&-3\,\h(0,1,1;x)-48\,\h(0,0,0;x)+8\,\h(1,0,0;x)
+\mathcal{O}\left(\epsilon\right)\Bigr]\,.
\end{split}
\end{equation}

\section{$\phi$-DIS coefficient functions}
\label{app:phidiscoeffunc}
\noindent The coefficient functions for $\phi$-DIS structure functions derived by comparing them with the antennae as done in section~\ref{sec:discoeff} are given by:

\begin{align}
c^{\l(1\r)}_{\phi,g}=&\Bigl[\delta(1-z) \bigl(\frac{67 \ca}{9}-\frac{10 \nf}{9}-\frac{2 \ca \pi ^2}{3}\bigr)+\bigl(-\frac{11 \ca}{3}+\frac{2\nf}{3}\bigr)\d_{0}+4\ca\d_{1}+8\ca\h(0;z)\nl
&+8 \ca\h(1;z)-\frac{4 \ca \h(0;z)}{1-z}+\frac{1}{z}\bigl(-\frac{11 \ca}{3}+\frac{2 \nf}{3}-4\ca\h(0;z)-4\ca\h(1;z)\bigr)\nl
&-z\bigl(4 \ca \h(0;z)+4 \ca \h(1;z)\bigr)+z^2 (4 \ca \h(0;z)+4 \ca \h(1;z))\Bigr]
\,,
\end{align}

\begin{align}
a^{\l(1\r)}_{\phi,g}=&\Bigl[-2\ca+\ca\pi^2+\delta(1-z)\bigl(-\frac{148\ca}{27}+\frac{28 \nf}{27}+\frac{11 \ca \pi ^2}{24}-\frac{\nf\pi^2}{12}\bigr)+\bigl(\frac{67 \ca}{18}-\frac{5 \nf}{9}\nl
&-\frac{\ca \pi ^2}{2}\bigr) \d_{0}+\bigl(-\frac{11 \ca}{6}+\frac{\nf}{3}\bigr)\d_{1}+\ca\d_{2}-4\ca\h(0,0;z)-4\ca\h(0,1;z)\nl
&-4 \ca \h(1,0;z)-4\ca\h(1,1;z)+\frac{1}{1-z}\bigl(\frac{11}{6}\ca \h(0;z)-\frac{1}{3} \nf \h(0;z)+2 \ca \h(0,0;z)\nl
&+2 \ca \h(0,1;z)+2\ca\h(1,0;z)\bigr)+\frac{1}{z}\bigl(\frac{67\ca}{18}-\frac{5 \nf}{9}-\frac{\ca \pi ^2}{2}+\frac{11}{6} \ca \h(0;z)\nl
&-\frac{1}{3} \nf \h(0;z)+\frac{11}{6}\ca \h(1;z)-\frac{1}{3} \nf \h(1;z)+2 \ca \h(0,0;z)+2 \ca \h(0,1;z)\nl
&+2\ca\h(1,0;z)+2 \ca \h(1,1;z)\bigr)+z \bigl(-\frac{\ca \pi ^2}{2}+2 \ca\h(0,0;z)+2 \ca \h(0,1;z)\nl
&+2 \ca \h(1,0;z)+2 \ca \h(1,1;z)\bigr)+z^2 \bigl(\frac{\ca \pi ^2}{2}-2 \ca \h(0,0;z)-2 \ca\h(0,1;z)\nl
&-2 \ca \h(1,0;z)-2 \ca \h(1,1;z)\bigr)\Bigr]
\,,
\end{align}

\begin{align}
c^{\l(1\r)}_{\phi,q}=&\Bigl[4\cf\h(0;z)+4\cf\h(1;z)-\frac{1}{z}\bigl(3\cf+4\cf\h(0;z)+4\cf\h(1;z)\bigr)\hspace{3cm}\nl
&+z\bigl(2\cf-2\cf\h(0;z)-2\cf\h(1;z)\bigr)\Bigr]
\,,
\end{align}

\begin{align}
a^{\l(1\r)}_{\phi,q}=&\Bigl[-2 \cf+\frac{\cf \pi ^2}{2}-2 \cf \h(0,0;z)-2 \cf \h(0,1;z)-2 \cf \h(1,0;z)-2\cf \h(1,1;z)\nl
&+\frac{1}{z}\bigl(\frac{7 \cf}{2}-\frac{\cf \pi ^2}{2}+\frac{3}{2}\cf \h(0;z)+\frac{3}{2} \cf \h(1;z)+2 \cf \h(0,0;z)+2 \cf \h(0,1;z)\nl
&+2\cf\h(1,0;z)+2 \cf \h(1,1;z)\bigr)+z\bigl(-\frac{\cf \pi ^2}{4}-\cf \h(0;z)-\cf \h(1;z)\nl
&+\cf \h(0,0;z)+\cf\h(0,1;z)+\cf \h(1,0;z)+\cf \h(1,1;z)\bigr)\Bigr]
\,,
\end{align}

\begin{align}
c^{\l(2\r)}_{\phi,g}=&\Bigl[\frac{5150\ca^2}{27}-\frac{566 \ca \nf}{27}+\frac{521 \cf \nf}{9}-\frac{266 \ca^2 \pi ^2}{9}+\frac{20}{9}\ca\nf\pi^2-2\cf\nf\pi^2\nl
&-160\ca^2 \zeta_{3}-8 \cf \nf \zeta_{3}+\delta(1-z)
\bigl(\frac{30425 \ca^2}{162}-\frac{4112 \ca \nf}{81}-\frac{63 \cf \nf}{2}+\frac{100 \nf^2}{81}\nl
&-\frac{389 \ca^2\pi ^2}{27}+\frac{28}{9} \ca \nf \pi ^2-\frac{4 \nf^2 \pi ^2}{27}+\frac{101 \ca^2 \pi ^4}{180}-\frac{242}{3}\ca^2\zeta_{3}-\frac{28}{3}\ca\nf\zeta_{3}\nl
&+24 \cf \nf \zeta_{3}\bigr)+\bigl(-\frac{2570 \ca^2}{27}+\frac{224\ca \nf}{9}+2\cf\nf-\frac{40 \nf^2}{27}+\frac{88 \ca^2 \pi ^2}{9}-\frac{16}{9} \ca \nf \pi ^2\nl
&+68 \ca^2 \zeta_{3}\bigr)\d_{0}+\bigl(\frac{778}{9}\ca^2-\frac{56}{3}\ca\nf+\frac{8}{9}\nf^2-\frac{20}{3}\ca^2\pi^2\bigr)\d_{1}-\bigl(\frac{88}{3}\ca^2-\frac{16}{3}\ca\nf\bigr)\d_{2}\nl
&+8\ca^2\d_{3}+\frac{32}{3}\ca^2\pi^2\h(-1;z)+\frac{449}{9} \ca^2 \h(0;z)-\frac{242}{9}\ca\nf\h(0;z)\nl
&+\frac{86}{3} \cf \nf \h(0;z)-\frac{32}{3}\ca^2\pi^2\h(0;z)-\frac{8}{3}\cf\nf\pi^2 \h(0;z)+\frac{1192}{9} \ca^2 \h(1;z)\nl
&-\frac{256}{9} \ca \nf \h(1;z)+\frac{62}{3}\cf \nf \h(1;z)-\frac{40}{3} \ca^2 \pi ^2 \h(1;z)+\frac{532}{3} \ca^2 \h(0,1;z)\nl
&-\frac{40}{3}\ca \nf \h(0,1;z)+12 \cf \nf \h(0,1;z)+64 \ca^2 \h(0,0,1;z)+16 \cf \nf \h(0,0,1;z)\nl
&-96\ca^2 \h(0,-1,0;z)+16 \ca^2 \h(-1,0;z)-32 \ca^2 \h(-1,0,1;z)+202 \ca^2 \h(0,0;z)\nl
&-12\ca \nf \h(0,0;z)+14 \cf \nf \h(0,0;z)+136 \ca^2 \h(1,0;z)-16\ca\nf\h(1,0;z)\nl
&+4 \cf \nf \h(1,0;z)+136 \ca^2 \h(1,1;z)-16 \ca \nf \h(1,1;z)+4\cf \nf \h(1,1;z)\nl
&+112 \ca^2 \h(1,0,1;z)+64 \ca^2 \h(0,1,0;z)+8 \cf \nf \h(0,1,0;z)+80\ca^2 \h(0,1,1;z)\nl
&+8 \cf \nf \h(0,1,1;z)+64 \ca^2 \h(-1,-1,0;z)-80 \ca^2 \h(-1,0,0;z)\nl
&+20\cf \nf \h(0,0,0;z)+80 \ca^2 \h(1,0,0;z)+112 \ca^2 \h(1,1,0;z)+96 \ca^2 \h(1,1,1;z)\nl
&+\frac{1}{1+z}\bigl(28 \ca^2 \zeta_{3}-\frac{16}{3} \ca^2 \pi ^2 \h(-1;z)+\frac{4}{3} \ca^2 \pi ^2 \h(0;z)-8\ca^2 \h(0,0,1;z)\nl
&+24 \ca^2 \h(0,-1,0;z)+16 \ca^2 \h(-1,0,1;z)-32 \ca^2 \h(-1,-1,0;z)\nl
&+40\ca^2 \h(-1,0,0;z)-12 \ca^2 \h(0,0,0;z)\bigr)+\frac{1}{1-z}\bigl(-\frac{778}{9}\ca^2\h(0;z)\nl
&+\frac{56}{3} \ca \nf \h(0;z)-\frac{8}{9} \nf^2 \h(0;z)+\frac{28}{3}\ca^2\pi^2 \h(0;z)-\frac{176}{3} \ca^2 \h(0,1;z)\nl
&+\frac{32}{3}\ca\nf\h(0,1;z)-56 \ca^2 \h(0,0,1;z)+24 \ca^2 \h(0,-1,0;z)-44 \ca^2 \h(0,0;z)\nl
&+8\ca\nf\h(0,0;z)-\frac{176}{3} \ca^2 \h(1,0;z)+\frac{32}{3} \ca \nf \h(1,0;z)-56 \ca^2 \h(1,0,1;z)\nl
&-48 \ca^2 \h(0,1,0;z)-56\ca^2 \h(0,1,1;z)-28 \ca^2 \h(0,0,0;z)-40 \ca^2 \h(1,0,0;z)\nl
&-56 \ca^2 \h(1,1,0;z)\bigr)+\frac{1}{z}\bigl(-\frac{2743 \ca^2}{9}+\frac{1312\ca\nf}{27}+\frac{361\cf\nf}{27}\nl
&-\frac{40 \nf^2}{27}+\frac{176 \ca^2 \pi ^2}{9}-\frac{16}{9}\ca\nf\pi^2-\frac{8}{9} \cf \nf \pi ^2+40 \ca^2 \zeta_{3}+\frac{16}{3} \ca^2 \pi^2 \h(-1;z)\nl
&-\frac{1556}{9} \ca^2 \h(0;z)+\frac{88}{3}\ca\nf\h(0;z)+\frac{64}{9} \cf \nf \h(0;z)-\frac{8}{9} \nf^2 \h(0;z)+8 \ca^2\pi^2 \h(0;z)\nl
&-\frac{340}{3} \ca^2 \h(1;z)+\frac{88}{3} \ca \nf \h(1;z)+\frac{64}{9}\cf\nf\h(1;z)-\frac{8}{9} \nf^2 \h(1;z)+\frac{20}{3} \ca^2 \pi ^2 \h(1;z)\nl
&-88 \ca^2 \h(0,1;z)+\frac{32}{3}\ca\nf \h(0,1;z)+\frac{16}{3} \cf \nf \h(0,1;z)-48 \ca^2 \h(0,0,1;z)\nl
&+\frac{88}{3} \ca^2\h(-1,0;z)-16 \ca^2 \h(-1,0,1;z)-\frac{220}{3} \ca^2 \h(0,0;z)+8 \ca \nf \h(0,0;z)\nl
&+\frac{16}{3}\cf \nf \h(0,0;z)-\frac{352}{3} \ca^2 \h(1,0;z)+\frac{32}{3} \ca \nf \h(1,0;z)+\frac{16}{3}\cf \nf \h(1,0;z)\nl
&-\frac{352}{3} \ca^2 \h(1,1;z)+\frac{32}{3} \ca \nf \h(1,1;z)+\frac{16}{3}\cf \nf \h(1,1;z)-56 \ca^2 \h(1,0,1;z)\nl
&-48 \ca^2 \h(0,1,0;z)-56 \ca^2 \h(0,1,1;z)+32\ca^2 \h(-1,-1,0;z)-40 \ca^2 \h(-1,0,0;z)\nl
&-16 \ca^2 \h(0,0,0;z)-40 \ca^2 \h(1,0,0;z)-56\ca^2 \h(1,1,0;z)-48 \ca^2 \h(1,1,1;z)\bigr)\nl
&+z \bigl(-\frac{4486 \ca^2}{27}+\frac{580 \ca \nf}{27}-\frac{386\cf \nf}{9}+\frac{136 \ca^2 \pi ^2}{9}-\frac{16}{9} \ca \nf \pi ^2-\frac{4}{3} \cf \nf \pi ^2\nl
&+72 \ca^2 \zeta_{3}-8 \cf \nf \zeta_{3}+\frac{16}{3}\ca^2 \pi ^2 \h(-1;z)-\frac{1477}{9} \ca^2 \h(0;z)+\frac{196}{9} \ca \nf \h(0;z)\nl
&-\frac{2}{3}\cf \nf \h(0;z)+\frac{56}{3} \ca^2 \pi ^2 \h(0;z)-\frac{8}{3} \cf \nf \pi ^2 \h(0;z)-\frac{662}{9}\ca^2 \h(1;z)\nl
&+\frac{182}{9} \ca \nf \h(1;z)-\frac{38}{3} \cf \nf \h(1;z)+\frac{20}{3}\ca^2 \pi ^2 \h(1;z)-\frac{224}{3} \ca^2 \h(0,1;z)\nl
&+\frac{32}{3} \ca \nf \h(0,1;z)+8 \cf\nf \h(0,1;z)-112 \ca^2 \h(0,0,1;z)+16 \cf \nf \h(0,0,1;z)\nl
&+16 \ca^2 \h(-1,0;z)-16\ca^2 \h(-1,0,1;z)-66 \ca^2 \h(0,0;z)+12 \ca \nf \h(0,0;z)\nl
&+10 \cf \nf \h(0,0;z)-92\ca^2 \h(1,0;z)+8 \ca \nf \h(1,0;z)-4 \cf \nf \h(1,0;z)\nl
&-92 \ca^2 \h(1,1;z)+8\ca \nf \h(1,1;z)-4 \cf \nf \h(1,1;z)-56 \ca^2 \h(1,0,1;z)\nl
&-80 \ca^2 \h(0,1,0;z)+8\cf \nf \h(0,1,0;z)-88 \ca^2 \h(0,1,1;z)+8 \cf \nf \h(0,1,1;z)\nl
&+32 \ca^2 \h(-1,-1,0;z)-40\ca^2 \h(-1,0,0;z)-96 \ca^2 \h(0,0,0;z)\nl
&+20 \cf \nf \h(0,0,0;z)-40 \ca^2 \h(1,0,0;z)-56\ca^2 \h(1,1,0;z)-48 \ca^2 \h(1,1,1;z)\bigr)\nl
&+z^2 \bigl(\frac{5747 \ca^2}{27}-\frac{554 \ca \nf}{27}-\frac{496\cf \nf}{27}-\frac{242 \ca^2 \pi ^2}{9}+\frac{4}{3} \ca \nf \pi ^2+\frac{8}{9} \cf \nf \pi ^2\nl
&-96\ca^2 \zeta_{3}+\frac{16}{3}\ca^2 \pi ^2 \h(-1;z)+\frac{242}{9}\ca^2 \h(0;z)-\frac{176}{9} \ca \nf \h(0;z)-\frac{136}{9}\cf \nf \h(0;z)\nl
&-\frac{32}{3} \ca^2 \pi ^2 \h(0;z)+\frac{778}{9}\ca^2 \h(1;z)-\frac{176}{9}\ca \nf \h(1;z)-\frac{136}{9} \cf \nf \h(1;z)\nl
&-\frac{20}{3} \ca^2 \pi ^2 \h(1;z)+\frac{484}{3}\ca^2 \h(0,1;z)-8 \ca \nf \h(0,1;z)-\frac{16}{3} \cf \nf \h(0,1;z)\nl
&+64 \ca^2 \h(0,0,1;z)-48\ca^2 \h(0,-1,0;z)+\frac{88}{3} \ca^2 \h(-1,0;z)-16 \ca^2 \h(-1,0,1;z)\nl
&+\frac{572}{3} \ca^2\h(0,0;z)-8 \ca \nf \h(0,0;z)-\frac{16}{3} \cf \nf \h(0,0;z)+\frac{308}{3} \ca^2\h(1,0;z)\nl
&-8 \ca \nf \h(1,0;z)-\frac{16}{3} \cf \nf \h(1,0;z)+\frac{308}{3} \ca^2\h(1,1;z)-8 \ca \nf \h(1,1;z)\nl
&-\frac{16}{3} \cf \nf \h(1,1;z)+56 \ca^2 \h(1,0,1;z)+48\ca^2 \h(0,1,0;z)+56 \ca^2 \h(0,1,1;z)\nl
&+32 \ca^2 \h(-1,-1,0;z)-40 \ca^2 \h(-1,0,0;z)+40\ca^2 \h(0,0,0;z)+40 \ca^2 \h(1,0,0;z)\nl
&+56 \ca^2 \h(1,1,0;z)+48 \ca^2 \h(1,1,1;z)\bigr)\Bigr]
\,,
\end{align}

\begin{align}
c^{\l(2\r)}_{\phi,q}=&\Bigl[\frac{3086 \ca \cf}{27}+104 \cf^2-\frac{332 \cf \nf}{27}-\frac{176}{9}\ca\cf\pi^2-\frac{10\cf^2\pi^2}{3}+\frac{8}{9}\cf\nf\pi^2\nl
&-64\ca \cf \zeta_{3}-8 \cf^2 \zeta_{3}+\frac{8}{3} \ca \cf \pi ^2 \h(-1;z)+\frac{8}{3}\cf^2\pi^2\h(-1;z)-\frac{20}{3}\ca\cf\h(0;z)\nl
&+15 \cf^2 \h(0;z)-\frac{52}{3}\cf\nf\h(0;z)+\frac{8}{3} \ca \cf \pi ^2 \h(0;z)-\frac{16}{3} \cf^2 \pi ^2 \h(0;z)\nl
&+\frac{242}{9}\ca\cf\h(1;z)+56 \cf^2 \h(1;z)-\frac{116}{9} \cf \nf \h(1;z)-\frac{20}{3} \ca\cf\pi^2 \h(1;z)\nl
&+\frac{352}{3} \ca \cf \h(0,1;z)+20 \cf^2 \h(0,1;z)-\frac{16}{3}\cf\nf\h(0,1;z)-16 \ca \cf \h(0,0,1;z)\nl
&+32 \cf^2 \h(0,0,1;z)-16 \ca \cf \h(0,-1,0;z)-32\cf^2 \h(0,-1,0;z)\nl
&+32 \ca \cf \h(-1,0;z)-16 \cf^2 \h(-1,0;z)-16 \ca \cf \h(-1,0,1;z)\nl
&+\frac{536}{3}\ca \cf \h(0,0;z)-12 \cf^2 \h(0,0;z)-\frac{32}{3} \cf \nf \h(0,0;z)+\frac{184}{3}\ca\cf\h(1,0;z)\nl
&+32 \cf^2 \h(1,0;z)-\frac{16}{3} \cf \nf \h(1,0;z)+\frac{140}{3}\ca \cf \h(1,1;z)+44 \cf^2 \h(1,1;z)\nl
&-\frac{8}{3} \cf \nf \h(1,1;z)+40\ca\cf \h(1,0,1;z)+16 \cf^2 \h(1,0,1;z)+24 \cf^2 \h(0,1,0;z)\nl
&+8\ca\cf \h(0,1,1;z)+24\cf^2 \h(0,1,1;z)+32 \cf^2 \h(-1,-1,0;z)\nl
&-24 \ca \cf \h(-1,0,0;z)-16 \cf^2 \h(-1,0,0;z)-40\ca\cf\h(0,0,0;z)\nl
&+20 \cf^2 \h(0,0,0;z)+40 \ca \cf \h(1,0,0;z)+40\ca\cf\h(1,1,0;z)+16 \cf^2 \h(1,1,0;z)\nl
&+40 \ca \cf \h(1,1,1;z)+8 \cf^2 \h(1,1,1;z)+\frac{1}{z}\bigl(-\frac{13457 \ca \cf}{54}-\frac{59 \cf^2}{2}+\frac{737\cf \nf}{27}\nl
&+\frac{106}{9} \ca \cf \pi ^2+6 \cf^2 \pi ^2-\frac{8}{9} \cf \nf \pi ^2+72 \ca \cf \zeta_{3}-32 \cf^2 \zeta_{3}+\frac{8}{3}\ca\cf\pi^2 \h(-1;z)\nl
&+\frac{8}{3} \cf^2 \pi ^2 \h(-1;z)-\frac{454}{3} \ca \cf \h(0;z)-10\cf^2 \h(0;z)+\frac{64}{3} \cf \nf \h(0;z)\nl
&+\frac{16}{3} \ca \cf \pi ^2 \h(0;z)+\frac{8}{3}\cf^2 \pi ^2 \h(0;z)-32 \ca \cf \h(1;z)-64 \cf^2\h(1;z)\nl
&+\frac{116}{9} \cf\nf\h(1;z)+\frac{20}{3} \ca \cf \pi ^2 \h(1;z)-\frac{124}{3} \ca \cf \h(0,1;z)-36\cf^2\h(0,1;z)\nl
&+\frac{16}{3} \cf \nf \h(0,1;z)-32 \ca \cf \h(0,0,1;z)-16 \cf^2 \h(0,0,1;z)+\frac{88}{3}\ca\cf\h(-1,0;z)\nl
&-16 \ca \cf \h(-1,0,1;z)-\frac{212}{3} \ca \cf \h(0,0;z)+\frac{32}{3}\cf\nf\h(0,0;z)\nl
&-\frac{212}{3} \ca \cf \h(1,0;z)-36 \cf^2 \h(1,0;z)+\frac{16}{3}\cf \nf \h(1,0;z)-56 \ca \cf \h(1,1;z)\nl
&-48 \cf^2 \h(1,1;z)+\frac{8}{3}\cf\nf\h(1,1;z)-40 \ca \cf \h(1,0,1;z)-16 \cf^2 \h(1,0,1;z)\nl
&-32 \ca \cf \h(0,1,0;z)-16\cf^2 \h(0,1,0;z)-40\ca\cf \h(0,1,1;z)-16\cf^2 \h(0,1,1;z)\nl
&+32 \cf^2 \h(-1,-1,0;z)-24\ca \cf \h(-1,0,0;z)-16 \cf^2 \h(-1,0,0;z)\nl
&-16 \ca \cf \h(0,0,0;z)-40\ca\cf \h(1,0,0;z)-40 \ca \cf \h(1,1,0;z)\nl
&-16 \cf^2 \h(1,1,0;z)-40 \ca \cf \h(1,1,1;z)-8\cf^2 \h(1,1,1;z)\bigr)+z \bigl(\frac{38 \ca \cf}{27}-15\cf^2\nl
&+\frac{28 \cf \nf}{27}-\frac{14}{9} \ca \cf \pi ^2+\frac{4 \cf^2 \pi ^2}{3}-\frac{4}{9} \cf\nf\pi^2+44\ca\cf\zeta_{3}-12 \cf^2 \zeta_{3}\nl
&+\frac{4}{3} \ca \cf \pi ^2 \h(-1;z)+\frac{4}{3} \cf^2 \pi ^2 \h(-1;z)-57 \ca \cf \h(0;z)+18\cf^2 \h(0;z)\nl
&+\frac{22}{3}\cf\nf\h(0;z)+\frac{16}{3} \ca \cf \pi ^2 \h(0;z)+\frac{8}{3}\cf^2 \pi ^2 \h(0;z)-\frac{127}{9} \ca \cf \h(1;z)\nl
&+3 \cf^2 \h(1;z)+\frac{58}{9}\cf\nf \h(1;z)+\frac{10}{3} \ca \cf \pi ^2 \h(1;z)+\frac{64}{3} \ca \cf \h(0,1;z)\nl
&-24\cf^2 \h(0,1;z)+\frac{8}{3} \cf \nf \h(0,1;z)-32 \ca \cf \h(0,0,1;z)-16 \cf^2 \h(0,0,1;z)\nl
&+12\ca \cf \h(-1,0;z)-16 \cf^2 \h(-1,0;z)-8 \ca \cf \h(-1,0,1;z)+\frac{26}{3} \ca\cf \h(0,0;z)\nl
&+\cf^2 \h(0,0;z)+\frac{16}{3} \cf \nf \h(0,0;z)+\frac{4}{3} \ca \cf\h(1,0;z)-14 \cf^2 \h(1,0;z)\nl
&+\frac{8}{3} \cf \nf \h(1,0;z)+\frac{26}{3}\ca\cf\h(1,1;z)-24 \cf^2 \h(1,1;z)+\frac{4}{3} \cf \nf \h(1,1;z)\nl
&-20\ca\cf\h(1,0,1;z)-8\cf^2 \h(1,0,1;z)-24 \ca \cf \h(0,1,0;z)\nl
&-12 \cf^2 \h(0,1,0;z)-28 \ca \cf \h(0,1,1;z)-12\cf^2 \h(0,1,1;z)+16 \cf^2 \h(-1,-1,0;z)\nl
&-12 \ca \cf \h(-1,0,0;z)-8 \cf^2 \h(-1,0,0;z)-28\ca \cf \h(0,0,0;z)\nl
&-10 \cf^2 \h(0,0,0;z)-20 \ca \cf \h(1,0,0;z)-20 \ca \cf\h(1,1,0;z)-8 \cf^2 \h(1,1,0;z)\nl
&-20 \ca \cf \h(1,1,1;z)-4 \cf^2 \h(1,1,1;z)\bigr)+z^2 \bigl(\frac{608\ca \cf}{27}-\frac{8}{3} \ca \cf \pi ^2\nl
&-\frac{352}{9} \ca \cf \h(0;z)-\frac{176}{9}\ca\cf\h(1;z)+16 \ca \cf \h(0,1;z)+\frac{16}{3} \ca \cf \h(-1,0;z)\nl
&+\frac{64}{3}\ca\cf\h(0,0;z)+\frac{16}{3} \ca \cf \h(1,0;z)+\frac{16}{3} \ca \cf \h(1,1;z)\bigr)\Bigr]
\,.
\end{align}
They are in full agreement with~\cite{moch2}.

\end{document}